\documentclass[a4paper,12pt]{article}
\pdfoutput=1
\usepackage{amsfonts,amsmath,amssymb,youngtab}
\usepackage[paper=letterpaper,margin=1.0in]{geometry}
\usepackage{graphicx}
\usepackage{cite}
\usepackage{hyperref}

\newcommand{\be}{\begin{equation}}
\newcommand{\ee}{\end{equation}}
\newcommand{\bea}{\begin{eqnarray}\displaystyle}
\newcommand{\eea}{\end{eqnarray}}

\newcommand{\ba}{\begin{array}}
\newcommand{\ea}{\end{array}}
\newcommand{\ben}{\begin{enumerate}}
\newcommand{\een}{\end{enumerate}}
\newcommand{\bi}{\begin{itemize}}
\newcommand{\ei}{\end{itemize}}
\newcommand{\bc}{\begin{center}}
\newcommand{\ec}{\end{center}}
\newcommand{\bfig}{\begin{figure}}
\newcommand{\efig}{\end{figure}}
\newcommand{\bq}{\begin{quotation}}
\newcommand{\eq}{\end{quotation}}
\newcommand{\bt}{\begin{table}}
\newcommand{\et}{\end{table}}
\newcommand{\btab}{\begin{tabular}}
\newcommand{\etab}{\end{tabular}}
\newcommand{\bmi}{\begin{minipage}}
\newcommand{\emi}{\end{minipage}}
\newcommand{\bs}{\begin{slide}}
\newcommand{\es}{\end{slide}}

\def\wt{ \widetilde }

\newcommand{\pa}{\partial}
\newcommand{\pd}{\partial}

\def\L{ \Lambda } 

\def\tr{ {\rm tr } } 
\def\Tr{ {\rm Tr } } 
\def\Dim{ {\rm Dim} }
\def\Sym{ {\rm Sym} }
 
\def\d{ { \rm d}  } 
\def\End{ {\rm End} }
\def\Ker{ {\rm Ker} }
\def\Fun{ {\rm Fun} }
\def\Im{ {\rm Im} }

  \def\cF{{\cal F}}
\def\cG{{\cal G}} \def\cH{{\cal H}} \def\cI{{\cal I}}
  
\def\cM{{\cal M}} \def\cN{{\cal N}} \def\cO{{\cal O}}
\def\cP{{\cal P}}

 \def\cZ{{\cal Z}}

\newcommand{\mbC}{{\mathbb C}}

\newcommand{\mbR}{{\mathbb R}}
\newcommand{\mbP}{{\mathbb P}}

\newcommand{\mC}{ \mathbb{C} } 
\newcommand{\mR}{ \mathbb{R}  }
\newcommand{\mP}{ \mathbb{P} }

\newcommand{\CP}{\mathbb{CP}}
\newcommand{\bz}{\bar{z}}

\makeatletter
\@addtoreset{equation}{section}
\makeatother

\newcommand{\comment}[1]{}

\begin{document}
\rightline{QMUL-PH-11-17}

\vskip 1cm

\centerline{{\LARGE \bf  Quantum states to brane geometries   }}
 \centerline{{\LARGE \bf  via fuzzy moduli spaces of giant gravitons  }
} 
\medskip

\vspace{.4cm}

\centerline{   {\large \bf Jurgis Pasukonis}\footnote{j.pasukonis@qmul.ac.uk}
{ \bf  and }  {\large \bf Sanjaye Ramgoolam}\footnote{s.ramgoolam@qmul.ac.uk}  } 
\vspace*{4.0ex}
\begin{center}
{\large Department of Physics\\
Queen Mary, University of London\\
Mile End Road\\
London E1 4NS UK\\
}
\end{center}

\vspace*{5.0ex}

\centerline{\bf Abstract} \bigskip

Eighth-BPS local operators in $\cN=4$ SYM are dual to 
quantum states arising from the quantization of a moduli space of
giant gravitons in $AdS_5 \times S^5$. 
Earlier results on the quantization of this moduli space
give a Hilbert space of multiple harmonic oscillators in 3 dimensions. 
We use these  results, along with techniques from fuzzy geometry, 
to develop a map between quantum states and brane geometries. 
In particular there is a map between the oscillator states and points 
in a discretization of the base space in the toric fibration of 
the moduli space. We obtain a geometrical decomposition of 
the space of BPS states with labels  consisting of  $U(3)$ representations 
along with  $U(N)$ Young diagrams and associated group theoretic multiplicities. Factorization properties in the counting of BPS states  
lead to predictions for BPS world-volume excitations of 
specific brane geometries. Some of our results suggest an intriguing 
complementarity between localisation in the moduli space of branes  
and localisation in space-time.

\newpage

\tableofcontents

\newpage

\setcounter{footnote}{0}


\section{Introduction}

The study of BPS states in the non-planar regime of four-dimensional $\cN=4$, $U(N)$ Yang-Mills 
theory (SYM) has been a very rich area of research, allowing investigations of 
the AdS/CFT correspondence \cite{malda,gkp,witten} 
beyond the supergravity approximation. 
In the regime of energies of order $N$, 
the states correspond to D3 brane geometries in the bulk, also
known as ``giant gravitons'' \cite{mst}. 
For the case of half-BPS giant gravitons, the gauge theory duals 
were found in terms of operators associated with Young 
diagrams \cite{bbns,cjr}.  The operators are related to states by 
the operator-state correspondence in the radial quantization of 
the conformal field theory. An elegant description of the moduli space of 
eighth-BPS giant gravitons was found in terms of holomorphic surfaces 
in $\mC^3$ \cite{mikhailov}. The  construction of  gauge theory operators 
associated with this general class of giant graviton geometries 
has been a long-standing problem.

The correspondence between  SYM operators and brane geometries
for the half-BPS sector has been particularly illuminating, shedding 
light on the emergence of the $AdS_5 \times S^5$ background 
where the dual strings propagate.   
The giant gravitons in the half-BPS sector are systems of multiple spherical D3 branes, expanding either in $S^5$ (sphere giants) or in $AdS_5$ (AdS giants) \cite{hasitzh,myers}. The Schur basis for 
multi-trace  operators in the gauge theory constructed from one 
complex matrix are associated with 
Young diagrams of $U(N)$ (with no more than $N$ rows). 
They  offer natural candidates for duals of these brane geometries \cite{bbns} \cite{cjr}.  Operators with order $1$  long columns 
(length order $N$) correspond to sphere giants 
while those with order $1$ long rows (length order $N$) correspond to AdS 
giants. 
This basic picture  has been confirmed  by constructing 
modifications of the Schur operators, which correspond  to attaching 
strings to the branes \cite{Balasubramanian:2002sa, Aharony:2002nd, Berenstein:2003ah, bbfh, dMSSI,dMSSII,dMSSIII}. 
The prescription can be described in terms of 
``restricted Schur'' operators constructed using restrictions of 
symmetric group representations to their subgroups. It leads 
 to evidence of integrability 
beyond the setting of the usual planar limit 
\cite{npint1103,npint1106,rdgm}.  
 The open string excitations include vibrations of the branes, which have been studied from the world-volume perspective in \cite{Das:2000st}.

The sector of quarter-BPS or eighth-BPS states, annihilated by 8 or 4 supercharges respectively, is far less understood. The eighth-BPS brane geometries extending in $S^5$ were constructed in \cite{mikhailov}: the moduli space of these solutions is the moduli
space of intersections of a four-dimensional holomorphic 
surface in $\mbC^3$, with $S^5$. These configurations are more rich than those in half-BPS sector, containing intersecting branes, vibrating branes and other intricate surfaces. There has been significant 
progress in constructing quarter- or eighth-BPS operators in field theory 
\cite{bhrI,tomquarter,yusuke,pgp,Kimura:2011df}. However the problem of 
finding precise duals to these brane geometries is still unsolved.

Eighth-BPS operators in the $U(3)$ sector of SYM are constructed 
from holomorphic gauge invariant functions of three complex scalar matrices, 
subject to the condition that they are annihilated by the 
one-loop dilatation operator\footnote{There is a more general 
 class of eighth-BPS 
operators involving fermionic highest weights which form the $U(3|2)$ sector. 
In this paper we will always work in the $U(3)$ subsector and 
eighth-BPS or simply BPS will refer to this. If desired the considerations can be restricted to the $U(2)$ or quarter-BPS sector.}. It is in principle possible to do this systematically \cite{pgp} and calculate an orthogonal basis of eighth-BPS operators for any fixed charges in terms of representation theory. In practice, however, the procedure is computationally difficult due to calculation of Clebsch-Gordan coefficients, and only possible for charges of order $O(1)$. In order to study duals of brane geometries we need a basis at energies of $O(N)$ with large $N$, and no such basis has been explicitly constructed.

One description of the eighth-BPS sector in SYM is provided by the chiral ring, where the states built from three chiral scalars are identified up to F-terms. It was used in \cite{index} to calculate the exact spectrum of eighth-BPS states. The number of 
states as a function of the three $U(1)$ charges was found to agree with the counting of states of $N$ bosons in a 3D harmonic oscillator. The structure of the chiral ring is, however, not enough to calculate operator two-point functions 
which provide an inner product. The explicit construction of 
gauge invariant operators, which are annihilated by  the one-loop 
dilatation operator,  would allow the calculation of this inner product  
and  would help  find duals of the brane geometries.

 The eighth-BPS spectrum can be constructed either by quantizing the 
moduli space of giant gravitons large in the $AdS_5$ directions 
\cite{Mandal:2006tk} or alternatively by quantizing   \cite{beasley,minwalla}
the giants which are large in the $S^5$ directions \cite{mikhailov}.  
Our main interest in this paper will be the giants which are large in $S^5$
It was argued that the quantization of 
this moduli space is equivalent to the geometric quantization of 
a complex projective space. This physical moduli space is  
related to 
the moduli space of polynomials in three complex variables, also a projective space, through a non-trivial procedure. This procedure 
involves, among other things, the shrinking of holes 
in the  moduli space of polynomials associated with polynomials 
whose zero set does not intersect the $S^5$. 
The partition function over the resulting Hilbert space 
exactly matched the one counting elements of the chiral ring. 
This construction thus gives additional structure to the states in
the chiral ring: it maps them to states in a Hilbert space, which, 
furthermore, has structure carried over from the moduli space of branes.

In this paper, we initiate a systematic 
study of the correspondence between quantum states 
in the geometric quantization of the physical moduli space and 
explicit eighth-BPS brane geometries. 
This can be viewed as an intermediate step 
in  connecting quantum states associated to 
gauge theory operators (by the operator state correspondence) 
with the geometries. 

Let us state that our main focus is 
{ \it not }  the construction of an overcomplete basis 
of coherent states associated with arbitrary points on the moduli space,
but rather the association of subspaces of moduli space to a complete 
basis of orthogonal energy eigenstates. We expect that 
combining the constraints associated with orthogonality, completeness and 
symmetries can be a powerful guide in finding how  gauge theory 
local operators  map to branes in the dual space-time. 
In studying the map between geometric quantization states and geometries, 
we find that fuzzy geometry provides the ideal set of tools. 
A lot of work has been done on fuzzy projective spaces
with a view to modeling fuzzy space in string theory 
or with a view to regulating continuum field theories. 
We are able to draw on and apply this existing literature 
to clarify how the oscillator basis corresponds to the geometry of 
the moduli space of branes.

This allows us  to predict some physical properties of various brane configurations, such as the BPS open string spectrum:  
a check that has been crucial in the study of half-BPS states.
The understanding of this Hilbert space lets us make predictions about the structure of Hilbert space of BPS operators in the gauge theory. 
 In particular, we develop a new group-theoretic labelling of the states
which relies on the decomposition of the moduli space of 
giant gravitons according to the degree of the polynomials 
appearing in the Mikhailov description. 
From the AdS/CFT correspondence, this  
geometric and group theoretic labelling will apply equally to the 
gauge theory construction of operators and can be expected to 
provide a valuable guide in this construction.

We start the paper in Section~\ref{sec:review} by reviewing the eighth-BPS brane geometries found in \cite{mikhailov} and their quantization according to \cite{minwalla}.

In Section~\ref{sec:cp} we review some fuzzy geometry techniques and apply them to study the correspondence between Hilbert space and classical geometries.
Once we have a space of states $\cH$, 
we also have a space of operators forming the 
endomorphisms $ \End ( \cH ) $. In the case at hand, by considering the 
space of states coming from quantized projective spaces $ \mC \mP^{n} $, 
the corresponding algebra of operators can be identified with 
a fuzzy $ \mC \mP^{n} $, with fuzziness $1/N$.   
We will also recall the structure of  $ \mC \mP^{n} $ 
as a toric manifold with a $T^{n}$ fiber over a simplex in 
 $\mR^{n} $. 
By using fuzzy geometry constructions, we find that the  states 
are uniformly spread out along the tori but localized 
at points in the simplex.   
This makes contact with recent literature on fuzzy projective spaces 
as models of space in string theory \cite{bdlmo,abiy,gs99,saefuztor,vaidtriv}.
From this point of view, distinguished states lie at the corners (vertices)
of the simplex. 

In Section~\ref{complete-group-labels}, we use the connection between BPS states and 
quantization of projective spaces, to give a group theoretic 
labelling of the states in terms of $U(3)$ 
representations $ \Lambda $ along with a Young diagram $ Y $ of $U(N)$. 
States with specified $ \Lambda , Y $ are generically 
not unique, but all additional 
multiplicities are described in terms of other group theory data
such as Littlewood-Richardson numbers.  
This is one of our main new results. We discuss the geometry of the 
above labelling of states in terms of fibration structures of 
projective spaces and of the simplices in the base of their
toric fibrations.

We study the physics of the states at the corners of the base simplex 
in Section~\ref{sec:states2geom}. We find that 
the polynomial equation in the Mikhailov description 
of these giants is actually a monomial equation, simply setting to zero a 
monomial. For these corner states, the subtleties of the map between the 
physical moduli space and the moduli space of Mikhailov polynomials can be obviated by using 
symmetries. These branes share the property of being static with the familiar 
maximal giants of the half-BPS sector, although they are not the most
general static configurations, these being general homogeneous polynomials. 
We interpret these corner states described by monomial equations 
as composites of maximal half-BPS giants with angular momentum 
in different directions, along similar lines to \cite{BBJS}
 Hence we will refer to these corner states as 
maximal giants.   We consider the states  in the Hilbert space which are 
near  the states for these maximal 
brane geometries, and disentangle them into a tensor product 
of bulk closed string excitations and world-volume open string excitations. 
The spectrum of world-volume excitations is consistent with the 
interpretation of the brane as a composite of half-BPS maximal giants. 
It gives specific predictions which should be testable by construction of operators in the gauge theory or by world-volume calculations for branes in the bulk 
 space-time.

In Section~\ref{sec:single_partition}, we display the tensor product
structure of open and closed 
string excitations in the form of factorization properties of the 
partition function. A  very useful strategy in order to exhibit 
this in the simplest way is to focus on states which are near the 
stringy exclusion principle \cite{malstrom} cut-off, i.e states which exist 
in the Hilbert space for rank $N$, but not for rank $N -1$. 
The phenomenon of the stringy exclusion principle \cite{malstrom} 
and its explanation by the growth of a brane \cite{mst} 
is a remarkable example of how classical geometry explains 
the disappearance of specific quantum states as the rank of the 
gauge group is changed. This is in fact one of 
the key ingredients in the map between Young diagram operators
for half-BPS in the gauge theory  and brane geometries \cite{bbns,cjr}. 
It is therefore no surprise that the stringy exclusion principle continues 
to be illuminating in the eighth-BPS sector. 

In Section~\ref{sec:single_symplectic} we consider Hilbert space of ``nearby states''
directly by going to the Mikhailov's polynomials, without assuming 
that the global structure of the physical moduli space is given 
by projective spaces as argued by \cite{minwalla}. This is done by conducting 
a local analysis of the symplectic form near the 
points on the moduli space, corresponding to geometries of 
interest. We do the case of perturbations around a single giant graviton.  
We find agreement with the discussion in Sections~\ref{sec:states2geom}
and \ref{sec:single_partition}.

In Section~\ref{sec:non-maximal}, we extend the discussion of the correspondence between 
states and geometries beyond the maximal branes. This is substantially more 
subtle, but allows some geometrical understanding of the multiplicities 
encountered in the analysis of the world-volume excitations of 
the maximal branes.

In Section~\ref{sec:discussion}, we discuss the implications of our results 
for the construction of gauge theory operators. We observe that some of 
our results can be interpreted, at a qualitative level, 
in terms of a complementarity between localization in space-time 
and localization of branes in space-time. 
We also consider implications of the lessons we have learned 
for the broader discussion of states and geometries 
in the context of bulk deformations of AdS spacetime and black hole physics.


\section{Review of phase space and quantization}
\label{sec:review}

In this section we will review how the phase space of eighth-BPS Mikhailov's solutions in $AdS_5\times S^5$ is described by $\CP^n$. This phase space can be geometrically quantized to give a Hilbert space isomorphic to $N$ bosons in a three-dimensional harmonic oscillator. The material in this section is largely based on \cite{minwalla} and we refer the reader there for the more complete 
treatment. 

We first describe the moduli space of giant graviton solutions. 
The 3-brane action gives a symplectic form on this space, 
which gives it the structure of a phase space. We describe 
 the symplectic form  and then use the geometric quantization prescription \cite{woodhouse} to build the Hilbert space and operators. 

The starting point is the following construction by Mikhailov \cite{mikhailov}. We consider D3 branes wrapping surfaces $\Sigma\subset S^5$ in $AdS_5 \times S^5$ which preserve 1/8 of supersymmetries (eighth-BPS). Mikhailov showed that all such surfaces $\Sigma$ can be constructed by taking  holomorphic functions in $\mbC^3$
\begin{equation}
\label{eq:Pxyz_definition}
	P(x,y,z)=\sum_{n_1,n_2,n_3=0}^\infty c_{n_1,n_2,n_3}\,x^{n_1}y^{n_2}z^{n_3}
\end{equation}
and intersecting the four-dimensional surface $P(x,y,z)=0$ with the unit five-sphere $|x|^2+|y|^2+|z|^2=1$ embedded in $\mbC^3$. The intersection $\Sigma$ is generically a three-dimensional surface in $S^5$ on which we wrap the D3 brane. More precisely, the shape of the D3 brane is a time-dependent solution given by polynomial
\begin{equation}
\label{eq:Pxyz_timedep}
	P(e^{it}x,e^{it}y,e^{it}z)=\sum_{n_1,n_2,n_3=0}^\infty c_{n_1,n_2,n_3}\, e^{i(n_1+n_2+n_3)t} x^{n_1}y^{n_2}z^{n_3}
\end{equation}
That is, the time evolution keeps the shape of the D3 brane fixed, and it just rotates with a phase factor in all coordinates.

For the simplest example take the polynomial
\begin{equation}
	P(x,y,z) = c\,z - 1.
\end{equation}
Then the $P(x,y,z)=0$ surface is $z=1/c$ or time dependent $z(t)=e^{it}/c$ and intersection with $S^5$ is
\begin{equation}
\label{eq:cz_s3}
	|x|^2+|y|^2 = 1 - \frac{1}{|c|^2} .	
\end{equation}
This defines a $S^3\subset S^5$ with radius $r=\sqrt{1-1/|c|^2}$, which is the original half-BPS giant graviton of \cite{mst}.

We now analyze the phase space\footnote{Note that surface $\Sigma \subset S^5$ defines a point in \emph{phase space} rather than just configuration space, because it determines both position and velocity. This is a result of the BPS condition. See, for example, \cite{Das:2000st}.} $\cM$
of such eighth-BPS giants in $S^5$. Let us first consider the space $\cP$ of  holomorphic surfaces in $\mbC^3$ given by\footnote{We will often abbreviate $P(x,y,z)$ as $P(z)$, nevertheless, these are always polynomials of three complex coordinates.}
 $P(z)=0$. 
The points in $\cP$ are labelled by coefficients $\{c_{n_1,n_2,n_3}\}$. In fact, the coefficients are projective coordinates $\{c_{n_1,n_2,n_3}\} \sim \{\lambda c_{n_1,n_2,n_3}\}$, because multiplying them by a common factor $\lambda$ keeps the surface $P(z)=0$ unchanged. It is convenient to regularize the infinite-dimensional space $\cP$ by considering a \emph{finite}-dimensional subspace $\cP_C \subset \cP$ where only a subset $\{c_{n_1,n_2,n_3}\,|\,(n_1,n_2,n_3)\in C\}$ of coefficients are allowed to be non-zero. If $n_C$ is the number of elements in $C$, then we get a space spanned by $n_C$ complex projective coordinates, that is, topologically $\cP_C=\CP^{n_C-1}$. For example, we could take $C=\{(0,0,0),\,(1,0,0),\,(0,1,0),\,(0,0,1)\}$ for which $\cP_C$ is the space of linear polynomials (see (\ref{eq:P_linear}) in the next section), topologically $\CP^3$. In the end the full $\cP$ can be defined as a limit $\cP = \lim_{d\rightarrow\infty}\cP_{C_d}$, where $C_d$ is a sequence which includes ever more monomials $C_{d}\subset C_{d+1}$. For example, $C_{d}$ could be all coefficients that multiply monomials of degree up to $d$. The important aspect of this construction is that at every step we are dealing with a complex projective space $\CP^{n_C-1}$. The limiting case is $\cP = \CP^\infty$.

Next, the intersection of each $P(z)=0$ with $S^5$ is a surface $\Sigma(P) \subset S^5$ which defines the shape of a D3 brane and therefore labels a point in the phase space $\cM$. That is, there is a map
\begin{equation}
	\cP \rightarrow \cM
\end{equation}
The regularized subspace $\cP_C$ is mapped to $\cM_C$, which is a finite-dimensional subspace of $\cM$. It is argued that $\cM_C$ is also $\CP^{n_C-1}$. One problem that has to be dealt with is that the map is many-to-one, that is, different polynomials $P(z)=0$ can lead to the same intersection $\Sigma$. In fact, it was shown in \cite{minwalla} that two polynomials $P_1(z)$ and $P_2(z)$ have the same intersection with $S^5$ if and only if
\begin{equation}
	P_1(z) = p(z)r_1(z), \quad P_2(z) = p(z)r_2(z)
\end{equation}
where $r_1(z)=0$ and $r_2(z)=0$ do not intersect $S^5$. Therefore, in order to get the space $\cM_C$ from $\cP_C$, we need to identify
\begin{equation}
	P(z) \sim P(z)r(z)
\end{equation}
with any $r(z)$ that does not intersect $S^5$. Note that all polynomials $r(z)$ that do not intersect $S^5$ are themselves identified with a single polynomial $P(z)=1$, which is the vacuum point ($\Sigma=\emptyset$) in the phase space. It was also shown in \cite{minwalla} that these identifications can be performed smoothly and the resulting space $\cM_{C}$ is indeed still $\CP^{n_C-1}$. Let us denote the projective coordinates on $\cM_C$ by $\{w_{n_1,n_2,n_3}\}$, with indices running over the same set $C$. The map $\cP_C \rightarrow \cM_C$ then takes the form of functions
\begin{equation}
	w_{n_1,n_2,n_3} = w_{n_1,n_2,n_3}(c,\bar{c})
\end{equation}
They should be such that $w_{n_1,n_2,n_3}(c,\bar{c})=w_{n_1,n_2,n_3}(c',\bar{c}')$ whenever points $c_{n_1,n_2,n_3}$ and $c'_{n_1,n_2,n_3}$ should be identified.

Let us now turn to the discussion of the symplectic form on the phase space $\cM_C$, which is necessary for quantization. The starting point is the world-volume action on a single D3 brane with no world-volume field strength or fermions:
\begin{equation}
\label{eq:S_defined}
	S=S_{\rm BI} + S_{\rm WZ} =
	\frac{1}{(2\pi)^3(\alpha')^2g_s} \int_{\Sigma} \d^4 \sigma \sqrt{-\tilde{g}}
	+\int_{\Sigma} A
\end{equation}
Here $\tilde{g}$ is the induced metric, and $A$ is the four-form background gauge field, such that field strength $F=\d A$ is proportional to $S^5$ volume form. The symplectic form can then be written as
\begin{equation}
\label{eq:omega_defined}
\begin{split}
	\omega &= \int_\Sigma \d^3 \sigma \,
	\delta \left(\frac{\delta S}{\delta \dot{x}^\mu}\right)
	\wedge
	\delta x^\mu
	\\
	&= \frac{N}{2\pi^2} \int_\Sigma \d^3 \sigma \,
		\delta\left(
			\sqrt{-g}g^{0\alpha} \frac{\pd x^\nu}{\pd \sigma^\alpha}
			G_{\mu\nu}
		\right) \wedge \delta x^\mu
	+
		\frac{2N}{\pi^2} \int_\Sigma \d^3 \sigma
		\frac{\delta x^\lambda \wedge \delta x^\mu}{2}
		\left( 
			\frac{\pd x^\nu}{\pd \sigma^1}
			\frac{\pd x^\rho}{\pd \sigma^2}
			\frac{\pd x^\sigma}{\pd \sigma^3}
		\right)
		\epsilon_{\lambda \mu \nu \rho \sigma}
\end{split}	
\end{equation}
Now the metric $G_{\mu\nu}$ and the induced metric $g_{\alpha\beta}=G_{\mu\nu}\partial_\alpha{x^\mu}\partial_\beta{x^\nu}$ is taken on a unit radius $S^5$ ($g$ is related to $\tilde{g}$ by rescaling). This symplectic form is defined on the phase space $\cM_{\rm full}$ of \emph{all} configurations of a D3 brane, supersymmetric or not. Space $\cM_{\rm full}$ is, of course, much larger than the supersymmetric subspace $\cM\subset \cM_{\rm full}$. The ``coordinates'' on $\cM_{\rm full}$ are fields $\{x^\mu(\sigma), \dot{x}^\mu(\sigma)\}$, whereas $\cM$ is parametrized by ``collective coordinates'' $\{w_{n_1,n_2,n_3}\}$. In any case, we have a map $\cM_C \rightarrow\cM\rightarrow \cM_{\rm full}$ and the pullback of (\ref{eq:omega_defined}) defines a symplectic form on $\cM$ or $\cM_C$. In fact, since we have a map $\cP_C\rightarrow\cM_C$ we can also take a pullback of $\omega$ on the space of holomorphic polynomials $\cP_C$. This pullback will inevitably be degenerate and have singularities, but it can nevertheless be convenient for explicit calculations. 

Crucially, it was argued in \cite{minwalla}, that not only $\cM_C$ is topologically $\CP^{n_C-1}$, but also that the symplectic form $\omega$ is globally well defined, closed, and in the same cohomology class as $2\pi N \omega_{FS}$.
This implies it is always possible to find such coordinates $w_{n_1,n_2,n_3}$ that the pullback of (\ref{eq:omega_defined}) becomes \emph{proportional to the Fubini-Study form}, with coefficient $2\pi N$:
\begin{equation}
\label{eq:omega_fs}
\begin{split}
	\omega &= 2\pi N \omega_{FS} =  2N \left[
		\frac{1}{1+|w|^2} 
		\frac{\d \bar w_I \wedge \d w_I}{2i}
		-
		\frac{1}{(1+|w|^2)^2}
		\frac{w_I \bar w_J \, \d \bar w_I \wedge \d w_J}{2i}
	\right] 
\end{split}
\end{equation}
 Here $|w|^2\equiv w_I \bar{w}_I$ and we use shorthand $w_I$ for \emph{inhomogeneous} coordinates on $\CP^{n_C-1}$. For example in the patch $w_{0,0,0}=1$ the index $I$ runs over $n_C-1$ remaining $(n_1,n_2,n_3)$ tuples in $C$. 

Once we have the phase space manifold as $\CP^{n_C-1}$ with Fubini-Study form as the symplectic form, the geometric quantization is well known. The Hilbert space $\cH_C$ is spanned by wavefunctions, which are holomorphic polynomials of the $n_C$ projective coordinates $w_{n_1,n_2,n_3}$ of degree $N$
\begin{equation}
\label{eq:HC_definition}
	\cH_C = \left\{ \prod_{(n_1,n_2,n_3)\in C} (w_{n_1,n_2,n_3})^{k_{n_1,n_2,n_3}}\; \left| \; \sum k_{n_1,n_2,n_3} = N \right.\right\}
\end{equation}
or equivalently polynomials of $n_C-1$ inhomogeneous coordinates $w_{n_1,n_2,n_3}$ of degree up to $N$ (if we take e.g. $w_{0,0,0}=1$ in the patch). It is important to note how $N$ enters the definition of Hilbert space purely through setting the scale of $\omega$, which controls the effective Planck constant $1/N$ or the area in phase space that a single quantum state occupies. As we increase $N$, the area occupied by a state decreases, and we get more states in $\cH_C$.

Finally, we need to discuss the conserved charges in the system. There is a natural $U(3)$ symmetry acting on the coordinates $(x,y,z)$ which preserves the shape of $\Sigma$. The Cartan subgroup $U(1)^3$ rotating each coordinate by a phase will give three commuting charges $L_i$ that we can use to label the states. The action $(x,y,z)\rightarrow(e^{i\alpha_1}x,e^{i\alpha_2}y,e^{i\alpha_3}z)$ induces transformation
\begin{equation}
	c_{n_1,n_2,n_3} \rightarrow e^{i n_1 \alpha_1} e^{i n_2 \alpha_2} e^{i n_3 \alpha_3} c_{n_1,n_2,n_3}
\end{equation}
on $\cP_C$, as seen from (\ref{eq:Pxyz_definition}). Now we also need to use the fact argued in \cite{minwalla} that the map  $c_{n_1,n_2,n_3} \rightarrow w_{n_1,n_2,n_3}$ can be done in a $U(3)$ invariant way, so that the action on the final $\cM_C\sim \CP^{n_C-1}$ phase space coordinates is also $w_{n_1,n_2,n_3} \rightarrow e^{i n_1 \alpha_1} e^{i n_2 \alpha_2} e^{i n_3 \alpha_3} w_{n_1,n_2,n_3}$.
That means we have three vector fields on $\cM_C$ generated by $L_i$
\begin{equation}
	V_{L_i} = \sum_{n_1,n_2,n_3} i\,n_i w_{n_1,n_2,n_3} \partial_{n_1,n_2,n_3} - i\,n_i \bar{w}_{n_1,n_2,n_3} \bar\partial_{n_1,n_2,n_3}
\end{equation}
We have used the abbreviation 
\begin{equation}
\begin{split} 
 & \partial_{n_1,n_2,n_3} \equiv { \partial \over \partial w_{n_1, n_2, n_3}  } \\ 
 & \bar \partial_{n_1,n_2,n_3} \equiv { \partial \over \partial \bar w_{n_1, n_2, n_3} }  
\end{split}
\end{equation}
Upon geometric quantization these become operators on the Hilbert space
\begin{equation}
\label{eq:Li_definition}
	\hat{L}_i = \sum_{n_1,n_2,n_3} n_i w_{n_1,n_2,n_3} \partial_{n_1,n_2,n_3}
\end{equation}
So that the charge of each excitation $w_{n_1,n_2,n_3}$ is simply $n_i$ under each of the respective $U(1)$, and the total charge of a state $\Psi$ is the sum of all excitation charges. Note that the time evolution is given by an overall $U(1)$, generated by Hamiltonian
\begin{equation}
	\hat{H} = \hat{L}_1 + \hat{L}_2 + \hat{L}_3
\end{equation}
This also reflects the BPS condition. Given the charge assignments we can write a partition function over the Hilbert space (\ref{eq:HC_definition})
\begin{equation}
\label{eq:ZC_defined}
	\cZ_C(x_1,x_2,x_3) = \Tr_{\cH_C}\left( x_1^{L_1}x_2^{L_2}x_3^{L_3} \right)
	= \left[
	\prod_{n_1,n_2,n_3\in C} \frac{1}{1-\nu x_1^{n_1}x_2^{n_2}x_3^{n_3}}
	\right]_{\nu^N}
\end{equation}
The notation $[\ldots]_{\nu^N}$ denotes the coefficient of $\nu^N$, which enforces the degree $N$ of wavefunction. This matches the partition function over the chiral ring in $\cN=4$, and so reproduces the correct supersymmetric spectrum from quantizing giant gravitons.

A comment needs to be made on the validity of the D3 world-volume action (\ref{eq:S_defined}). It certainly is a good description for large branes of energy $O(N)$, but not for small ones with high curvature. However, the spectrum of BPS gravitons at energies $O(1)$ is still correctly reproduced by $\cH_C$ derived from $\omega$, and that part of the spectrum comes precisely from very small D3 branes, where $\omega$ should not be valid. This may be a result of the fact that the full symplectic form, corrected for small branes, is still in the same cohomology class as $\omega$ and also $U(3)$ invariant. 

\subsection{Example: single half-BPS giant}\label{sec:ex-sphere}

In order to illustrate various concepts in this section, let us quickly go through an example of linear polynomials. It will also serve as a starting point for further calculations in this paper. Take $C=\{(0,0,0),\,(1,0,0),\,(0,1,0),\,(0,0,1)\}$, then $\cP_C\sim \CP^3$ is the space of hyperplanes
\begin{equation}
\label{eq:P_linear}
	P(z) = c_{1,0,0}\,x + c_{0,1,0}\,y + c_{0,0,1}\,z + c_{0,0,0} = 0 .
\end{equation}
We abbreviate $c_0=c_{0,0,0},\,c_1=c_{1,0,0},\,c_2=c_{0,1,0},\,c_3=c_{0,0,1}$. For inhomogeneous coordinates we set $c_0=1$.

Intersection with $S^5$ yields an $S^3$ of radius
\begin{equation}
	r = \sqrt{1 - \frac{1}{|c_1|^2+|c_2|^2+|c_3|^2}} \equiv \sqrt{1-\frac{1}{|c|^2}}
\end{equation}
-- the same as in (\ref{eq:cz_s3}), only $U(3)$-rotated. The energy and momenta of this solution are:  
\begin{equation}
\label{eq:EL_sphere}
	E = N\frac{|c|^2-1}{|c|^2}, \quad L_i = N |c_i|^2 \frac{|c|^2-1}{|c|^4}
\end{equation}
As typical, the map $\cP_C\rightarrow \cM_C$ is not one-to-one, the region $|c|^2\le 1$ does not intersect $S^5$ and so maps to a single point: the vacuum. Good coordinates on $\cM_C$ as $\CP^3$ can be constructed by rescaling:
\begin{equation}  
	w_i = 
	\begin{cases}
 	 \sqrt{\frac{|c|^2-1}{|c|^2}} c_i & \text{if $|c|^2\ge1$ } \\
 	 0 & \text{if $|c|^2\le1$ }
 \end{cases}	 
\end{equation}
As explained in detail in \cite{minwalla}, this smoothly contracts ``the hole'' at $|c|^2\le1$ to a point $w_i=0$.

The symplectic form, which in this case can be calculated explicitly using (\ref{eq:omega_defined}), takes the following form in $c_i$ coordinates:
\begin{equation}
\label{eq:w_ci}
	\omega = 2N \left[ 
		\left( \frac{1}{|c|^2} - \frac{1}{|c|^4} \right)
		\frac{\d \bar c_i \wedge \d c_i}{2i}
		-
		\left( \frac{1}{|c|^4} - \frac{2}{|c|^6} \right)
		\frac{c_i \bar c_j \, \d \bar c_i \wedge \d c_j}{2i}
	\right]
\end{equation}
as long as $|c|^2\ge1$. In $w_i$ coordinates this becomes
\begin{equation}
\label{eq:omega_wi_sphere}
	\omega = 2N \left[
		\frac{1}{1+|w|^2} 
		\frac{\d \bar w_i \wedge \d w_i}{2i}
		-
		\frac{1}{(1+|w|^2)^2} 
		\frac{w_i \bar w_j \, \d \bar w_i \wedge \d w_j}{2i}
	\right],
\end{equation}
precisely $2\pi N$ times Fubini-Study form on $\mbC\mbP^3$, with perfectly good behavior at $|c|^2=1\,\sim\,|w|^2=0$. 

This $\CP^3$ can now be geometrically quantized to a Hilbert space $\cH_C$ of wavefunctions
\begin{equation}
\label{eq:psi_wi}
	\Psi_{k_1,k_2,k_3} = (w_1)^{k_1} (w_2)^{k_2} (w_3)^{k_3} (w_0)^{N-k_1-k_2-k_3}
\end{equation}
or in terms of only inhomogeneous coordinates (setting $w_0=1$)
\begin{equation}
	\Psi_{k_1,k_2,k_3} = (w_1)^{k_1} (w_2)^{k_2} (w_3)^{k_3}, \quad \sum k_i \le N
\end{equation}
The momenta are\footnote{
Remember (\ref{eq:Li_definition}) e.g. $L_1 = \sum n_1 w_{n_1,n_2,n_3}\partial_{n_1,n_2,n_3} = w_{1,0,0}\partial_{1,0,0} \equiv w_1\partial_1$}
\begin{equation}
	\hat{L}_i\,\Psi_{k_1,k_2,k_3} = k_i\,\Psi_{k_1,k_2,k_3}
\end{equation}
and total energy
\begin{equation}
	\hat{E} = k_1 + k_2 + k_3
\end{equation}
Note the maximum energy of a state in this $\cH_C$ is $E=N$, that of a maximal sphere giant, corresponding to $c_i \rightarrow \infty$ in (\ref{eq:EL_sphere}).

Finally, let us emphasize one point which will be important later on: (\ref{eq:omega_wi_sphere}) is written in inhomogeneous coordinates where $w_1=w_2=w_3=0$ corresponds to the vacuum point with $E=0$. But we can equally well take a different coordinate patch in $\CP^3$, for example where $w_3=1$ and $(w_0,w_1,w_2)$ parametrize the point. The new inhomogeneous coordinates are expressed in terms of the old ones as
\begin{equation}
	w'_0 = \frac{1}{w_3}, \quad w'_1 = \frac{w_1}{w_3}, \quad w'_2 = \frac{w_2}{w_3} 
\end{equation}
The symplectic form (\ref{eq:omega_wi_sphere}) has the same form in terms of $(w'_0,w'_1,w'_2)$.  But now the point $w'_0=w'_1=w'_2=0$ corresponds to $w_3\rightarrow \infty$, $c_3\rightarrow\infty$, which is the maximal giant arising from polynomial
\begin{equation}
	P(z) = z = 0
\end{equation}
We can choose to write the wavefunctions in terms of these coordinates
\begin{equation}
	\Psi'_{k_0',k_1',k_2'} = (w'_0)^{k'_0} (w'_1)^{k'_1} (w'_2)^{k'_2} 
\end{equation}
which is, of course, still the same Hilbert space as in (\ref{eq:psi_wi}),
 with a map  $\Psi'_{k_0',k_1',k_2'} \rightarrow  \Psi_{k_1=k_1',k_2=k_2',k_3=N-k_0'-k_1'-k_2'}$. One difference, though, is that now the vacuum $\Psi'_{0,0,0}=1$ has $E=L_3=N$, and the excitations can have negative charges:
\begin{equation}
\begin{split}
	\hat{L}_1 = w'_1 \partial'_1, \quad
	\hat{L}_2 = w'_2 \partial'_2, \quad
	\hat{L}_3 = - w'_0 \partial'_0 - w'_1 \partial'_1 - w'_2 \partial'_2, \quad\quad
	\hat{H} = - w'_0 \partial'_0
\end{split}
\end{equation}
Physically $w'_1$ and $w'_2$ quanta keep the giant energy the same, just rotate it in $U(3)$, while $w'_0$ takes the giant away from maximal by decreasing energy and $L_3$.


\section{Fuzzy \texorpdfstring{$\CP$}{CP} and giants as points on a simplex }
\label{sec:cp}

The Hilbert space $\cH_C$ arising from geometric quantization of $\CP^{n_C-1}$ 
is closely related to  \emph{fuzzy} or non-commutative $\CP_N^{n_C-1}$. 
We will review this relation and  use it to show how 
the holomorphic basis of the Hilbert space, is related to 
a discretization of the base in a description of 
  $\CP_N^{n_C-1}$  as a toric fibration  over a simplex in $\mR^{ n_C-1}$ \cite{bdlmo,abiy,gs99,saefuztor,vaidtriv,heckver,furok}. 
The wavefunctions are localized at points on the toric base
and spread out in the torus fibers.
 With the fuzzy $\CP$ 
technology in hand, we demonstrate this nice geometrical character 
of the states,  using 
elementary calculations of expectation values of $SU(n_C)$ 
Lie algebra elements and their products, evaluated on  states of $\cH_C$. 
From this point of view, we find that the corners of
the simplex, where the tori degenerate, correspond to distinguished 
states. We will return to these states in section (\ref{sec:states2geom}). 
We will show that  they  correspond to maximal giants, where
the Mikhailov polynomials become monomials.

In the case of half-BPS giant gravitons, 
this will allow us to relate the Young diagram labels
 which arise in the construction of 
corresponding operators in the dual SYM theory, to the coordinates of 
points in the discretized toric base, which as we will explain 
is a simplex in $\mbR^{n_C-1}$.  The Young diagram labels 
have a physical interpretation in terms of brane multiplicities 
for branes of different angular momenta.

This shows that  fuzzy geometry can be a powerful tool in 
providing a precise connection between quantum states  and localization, 
with its complementary non-locality due to quantum uncertainty, 
in the moduli space of solutions. 

The geometry of the discretized 
simplex will also play an important role in subsequent sections, 
where we will  develop the states-geometries connection 
further.

Going beyond the half-BPS sector to the eighth-BPS sector, 
a complete group  theoretic basis 
including $U(3)$ representation  labels, 
along with $U(N)$ Young diagram labels, will be developed in Section 
\ref{complete-group-labels}.

\subsection{Fuzzy \texorpdfstring{$\CP$}{CP}  from  operators on  Hilbert space of giant states}\label{opsandfuzzyCP}  

The homogeneous coordinates for projective space $\mbC \mbP^{ n_C-1}$
are $W_0 , W_1 , \cdots , W_{n_C-1}$. A complete basis for rational 
functions is provided by 
\bea 
{ 
W_{I_1} W_{I_2} \cdots W_{I_n} \bar W_{J_1}  
\bar W_{J_2} \cdots   \bar W_{J_n} \over |W|^{2n}  } 
\eea 
where $|W|^2 = \sum_{ i=0}^{n_C-1} W_I\bar W_I $.  The denominator 
ensures that these functions are invariant under  scaling by a complex number 
$W_{ I } \rightarrow \lambda W_I$. These functions span the function space for 
$\mbC \mbP^{n_C-1} $, which we will denote as  $ \Fun ( \mbC \mbP^{n_C-1} )$. 
This  decomposes into irreducible representations of   $SU(n_C)$ as 
\bea 
\Fun ( \mbC \mbP^{n_C-1} ) = \bigoplus_{n=0 }^{\infty}  \cF_{ n , \bar n   }  
\eea
where $ \cF_{ n , \bar n   } $ transforms as an irreducible representation corresponding to 
the Young diagram with $n$ columns of length $1$ and $n $ columns of 
length of $n_C-1$, which we denote as $[n,\bar n] $. 

As we reviewed in Section~\ref{sec:review}, the 
geometric quantization of giant gravitons for $AdS_5 \times S^5$ with 
$N$ units of flux, in a sector of polynomials of dimension $n_C$, 
leads to a quantization of the moduli space $\mbC \mbP^{ n_C-1}$ which 
produces a Hilbert space of holomorphic polynomials of degree $N$. This Hilbert space 
$\cH_C$ consists of polynomials of degree $N$ in the homogeneous coordinates 
$W_0, \cdots , W_{n_C-1}$. It can be viewed as the $N$-fold symmetric 
tensor product of the fundamental $V_{n_C}$ of $SU(n_C)$
\begin{equation}
	\cH_C = \Sym^N(V_{n_C})
\end{equation}
This is also isomorphic to a Hilbert space of oscillators 
\bea 
\label{eq:sunc_basis}
( a_{ n_C-1}^{\dagger} )^{n_{n_C-1} }   \cdots ( a_{2}^{\dagger})^{n_2}   ( a_{1}^{\dagger} )^{ n_1 }  ( a_{0}^{\dagger} )^{n_0}  | 0 \rangle 
\eea
with the constraint $ n_0 + n_1 + n_2 + \cdots + n_{ n_C -1 } = N $. 
Note also that this counting of oscillator states is equivalent to 
counting of Young diagrams of $U(N)$. The oscillator states 
characterized by $n_i$ can be mapped to the Young diagrams 
with $n_i$ rows of length $i$.

The dimension is 
\bea 
\Dim(\cH_C)=\binom{ N + n_C - 1}{N}
\eea 
Given this Hilbert space, it is natural to consider the algebra of 
operators, i.e the endomorphism algebra $ \End (\cH_C ) $.  
The decomposition into representations of $SU(n_C)$ is 
\begin{equation}\begin{split} 
 \End (\cH_C ) & =  \Sym^N( V_{n_C} ) \otimes  \Sym^N(\overline V_{n_C}) \\
              & =  \bigoplus_{n=0 }^{ N }  V_{ n , \bar n   }
\end{split}\end{equation}
where $V_{n,\bar n}$ transforms as the irreducible representation $[n,\bar n]$ described above. 
A basis for  $\End (\cH_C )$ is given by operators 
\bea 
  W_{I_1}\cdots W_{I_n}  \pa_{W_{J_1} } \cdots \pa_{W_{J_n} }
\eea
or in oscillator language 
\bea\label{oscillops}  
    a_{I_1}^{\dagger}   \cdots  a_{I_n}^{\dagger}     a_{J_1 }  \cdots 
 a_{ J_n }   
\eea
The indices on the oscillators range from $0$ to $n_C-1$, 
and 
\bea\label{heis}  
 [  a_{I} , a^{\dagger}_{ J } ] = \delta_{ I J } 
\eea

It is clear that the operators  in (\ref{oscillops}) 
  have a cut-off at $n=N$, since 
polynomials of degree $N$ will be annihilated by more than $N$ derivatives. 
The matrix algebra $\End ( \cH_C ) $ provides a finite dimensional 
approximation to $ \Fun ( \mbC \mbP^{n_C-1} ) $ with an $SU(n_C)$
invariant cutoff at $N$. The case $n_C =3$ has been much studied 
as a model for fuzzy brane world-volumes, fuzzy spatial directions for
Kaluza-Klein reduction in F-theory and elsewhere and as a model 
for quantum field theory with a UV cutoff that preserves spatial symmetries. 
Here we are finding  the fuzzy $\mC \mP$ in the set-up of quantizing a 
moduli space of giant gravitons.

The algebra $ \End ( \cH_{n_C} ) $  is generated by operators 
\bea\label{sugenw} 
E_{IJ} = W_{I} \pa_{W_J} 
\eea
or in oscillator language
\bea\label{sugena}  
E_{IJ} = a_{I}^{\dagger} a_J   
\eea
These form a basis for the  algebra of $SU(n_C) \oplus U(1) $ 
\bea
[  E_{IJ} ,  E_{KL} ] = \delta_{JK}  E_{IL} - 
 \delta_{IL}  E_{JK}
\eea
The traceless generators 
\bea 
 \tilde E_{IJ} = E_{IJ } - \delta_{ IJ} { N \over n_C } 
\eea
form the Lie algebra of $SU(n_C)$.

 Using (\ref{heis}) along with the constraint 
\bea 
\sum_I  a_I^{\dagger}  a_I = N ~~ \hbox{ in } \cH_{ n_C } 
\eea
we may also obtain the relations 
\begin{equation}\begin{split} 
&  E_{IJ }  E_{JK } = ( N + n_C -1 ) \hat E_{IK} \\
& E_{IJ }  E_{JI } = N ( N + n_C -1  ) \\
& \tilde E_{IJ} \tilde E_{JI} =  E_{IJ }  E_{JI } - { N^2 \over n_C } 
\end{split}\end{equation}
We also have 
\bea 
 E_{IJ }  E_{KL } =  a_{I}^{\dagger}  a_J^{\dagger} 
 a_K  a_L  + \delta_{ J K }  E_{IL }
\eea

If we define 
\begin{equation}\begin{split} 
  e_{IJ} &= { 1 \over N }  E_{IJ }\\
 e_{IJ;  KL } &= { 1 \over N^2 }   a^{\dagger}_{I} a^{\dagger}_J
 a_K  a_{L } \\
 e_{I_1 \cdots I_n ; J_1 \cdots J_n } &= { 1 \over N^n} 
     a^{\dagger}_{I_1} \cdots a^{\dagger}_{I_n} 
a_{J_1}  \cdots  a_{J_n} 
\end{split}\end{equation}
we have relations 
\begin{equation}\begin{split} 
[ e_{IJ} , e_{KL} ]  
& = {1 \over N } ( \delta_{JK} e_{IL} - \delta_{IL} e_{JK} ) \\
e_{IJ} e_{JI} & = 1 + { (n_C-1)\over N }  \\ 
e_{IJ} e_{KL} & = e_{IJ ; KL } + { \delta_{ JK } \over N } e_{IL} 
\end{split}\end{equation}
These relations simplify at large $N$. The $e_{IJ}$ generate a commutative algebra. We have 
\begin{equation}\begin{split} 
\label{relsees}  
[e_{IJ} , e_{ KL } ] & =  0 \\
e_{IJ} e_{JI} & =1 \\
e_{IJ} e_{KL} & = e_{IJ ; KL }  = e_{IL} e_{JK} 
\end{split}\end{equation}
There is a homomorphism from these generators of $\End ( \cH_{ C } ) $ 
to $\Fun ( \mbC \mbP^{n_C-1} ) $ given by 
\bea\label{homomorph}  
e_{I_1 \cdots I_n ; J_1 \cdots J_n } \leftrightarrow  { 1 \over N^n } 
{ W_{I_1} \cdots W_{I_n} \bar W_{J_1} \cdots \bar W_{J_n} \over |W|^{2n} } 
\eea
where  $|W|^2 = \sum_{ I =0 }^{ n_C -1 }  W_I  \bar W_{I}$. The homomorphism 
property is easily established by verifying that the functions on the RHS
of (\ref{homomorph}) obey relations (\ref{relsees}). 
At finite $N$, the algebra $ \End(\cH_C)$ is a fuzzy deformation of 
$\Fun ( \mbC \mbP^{n_C-1} ) $. This can be made precise by using the
map to define a star product on the classical algebra 
\cite{bdlmo} \cite{furok} \cite{abiy}.

\subsubsection{Toric geometry of \texorpdfstring{$\CP$}{CP} from the  Lie algebra embedding} 

The coordinates $e_{IJ}$ give a description of $\mbC \mbP^{n_C -1}$ 
as embedded in $\mbR^{ n^2_C -1 } \subset \mbR^{ n^2_C}$ which is the Lie algebra  of $SU(n_C) \subset U(n_C) $.  This is an example of a general construction of co-adjoint orbits 
\cite{bdlmo}. Another aspect of the geometry of  $\mbC \mbP^{n_C -1}$
will be of interest to us, namely the fact that it is 
a toric variety.  Let us describe this in the cases of 
$\mbC \mbP^2 $, which generalizes to the general $n_C > 3 $ case. 

Given the homogeneous coordinates $W_I $, we can impose the 
equivalence $ W_{I} \sim \lambda W_{I} $ by first setting 
\bea 
W_{I} \bar W_I = 1 
\eea 
and then modding out by a phase $ W_{I} \sim e^{ i \theta }  W_{I} $. 
This shows that $\mbC \mbP^{n_C-1}$ is the base space of a 
fibration of $ S^{2n_C -1 }$ with $S^1$ fiber.

\begin{figure}[h]
\begin{center}
\includegraphics{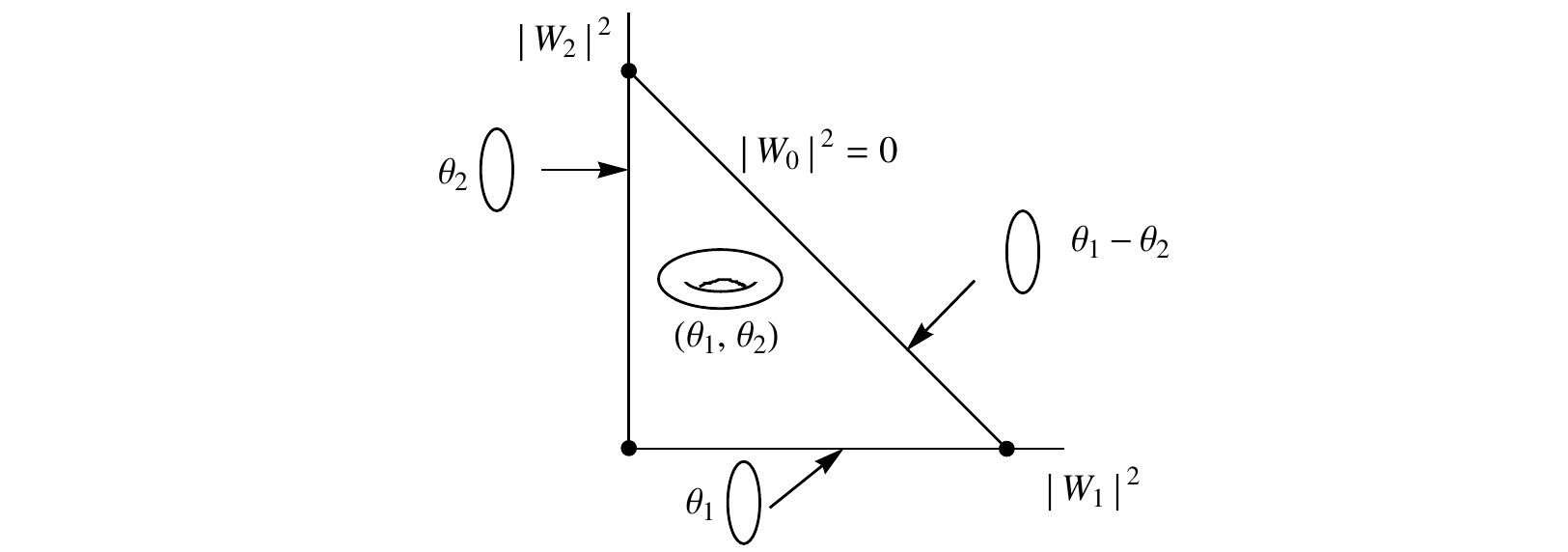}
\caption{$\CP^2$ as a toric fibration. The base is the triangle (2-simplex) parametrized by $|W_1|^2, |W_2|^2$ and the fiber is the torus $ (\theta_1, \theta_2)$. The fiber degenerates to a circle on the edges of the triangle and to a point in the corners.
}
\label{fig:cp2-toric}
\end{center}
\end{figure}

Consider the case $n_C =3 $ where we have $ \mbC \mbP^2 $. Lets us recall 
the toric description \cite{vafaiqmyst}.  
Keeping in mind that $|W_0|^2 = 1 - |W_1|^2 - |W_2|^2  \ge 0$, 
we can consider the quadrant parametrized by coordinates $|W_1|^2 , |W_2|^2 $. 
The allowed values of $|W_1|^2 , |W_2|^2 $ fall inside  a triangle
with vertices $ ( 0,0) , ( 0,1) , (1,0)$. For each chosen point inside the 
triangle, there is, in the $\mbC \mbP^2 $, a $T^2$ of phases given by $ (\theta_1 = \arg ~ W_1 , \theta_2  = \arg ~ W_2)$. The cycle parametrized by $\theta_1$ 
 collapses on the vertical axis ($|W_1|^2 =0$),
 the one parametrized by $\theta_2$    
on the horizontal axis ($|W_2|^2 =0 $), 
and the combination $ \theta_1 + \theta_2$ collapses 
on the line $|W_0|^2 = 0 $. See Figure~\ref{fig:cp2-toric}.

This generalizes straightforwardly to $\mbC \mbP^{n_C-1}$. 
The toric description has a base space which is a generalized 
tetrahedron or {\it simplex}  in $ \mbR^{n_C-1}$ (for more on simplices
see \cite{wiki-simplex}). 
There is a fiber $T^{n_C-1}$ 
related to angles of $W_I$ (modulo the overall $U(1)$).  

The identification 
\bea 
e_{IJ} = { W_I \bar W_J \over |W|^2 } 
\eea
from (\ref{homomorph}) 
shows that the diagonal $e_{II } $ 
are equal  to the 
coordinates used to parametrize the toric base. 
The off-diagonal $e_{IJ } $ are sensitive to the  angles. 
Their magnitudes are completely determined once the diagonal 
generators are known since 
\bea 
e_{IJ} e_{JI} = e_{II} e_{ JJ } 
\eea
We can write 
\bea 
e_{IJ } = \sqrt { e_{II} e_{JJ} } ~~ e^{ i ( \theta_I -  \theta_{ J }  ) }
\eea
Hence, the off-diagonal elements of the Lie algebra 
are associated with the angular variables of the toric description
and the diagonal ones with the base space.

\subsubsection{Giants: points on toric base and delocalized on fiber} 

For a state $ |\vec n \rangle  \equiv | n_0, n_1 , n_2 , \cdots n_{n_C-1}  \rangle  $ 
described by the monomial $W_{0}^{n_0 } W_{1}^{n_1} \cdots W_{n_C-1}^{ n_{n_C-1} } $ 
we can calculate 
\begin{equation}\begin{split} 
\frac{\langle \vec n |  e_{II} | \vec n \rangle}{\langle \vec n |  \vec n \rangle}  & = n_I \\
\langle \vec n | e_{II}^2 | \vec n \rangle - \langle \vec n | e_{II} | \vec n \rangle^2  & = 0 
\end{split}\end{equation}
This shows that the states $|\vec n \rangle  $ have definite locations 
on the toric base. For the off-diagonal coordinates in the Lie algebra 
\begin{equation}\begin{split} 
& X_{IJ} = { E_{IJ} + E_{JI} \over \sqrt { 2 }} \\ 
&  Y_{IJ} = - i  { E_{IJ} - E_{JI} \over \sqrt { 2 }}
\end{split}\end{equation}
we have 
\begin{equation}\begin{split}\label{uncertainties}  
& \langle  \vec n | X_{IJ} | \vec n \rangle  =
 0 ~~ ; ~~ \langle  \vec n | Y_{IJ} | \vec n \rangle = 0 
\\
&  { \langle  \vec n | X_{IJ}^2  | \vec n \rangle \over \langle 
 \vec n | \vec n \rangle } 
 =  \left ( n_I n_J + { n_I + n_J \over 2 } \right )   \\ 
& { \langle  \vec n | Y_{IJ}^2  | \vec n \rangle  \over \langle 
 \vec n | \vec n \rangle  } 
 =    \left ( n_I n_J + { n_I + n_J\over 2 }  \right )
\end{split}\end{equation}
The variances of these off-diagonal coordinates are non-zero 
and change along the base of the toric fibration parametrized by 
$ \langle  E_{II}  \rangle $. 

A related way to describe where these states are localized and 
where they are spread, is to note that for any operator 
 $ \cO$ in  $ \End ( \cH_{ C } ) $, 
\bea 
\langle \vec n | \cO | \vec n \rangle   = tr ~~ \cO P_{\vec n } 
\eea
where the projector $ P_{\vec n }   $ is $  | \vec n \rangle \langle \vec n | $. 
The trace is an $SU(n_C)$ invariant functional which becomes 
an integral $ \int \d \Omega $ over $\mC \mP^{ n_C -1 } $
in the large $N$ limit. The explicit 
form of the measure can be derived, and will not be important. 
The projector $ P_{\vec n }   $ maps, under the correspondence 
(\ref{homomorph})  
between  $ \End ( \cH_C ) $ and $ \Fun ( \mbC \mbP ) $ to 
\bea 
\prod_I  ( W_{I} \bar W_I )^{n_I} \over |W|^{ 2 \sum_I n_I  } 
\eea
This can be viewed as a density matrix associated with the 
state $ | \vec n \rangle  $. It is independent of the angular part of 
the $W_I$, which shows that these states are delocalized in the toric fiber.

To make the discussion more concrete let us take the set $C$ 
to be the set of  coefficients $c_{n_1 , n_2 , n_3 } $ with
 $ n_1 + n_2 + n_3 \le d $. We are now looking at polynomials of degree up to $d$.  
 The index $I$ in the above discussion 
runs over the triples $(n_1 , n_2 , n_3 )$ with $0\le  n_1 + n_2 + n_3 \le d $. 
The Hilbert space consists of polynomials in $w_{n_1, n_2 , n_3 } \equiv W_I $. 
The diagonal generators $W_I \partial_{W_I} $ of 
$U(n_C) \supset SU(n_C ) $ parametrize points
in the toric base. A state such as $w_{n_1, n_2 , n_3 }^N$
is an eigenstate  for the corresponding diagonal generator with 
maximal eigenvalue $N$, and has vanishing eigenvalue for the 
other generators. This defines a corner point on the base simplex 
for the toric fibration.   Consideration of (\ref{uncertainties})
shows that these states have distinguished localization 
properties. Indeed if a single $n_I $ is non-zero, and equal to 
$N$ as required by the condition $ \sum_I n_I = N$, then the uncertainty 
in the $X_{IJ}$ coordinates, given by $ \sqrt { \langle X_{IJ}^2 \rangle - \langle X_{IJ} \rangle } $ is order $\sqrt {  N } $. If a pair of $n_I , n_J $ 
are non-zero, then the uncertainty in $X_{IJ}$ is of  order $N$. 
These states, maximally localized at specific points 
in the physical moduli space of giants parametrized by $w_{n_1, n_2 , n_3 } $,  are natural candidates for the 
maximal (in the sense of being composites involving   large giants 
described by $x=0, y=0, z=0$) giants described by the polynomial $x^{n_1} y^{n_2} z^{n_3} = 0 $. Making this precise requires taking into account the fact that 
there is a non-trivial relation between the coefficients 
 $c_{n_1, n_2 , n_3 } $ of the polynomials and the 
 coordinates $w_{n_1,n_2,n_3}$ on the moduli space. 
This will be done in Section~\ref{maxwc} in the context of a 
discussion of quantum states near these corner points 
of the moduli space of giants in terms of physical (closed 
and open-string) excitations.

\subsection{Examples} \label{examples}

We now  describe how the connection between 
the discretized simplex at the toric base and giant graviton states 
works in concrete sectors, starting with the fuzzy  $\mbC \mbP^1$
for a single half-BPS giant. 

\subsubsection{Fuzzy \texorpdfstring{$\CP^1$}{CP1} from single half-BPS giant  }\label{sec:fuzzycp1} 

Let us start with the canonical example of a single giant in the half-BPS sector. The space of solutions can be parametrized by $c$ in the polynomial $P(z)=c z - 1 = 0$. The coordinate
$c$ is mapped to $w$ in  $\CP^1 = S^2$ as reviewed in Section~\ref{sec:review}. We will see how fuzzy $\CP^1$ arises in this context.

The Hilbert space $\cH$ 
 is $\{w_{0,0,0}^{n_0} w_{0,0,1}^{n_1} \,|\, n_0+n_1 = N\}$. This 
has an action of $SU(2) \subset U(2) $. The Lie algebra generators are 
$E_{00} , E_{11} , E_{01} , E_{10}$ given by (\ref{sugenw}) with $W_{0} = w_{0,0,0} , W_1 = w_{0,0,1} $. They generate the matrix algebra $ \End ( \cH ) $ 
which approaches the algebra of functions on the sphere in the large $N$
limit. 

To relate to  the usual description of fuzzy $S^2$ \cite{madore}, we can define 
\begin{equation}\begin{split} 
E_{00} - E_{11} = 2 J_3 \\ 
E_{01} + E_{10} = 2 J_1 \\ 
-i ( E_{10} + E_{01} ) = 2 J_2   
\end{split}\end{equation}
The $J_i$ obey the usual $SU(2)$ relations 
\bea 
[ J_i , J_j ] = i \epsilon_{ijk} J_k 
\eea
and the Casimir condition 
\begin{equation}
	J_1 J_1 + J_2J_2 + J_3J_3 = \frac{N(N+2)}{4}
\end{equation}

At $N \rightarrow \infty$ we rescale
$x_i = \frac{2 J_i}{N}$. Then $x_i x_i  \approx 1$ 
and the commutators have a factor of $1/N$, 
so in the limit $N\rightarrow \infty$ the product becomes commutative.

\begin{figure}[h]
\begin{center}
\resizebox{!}{6cm}{\includegraphics{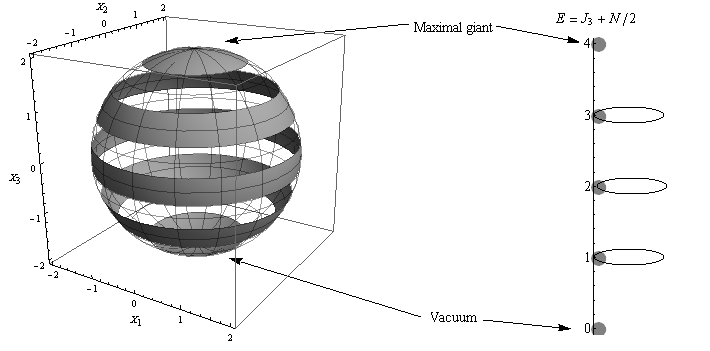}}
\caption{States on fuzzy $\CP^1_N$ with $N=4$. As wavefunctions on $S^2$ they are localized in the regions shown. If $\CP^1$ is viewed as a circle $T^1$ fibered over the line segment base, then we get the picture on the right: states are localized on the base and spread out in the fiber. Importantly, the diagram on the base is just the $SU(2)$ weight diagram for $V_{5}$.}
\label{fig:cp1_states}
\end{center}
\end{figure}

We can write the  expectation values 
\begin{equation}
\begin{split}
	\frac{\langle n | J_3 | n \rangle}{\langle n | n \rangle} &= n - \frac{N}{2} 
		~~~~~~~~~~~~~ \approx N \langle x_3 \rangle \\
	\frac{\langle n | J_{1,2} | n \rangle}{\langle n | n \rangle}  &= 0 
		~~~~~~~~~~~~~~~~~~~~ \approx N \langle x_{1,2} \rangle \\
	\frac{\langle n | (J_{1,2})^2 | n \rangle}{\langle n | n \rangle}  &= \frac{N}{4}+\frac{n(N-n)}{2}
		~~ \approx N^2 \langle (x_{1,2})^2 \rangle
\end{split}
\end{equation}
 These expectation values give us the picture of the states on fuzzy $S^2$ as shown in Figure~\ref{fig:cp1_states}.

Note that generically the states in this basis are spread
 out with $\langle \Delta x_{1,2}^2 \rangle \sim N^2$. However,
 there are two states which are maximally localized
 $\langle \Delta x_{1,2}^2 \rangle \sim N$: those at 
the north and south pole. Since they have energies $E=0$ 
and $E=N$, and are localized near classical points in phase 
space, they are  interpreted as the states corresponding 
to the vacuum and the maximal giant.

In the context of geometric quantization, the map from $\mC^\infty(S^2)$ to $\End(V_{N+1})$ is the way to go from classical functions on phase space to operators. This allows us to calculate expectation values of classical quantities on states in $\cH_C$ by using just $SU(2)$ algebra and representations.



\subsubsection{Fuzzy \texorpdfstring{$\CP^n$}{CPn} from \texorpdfstring{$n$}{n} half-BPS giants}
\label{fuzzycphalf}

Let us now generalize the picture to the case of two half-BPS giants $$P(z)=c_2 z^2 + c_1 z^1 - 1 = 0.$$ 
The phase space is $\CP^2$ and the Hilbert space  $\cH$ is 
$\{w_{0,0,0}^{n_0} , w_{0,0,1}^{n_1} w_{0,0,2}^{n_2} \,|\, n_0 + n_1 + n_2 =  N\}$. 
It is isomorphic to the symmetric $[N,0]$ representation 
of $SU(3)$. The space of operators $ \End ( \cH ) $ approaches 
the algebra of functions on  $\CP^2$ at large $N$. 
The action of $U(3) \supset SU(3) $ is given by (\ref{sugenw})
with the identification $ W_0= w_{0,0,0} , W_{1} = w_{0,0,1} , W_{2} = w_{0,0,2} $.  

It is useful to consider the relations 
\begin{equation}
\label{eq:U3_relations}
\begin{split}
	&E_{00}+E_{11}+E_{22} = N \\
	\qquad\qquad &E_{ij}E_{00} = E_{i0} E_{0j}, \qquad 1 \le i,j \le 2
\end{split}
\end{equation}
the latter being valid in the large $N$ limit.

For the construction of the fuzzy function algebra, the $SU(3)$ 
generators are again mapped to coordinates 
$E_{ij}\sim N e_{ij}$. The $\CP^2$ surface is 
embedded in this ambient 9-dimensional space. The following equations  
\begin{equation}\label{5constraints}
  e_{00}+e_{11}+e_{22} = 1 , \quad e_{ij}e_{00} = e_{i0}e_{0j}.
\end{equation}
allow us to formally express all the $e_{ij}$ in terms of two complex 
$e_{01} , e_{02} $, as expected for 5 real constraints in
 9 dimensional space. This  leaves the  4 dimensional $\CP^2$.

The dimensionality can also be understood
from the fact that the matrix algebra has dimension $ \sim N^4 $ 
at large $N$. This is a discrete geometry with coordinates such as 
$E_{ii}$ ranging over $N$. The vector space dimension of the 
function space is  $N^4$. A  discrete classical space with 
extent of order $N$ and $N^4$ elements has dimension $4$. 
This gives a  deduction of the dimensionality without explicitly
solving the constraints. We expect that such arguments 
based on state  counting, 
which is known from the BPS partition function, can be used to develop an understanding of giant gravitons for more general AdS/CFT duals. The space of 
quantum states in more general cases has been discussed from the 
point of view of dual giants, which are large in $AdS$ and classically 
point-like in the compact directions \cite{martsparks}. 
A treatment of quantum states for the case of giants large 
in the compact directions is not yet available
(for some efforts in this direction see \cite{hmp}).

\begin{figure}[h]
\begin{center}
\resizebox{!}{6cm}{\includegraphics{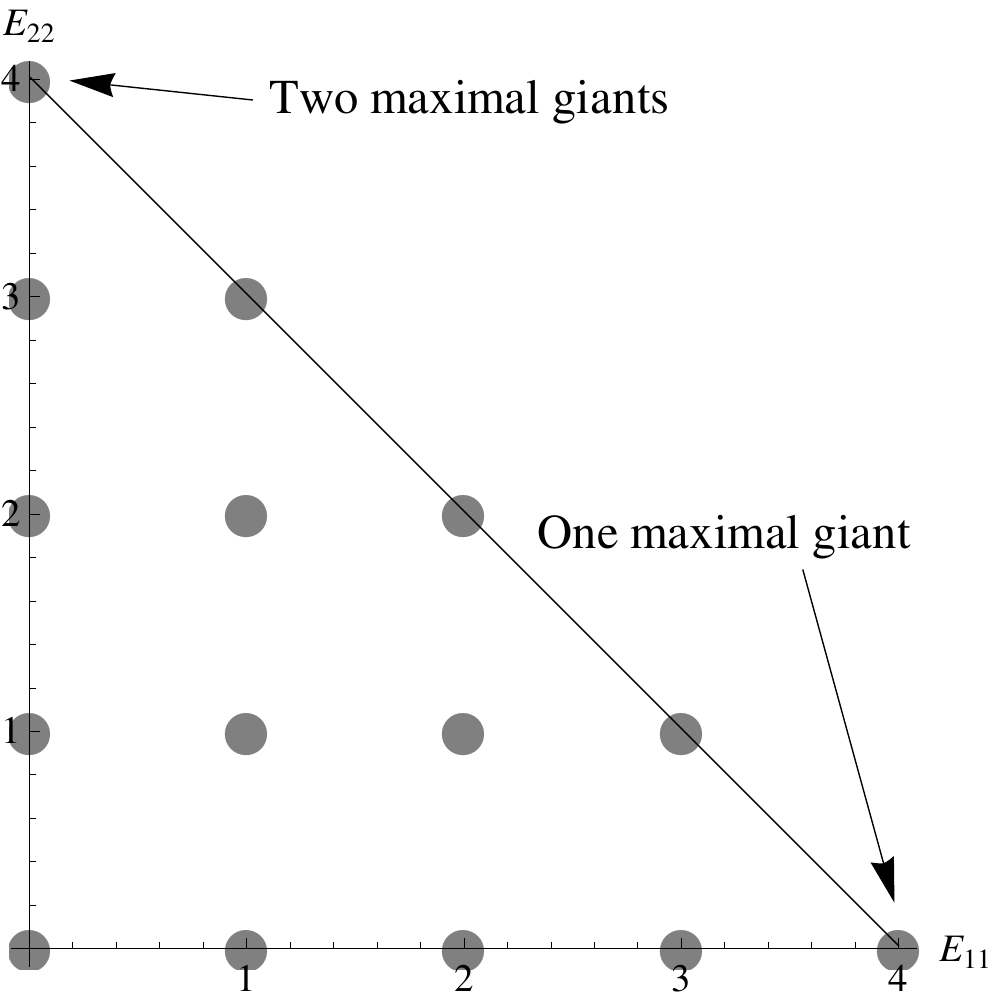}}
\caption{States on fuzzy $\CP^2_N$ with $N=4$. As wavefunctions on $\CP^2$ they are localized on the toric base shown and spread out over $T^2$ fiber. The diagram is also just the $U(3)$ weight diagram for $[4,0]$. The three corners of the base correspond to the points where $T^2$ shrinks to a point, and the states there are localized in all $\CP^2$ directions.}
\label{fig:cp2_states}
\end{center}
\end{figure}

The $\CP^2$ can be represented as a $T^2$ fibered over a toric base 
$E_{11}+E_{22}\le N$. The basis $|n_0 , n_1 , n_2 \rangle$ 
gives $\langle  E_{11} \rangle  , \langle  E_{22} \rangle    $ localized at points on the base 
base, and spread out over the $T^2$ fiber  see Figure~\ref{fig:cp2_states}. 

The special states in this basis $|N, 0,0\rangle, |0,N,0\rangle, |0,0,N\rangle$ again have a nice physical interpretation. They are localized near a point in phase space, which corresponds to vacuum, single maximal giant $z=0$, and two maximal giants $z^2=0$ respectively. We will discuss this more generally in Section~\ref{sec:states2geom}.

Using energies and charges, 
the labels $n_0 , n_1 , n_2 $ can be mapped to the Young diagram description 
of the sector of 2 half-BPS giants \cite{cjr}. The Young diagram has $n_1$ rows of length 
$1$ and $n_2$ rows of length $2$. The label $n_0$ is determined 
as $ N - n_1 - n_2$, and may be thought as the number of 
rows of length $0$. 

This generalizes to the case of $n$ half-BPS giants. 
Polynomials are $ c_0 + \sum_{i=1}^n c_i z_i = 0 $. 
The moduli space can be mapped to $\mbC \mbP^n $. 
The states in the Hilbert space map to points on the space 
described by $E_{11} , ... , E_{nn} $, by taking expectation values 
of these Lie algebra elements. These points provide a discretization 
of a simplex in $\mR^{ n }$.  The corresponding Young diagrams have 
$n_i$ rows of length $i$.

As we reviewed in Section~\ref{sec:review} 
it has been argued \cite{minwalla} that fuzzy $\CP^{n_C - 1 } $
will also arise in the eighth-BPS sector, when we consider 
polynomials of degree up to $d$ and identify 
\bea 
n_C = { ( d+1) (d+2) (d+3 ) \over 6 } 
\eea
In this case, we do not expect a straightforward identification 
of the coordinates of the 
points  on the discretized toric base with brane multiplicities. 
There is a $U(3)$ action which should be disentangled from 
the grading with degree of the polynomial for a more physical 
interpretation. We will present results in this direction 
in the next section. 

\section{A geometrical  group theoretic labelling for eighth-BPS states} 
\label{complete-group-labels}

The Young diagram labelling for half-BPS states \cite{cjr} has provided a powerful 
way to study perturbations of the half-BPS operators and to relate these 
to open strings connecting branes \cite{Balasubramanian:2002sa, Aharony:2002nd, Berenstein:2003ah, bbfh, dMSSI,dMSSII,dMSSIII}. Loosely speaking, 
corners of Young diagrams provide slots for anchoring strings to the operator. 
This has allowed the unraveling of integrable structure for 
strings connected to giant gravitons, giving a beautiful generalization 
of integrability in the planar limit \cite{rdgm}\cite{npint1106}.  
A group theoretic labelling for eighth-BPS operators
may be expected to have similar significance, 
for developing an explicit construction for BPS operators
and their excitations. In this section, 
we will derive such a labelling from the geometrical 
side of giant graviton quantization. 
We will demonstrate that, 
in the eighth-BPS case, there are $U(N)$ Young diagram 
labels along with the $U(3)$ global symmetry labels. 
These labels are supplemented by an  additional set of complete 
group theoretic labels, which include sequences of $U(3)$ Young diagrams and 
Littlewood-Richardson multiplicities for their tensor products. 
This labelling is arrived at by considering the decomposition of
the Hilbert space into a direct sum associated with states coming from 
polynomials of different degrees in the Mikhailov construction. 
We describe fibration structures of $\CP^{n_{C} -1}$ which are associated with these 
group theoretic labels in Section~\ref{sec:geometry_hilbert}.

In Section~\ref{sec:labels_discussion} we discuss the implications of the group theoretic labels
for the construction of operators at weak coupling.


\subsection{\texorpdfstring{$U(3)$}{U(3)} BPS  multiplicities and Young diagrams of \texorpdfstring{$U(N)$}{U(N)} } 

Consider the Hilbert space constructed from giant gravitons
associated to polynomials of degree up to $d$. 
Let us denote this as   $\cH^{I ;~   d}_N $. 
The $I$ indicates that we are dealing with inhomogeneous polynomials. 
The complex coordinates $x,y,z$ of $\mbC^3$ transform as the fundamental 
representation $V_3$ of $U(3)$. 
The holomorphic coordinates $w_{i_1,i_2,i_3} $ related to 
monomials $x^{i_1}y^{i_2} z^{i_3} $ of degree $k$, i.e.  
$i_1 + i_2 + i_3 = k $, transform in the irreducible 
representation     $\Sym^k ( V_3 )$ of $U(3)$ of dimension 
\begin{equation}
\label{eq:mk}
 m (k) = { (k+1)(k+2) \over 2}  \,.
\end{equation}
Polynomials of degree 
$N$ in these variables, when $k$ ranges from $0$ to $d$, form 
the Hilbert space $\cH^{I ;~  d}_N$. Hence 
\bea\label{decompdk} 
\cH^{I ;~  d}_N    = \Sym^N ( \oplus_{ k=0}^d   \Sym^k ( V_3 ) )
\eea

For concreteness, consider $d = 2$. 
There are three types of holomorphic variables (or oscillators in the Fock 
space language) generating the states in  the 
Hilbert space. We 
have $w_{0,0,0} $ of degree $0$,  
$w_{0,1,0} , w_{1,0,0} , w_{0,0,1} $ of degree $1$, 
and $ w_{2,0,0} , w_{0,2,0} , w_{0,0,2} , w_{1,1 , 0 } , w_{0,1,1} , w_{1,0,1} $ 
of degree $2$. These span the vector space
\bea 
\mC \oplus V_3 \oplus  \Sym^2 ( V_3 )
\eea 
A basis in 
\begin{equation}
\cH^{I ;  d=2}_N = \Sym^N ( \mC \oplus V_3 \oplus  \Sym^2 ( V_3 ) )
\end{equation}
can be chosen where a definite number of  holomorphic variables (or oscillators) come from each degree. 
There are three integers
$N_0, N_1 , N_2$ specifying how many there are in degree $0,1,2$. 
There is the constraint 
\bea 
N_0 + N_1 + N_2 = N 
\eea
We have seen in the half-BPS context that three 
integers with this constraint describe the points 
on the discretized toric base of fuzzy $\mC \mP^2$. And 
this set of integers also specifies a Young diagram with $3$ 
columns, with $N_i $ being the number of rows  of length $i$.

We may write 
\begin{equation}\begin{split} 
\label{gradNi}  
\cH^{I ;  d=2}_N  & =  \bigoplus_{ \substack { N_0 , N_1 , N_2 \\ 
 N_0 + N_1 + N_2 = N }  } 
  \cH_{N_0}^{h ;  k=0 }  \otimes  \cH_{N_1}^{ h ;  k=1}   \otimes 
\cH_{N_2}^{ h ; k=2} \\
 &=  \bigoplus_{ \substack { N_1 , N_2 \\ 
  N_1 + N_2 \le N }  } 
 \cH_{N_1}^{ h ;  k=1}   \otimes \cH_{N_2}^{ h ; k=2}
\end{split}\end{equation}
The superscript $h$ indicates homogeneous and 
\bea 
\cH^{ h ; k }_N = \Sym^N ( \Sym^k ( V_3) ) 
\eea
The formula (\ref{gradNi}) displays the inhomogeneous Hilbert 
space as a direct sum of tensor products of homogeneous Hilbert spaces 
parametrized by Young diagrams of $U(N)$ 
with at most two columns. The Hilbert space associated with degree zero 
is one-dimensional. This allows the rewriting in terms of just $N_1,N_2$. 

Each summand in (\ref{gradNi}) has well-defined $U(3)$  transformation 
properties
\bea 
\cH^{ h ; k }_N = \Sym^N ( \Sym^k ( V_3) ) = \bigoplus_{\substack { 
\Lambda : c_1 ( \Lambda ) \le 3 }  } V_{ \Lambda_k } \otimes V^{h ; k ; N }_{ \Lambda_k   }   
\eea
The space $ V^{h ; k ; N }_{ \Lambda  } $ is the vector space of 
multiplicities for $V_{\Lambda_k} $ in $\cH^{ h ; k }_N$.  
Its dimension gives the number of times the irreducible  
$V_{\Lambda_k } $ of $U(3)$ appears when the representation  
$\Sym^N ( \Sym^k ( V_3) )$ is decomposed into irreducible representations (irreps)
 We can obtain these multiplicities from a restriction 
to fixed degree of the oscillator partition function. 
\bea\label{genfunSNSk}
| V^{h ; k ; N }_{ \Lambda  }| = \left [ (x_1-x_2)(x_1-x_3)(x_2-x_3) 
\prod_{\substack{i_1, i_2 , i_3 \ge 0 \\ i_1 + i_2 + i_3 = k }} 
{ 1 \over ( 1 - \nu x_1^{i_1} x_2^{i_2} x_3^{i_3} ) }   \right ]_{ \nu^N  x_1^{\lambda_1 + 2} x_2^{\lambda_2 +1 } x_3^{\lambda_3}  }
\eea
It is  also the number of times the trivial irrep of 
the wreath product $S_N [ S_k] $ appears in the irrep $\Lambda $ of 
$S_{Nk} $. It is easy to see from (\ref{genfunSNSk}) 
that $ Sym^{N_k} ( Sym^k V_3 )$ only contains irreps of $U(3)$ 
with precisely $Nk$ boxes. This is a useful fact in working out 
the possible additional labels that go with a fixed $\Lambda$.

 By making the  $U(3)$ decomposition explicit, we can write : 
\bea 
\cH^{I ;  d=2}_N  & = &   \bigoplus_{ \substack {  N_1 , N_2 \\
   N_1 + N_2 \le  N } } \left (  \bigoplus_{\Lambda_1 }V_{ \Lambda_1}  \otimes  V^{h ; k =1 ; N_1 }_{ \Lambda_1   } \right )
\otimes \left (  \bigoplus_{\Lambda_2 }V_{ \Lambda_2} \otimes  V^{h ; k=2 ; N_2 }_{ \Lambda_2   } \right )
\eea
 We can take the $U(3)$ tensor products 
\bea\label{u3andUN}  
\cH^{I ;  d=2}_N  = 
 \bigoplus_{ \substack {  N_1 , N_2 \\
        N_1 + N_2 \le  N } } \bigoplus_{ \Lambda } V_{ \Lambda } \otimes \left ( \bigoplus_{ \Lambda_1 , \Lambda_2 }  V^{\Lambda }_{ \Lambda_1 , \Lambda_2 } \otimes  V^{h ; k =1 ; N_1 }_{ \Lambda_1   } \otimes  V^{h ; k=2 ; N_2 }_{ \Lambda_2   } \right )
\eea
The vector space  $V^{\Lambda }_{ \Lambda_1 , \Lambda_2 }$ has the dimension 
equal to the Littlewood-Richardson multiplicity  for getting 
$V_{ \Lambda }  $ in the $U(3)$ tensor product 
$ V_{ \Lambda_1 } \otimes  V_{ \Lambda_2 } $. 
The range of the direct sum is running over $ N_1 , N_2$ which parametrize 
points on the toric base of the fuzzy $\mC \mP^2$ or equivalently 
Young diagrams of $U(N)$ with up to two columns. Let us call these 
diagrams $Y $, which have $N_1 = N_1 ( Y)   $ rows of length $1$, 
$N_2= N_2 (Y) $ 
rows of length $2$ and a maximum of $N$ rows.   
\bea\label{d=2decomp}  
\cH^{I ;  d=2}_N  = 
 \bigoplus_{   Y ( N_1 , N_2 : N ) } \bigoplus_{ \Lambda } V_{ \Lambda } \otimes \left ( \bigoplus_{ \Lambda_1 , \Lambda_2 }  V^{\Lambda }_{ \Lambda_1 , \Lambda_2 } \otimes  V^{h ; k =1 ; N_1 }_{ \Lambda_1   } \otimes  V^{h ; k=2 ; N_2 }_{ \Lambda_2   } \right )
\eea
Another way to express the meaning of (\ref{d=2decomp})
 is that the states in the eighth-BPS 
sector, obtained from polynomials of degree no more than $2$ in $x,y,z$, 
can be written as 
\bea 
| \Lambda , M_{ \Lambda } , Y , \Lambda_1 , \Lambda_2 ,  a , b_1 , b_2 \rangle 
\eea
where $M_\L$ labels the state in $V_\L$, $b_k$ run over the multiplicity of $U(3)$ representation 
$ \Lambda_k $ in $ \Sym^{ N_k ( Y ) } ( \Sym^{k } (V_3)  ) $, and $a$ runs 
over the (Littlewood-Richardson) multiplicity of $U(3)$ irrep  
$\Lambda $ in $ \Lambda_1 \otimes \Lambda_2$. Explicit algorithms 
for  these are known \cite{fulhar}.

The steps above are easily written for general $d$, starting 
from the generalization of (\ref{gradNi}) 
\begin{equation}
\label{dgradNi}  
\cH^{I ;  d}_N  =  \bigoplus_{ \substack { N_1 , N_2 \cdots N_d  \\ 
  N_1 + N_2 + \cdots N_d \le N }  } 
 \cH_{N_1}^{ h ;  k=1}   \otimes \cH_{N_2}^{ h ; k=2} \otimes \cdots  \otimes 
\cH_{N_2}^{ h ; k=d}
\end{equation}
and leading to 
\begin{equation}\label{decompYNd}
\boxed{
\cH^{I ;  d }_N  = 
 \bigoplus_{ Y (N_1 , N_2 ,  \cdots , N_d ; N )    }
~~  \bigoplus_{ \Lambda   \in Reps ( U(3) ) } V_{ \Lambda} \otimes 
\left ( \bigoplus_{ \Lambda_1 \cdots \Lambda_d  } 
 V^{\Lambda }_{ \Lambda_1 , \cdots ,  \Lambda_d  }
 \otimes_{k=1}^d     V^{h ; k  ; N_k }_{ \Lambda_k } 
 \right )
} 
\end{equation} 
This means that there is a labelling to BPS states by 
\bea 
| \Lambda , M_\L , Y, \Lambda_1 , \cdots , \Lambda_d , a , b_1 , \cdots , b_d 
\rangle  
\eea
where $\Lambda , \Lambda_1 , \cdots ,  \Lambda_d $ are irreps of $U(3)$, 
$a$ is a Littlewood-Richardson multiplicity for $\Lambda $ appearing 
in $ \Lambda_1 \otimes \Lambda_2 \cdots \otimes \Lambda_d $. The labels 
$b_k$ (for $k=1 \cdots d $) run over the multiplicities of $\Lambda_k$ 
appearing in $\Sym^{N_k} ( \Sym^k ( V_3))$. The numbers $N_k $ obey 
$N_1 + \cdots + N_d \le  N $ and determine the Young diagram $Y$ of $U(N)$.

As we explained in Section~\ref{sec:cp} 
the labels $ N_1 , N_2 , \cdots , N_d $ 
also parametrize points on the toric base of $\mbC \mbP^d $, when 
we use the fuzzy $\mC \mP$ constructions. 
As a result we may also write 
\begin{equation}\label{fuzzycps}
\cH^{I ;  d }_N  = 
 \bigoplus_{ Fuzzy_N ( CP^d : N_1 , N_2 \cdots N_d   )   }
~~  \bigoplus_{ \Lambda   \in Reps ( U(3) ) } V_{ \Lambda} \otimes 
\left ( \bigoplus_{ \Lambda_1 \cdots \Lambda_d  } 
 V^{\Lambda }_{ \Lambda_1 , \cdots ,  \Lambda_d  }
 \otimes_{k=1}^d     V^{h ; k  ; N_k }_{ \Lambda_k } 
 \right )
\end{equation} 
Note that the homogeneous Hilbert spaces in (\ref{dgradNi}) 
come from homogeneous polynomials which correspond to static 
giant gravitons. 
The  decomposition at hand can therefore also be viewed as 
giving the full Hilbert space in terms of tensor products 
of states coming from fixed points of the classical Hamiltonian 
action. This is a key result of this paper and we discuss its 
geometrical and physical  aspects in the remainder of this section.

\subsection{Geometry of the Hilbert space  decomposition}
\label{sec:geometry_hilbert}
 
  We started by describing  
how oscillator Hilbert spaces arise from projective spaces 
which are moduli spaces of  giant gravitons, constructed 
using polynomials in three variables with degree up to $d$. 
The Hilbert space is the sector of the Fock space generated 
by oscillators associated with monomials $x^{m_1}y^{m_2} z^{m_3} $, having 
no more than $N$ excitations with all  $m_i$ non-zero.  
We then did a decomposition of the Hilbert space, 
 organizing them according to the degree $m_1 + m_2 + m_3 $ 
of the monomials  they come from. It will be interesting 
to understand the outcome (\ref{decompYNd}), (\ref{fuzzycps})
in terms of  the  geometry of the moduli spaces. 
We make some remarks in this direction.

\begin{itemize}

\item 
A projective space $\mC \mP^n $,  with homogeneous coordinates 
$ W_0 , W_1, W_2 , \cdots , W_n $ subject to $ W_I \simeq \lambda W_I$
for non-zero complex $ \lambda $, is the base of the Hopf fibration 
with $S^1 = U(1)$ fiber and total space $S^{2n+1}$. The sphere is defined 
by $ \sum_I W_I \bar W_I = 1 $. The projection identifies orbits of the $U(1)$ action $W_I \sim  e^{ i \theta } W_I$.
\bea 
    && S^{2n+1}    \leftarrow U(1) \cr 
    &&   ~~~ \downarrow \cr 
    && ~~ \mC \mP^{n}
\eea

When the $\mC \mP^{n_C(d)  -1 }  $ 
is constructed from polynomials of degree up to $d$,
 it is natural to label the homogeneous coordinates $W_{ k , i } $ 
with $k$ running from $0$ to $d$ and $i$ running from $1$ to $m(k)$ (\ref{eq:mk}).  

We can map $\mC \mP^{n_C ( d ) -1  } $ to  $\mR^d  $, by using 
\bea 
 ( \sum_{ i } | W_{1, i} |^2 , \sum_{ i } | W_{2, i} |^2 , \cdots ,
\sum_{ i } | W_{d, i} |^2  ) 
\eea 
The fiber at each point is a product of spheres.
In the $d=2$ case, we have  three monomials at $k=1$, so 
$W_{1,1} = w_{1,0,0} ,  W_{1,2} = w_{0,1,0} ,  W_{1,3} = w_{0,0,1}$. 
There are six monomials at $k=2$, so $W_{2,1} = w_{2,0,0} , W_{2,2} = w_{0,2,0} , W_{2,3} = w_{0,0,2} , W_{2,4} = w_{1,1,0} , W_{2,5} = w_{1,0,1} , W_{2,6} = w_{0,1,1}$. The $\mC \mP^2$ at degree $1$ lives in $S^5 \subset \mC^3 $. 
The $\mC \mP^5 $ at degree $2$    lives in $S^{11}  \subset \mC^6 $. 
This means we have the fibration structure
\bea 
&&   \CP^{9} \leftarrow \qquad S^{5} \times S^{11} \leftarrow S^1 \times S^1  \cr 
&&  ~ \downarrow ~~~~~ \quad \qquad \downarrow \cr 
&&  ~ \mR^2 \qquad \CP^2 \times \CP^{5} 
\eea

This is illustrated in Figure~\ref{fig:s5s11}. 
The sphere $S^5$ degenerates on the $y$-axis, the sphere $S^{11} $ degenerates 
on the $x$-axis. The $N_1,N_2$ can be viewed as discretizing 
the $\mR^2$ base. In the Hilbert space, they parametrize the 
factorization $ \cH^{(h: k=1)}_{ N_1} \otimes \cH^{(h: k=2)}_{ N_2}$, 
where $\cH^{(h: k=1)} $ comes from quantization of $\CP^2 $ associated 
with degree $1$ monomials, 
and  $\cH^{(h: k=2)} $  from $\mC \mP^5 $. When the size of the $\mC \mP^2 $
vanishes, then there  is no non-trivial multiplicity of states from it. 
For fixed $N_1 + N_2$, as we move towards larger $N_1$, we get 
more states from $\mC \mP^2 $, while the fibration above 
has a large $\mC \mP^2$. The variation of the sizes of the $\mC \mP^2 , \mC \mP^5$ 
are reflected in the number of states they contribute, indicating 
this is the right geometrical way to think about the decomposition of 
Hilbert spaces.

\begin{figure}[h]
\begin{center}
\includegraphics{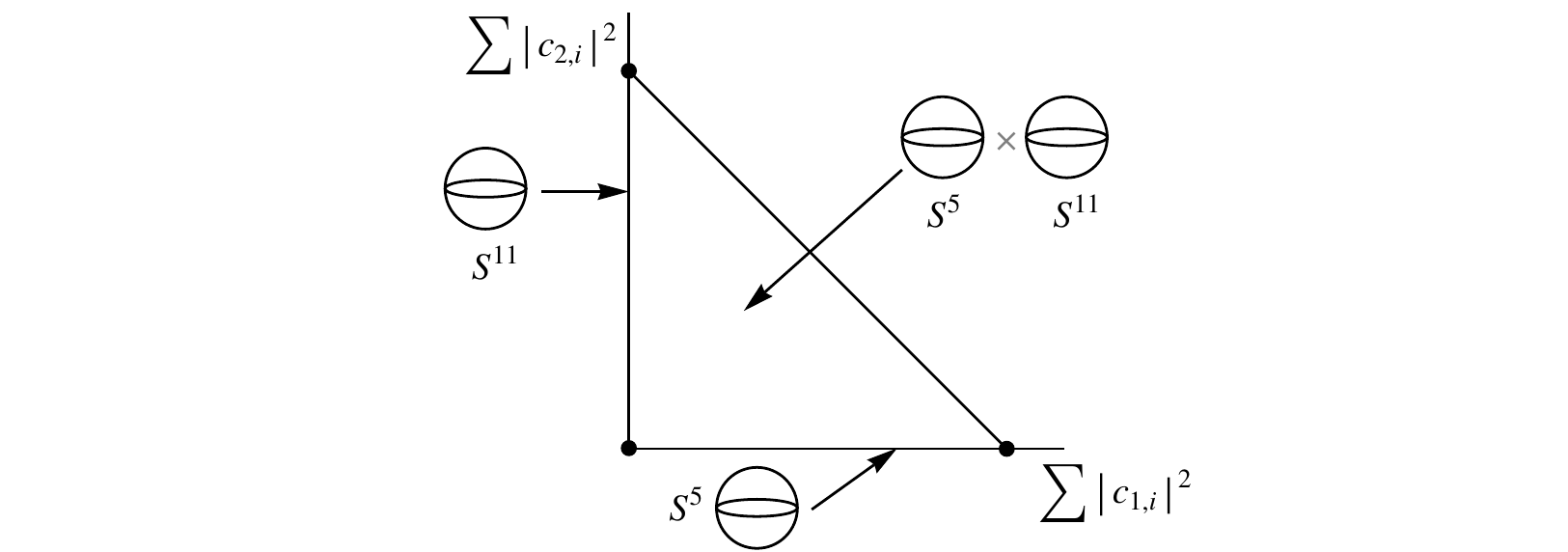}
\caption{$\CP^9$ as $S^5\times S^{11}$ fibered over a simplex in $\mR^2$. The sphere $S^5$ degenerates on the $y$-axis, the sphere $S^{11} $ degenerates 
on the $x$-axis.}
\label{fig:s5s11} 
\end{center}
\end{figure}

\item 
 As explained earlier, we can  think about states 
in $ \cH^{ I ; d }_N  $ is in terms of points on 
a simplex in $\mR^{n_C-1}$.  
The decomposition into states according to 
the degree of the monomials giving rise to the oscillators
 associates Young diagrams to partitions of $N$ across degrees.
The summation over $ Y $ is a sum 
over points in a simplex in $\mR^d$. 

The summands run over 
a product of discretized simplices 
$ \times_{k=1}^d   \Delta^{ m(k)}_{ N_k ( Y )} $, where $ \Delta^{m(k)  }_{N_k } $
denotes the simplex with $m(k)$ corners discretized so as to contain 
the coordinates of the Fock states in $\Sym^{N_k} ( \Sym^k ( V_3) ) $.  

 We may write 
\bea 
\Delta^{ n_C ( d )}_{   N }  = \amalg_{ Y   } \prod_{k=1 }^{ d } 
 \Delta^{ m(k)}_{ N_k ( Y ) }       
\eea
In geometry this is a fibration of a simplex with $n_C(d) $ corners
 in terms of a base simplex with $d$ corners, 
with the fiber being a product of simplices. 
\bea 
&& \Delta^{n_C ( d ) } \leftarrow \prod_{k=1 }^{ d } \Delta^{m(k)} \cr 
&& ~~~ \downarrow \cr
&&  ~~  \Delta^{ d  }
\eea
As a special case, consider the case $d=1$. 
The $k=0$ sector gives just one state for each $N$. The $k=1$ states give 
the  discretization of  a 3-simplex. The geometrical counterpart here is    a 
deconstruction of the  tetrahedron as a fibration of a triangle (3-simplex) over 
a 1-simplex.

\item 
It is known that the Hilbert spaces for homogeneous polynomials 
are related to line bundles over $ \mC \mP^2 $. 
\bea\label{symLinBun} 
\Sym^{ N } ( \Sym^{k} ( V_3 ) ) = 
\Sym^{N } \left ( H^0 ( \mC \mP^2 , \cO_{ \mC \mP^2 }   ( k )  )  \right )
\eea
This  $ \mC \mP^2 $ should be interpreted as the 
base space of the $S^5$ in the $AdS_5 \times S^5$ spacetime. 
The giant graviton world-volumes coming from 
homogeneous polynomials intersect this spacetime $\mbC \mbP^2$ 
on a genus $ (k+1) (k+2)/2 $ curve \cite{minwalla}. Given the role 
of the symmetric product (\ref{symLinBun}) in the decomposition 
(\ref{decompYNd}) it may be instructive to interpret these line bundles 
over the $\mC \mP^2$ embedded in spacetime,  in terms of 
quantization of branes.

\end{itemize}

\subsection{Discussion: labelling and construction of operators  } 
\label{sec:labels_discussion}

The above manipulations in the oscillator Hilbert space have 
some remarkable consequences, once we use the  AdS/CFT conjecture.  {\it They show that there is a  labelling  by Young diagrams 
$ Y \times \Lambda $  of $U(N) \times U(3) $ for eighth-BPS operators in SYM}.  
For the case of half-BPS operators there is a Young diagram 
$R$ which arises from the diagonalization of the inner product on 
the half-BPS operators \cite{cjr}. The above geometric 
group theoretic labels we have been constructing are also 
expected to be describing orthogonal states, as they use 
decomposition into degrees, and we expect states coming from different 
degree polynomials to be orthogonal. The case of quarter and eighth 
BPS operators in the gauge theory 
 is more tricky. One can diagonalize the inner product 
at zero coupling in a $U(3)$ covariant basis, 
obtaining $U(N)$ Young diagram labels $R$ (in addition to a Clebsch-Gordan 
inner product multiplicity $\tau$ of symmetric groups) \cite{bhrI,bhrII}.
The left action of $U(N)$ on free adjoint fields $X , Y , Z $  
is a symmetry of the action at zero coupling and its Casimirs 
are labelled by the Young diagrams $R$ \cite{ehs}. 
The same label $R$ appears in the restricted Schur basis for operators, 
where, in addition, 
 instead of $U(3)$ labels we have labels corresponding to a product of 
symmetric groups.   
However, $R$ does not give a quantum number at weak coupling. The one-loop 
dilatation operators mixes $R$. In the $U(3)$ covariant
basis, constraints on mixing of $R$ were obtained in \cite{tom-oneloop}.
In the context of restricted Schur basis, there has been extensive work 
on the mixing due to the action of one-loop dilatation operator 
in the context of integrability of strings attached to giant gravitons 
\cite{bbfh,dMSSI,dMSSII,dMSSIII,rdgm,npint1106,npint1103,bcv,
rdgm}.

What we are finding 
here is that there will nevertheless be some other Young diagram 
labels $Y$ which will appear once we solve for the kernel 
of the $\cH_2$ in the $U(3)$ sector. The relation of $Y$ to the free field 
labels $ R , \tau $ is a very important  problem for the future. 
At small $n$, where we are considering operators made from a 
just a few matrices, this relation can be worked out. 
Take for simplicity $\Lambda = [2 ,2 ] $. The possible 
$Y$'s in this case can be worked out to be 
 Young diagrams with rows $[2,2] $ or $[2,1,1]$, which 
each occur with multiplicity $1$. For  $Y=[2,2]$, the additional labels 
are fixed by   $\Lambda_2 = [2] , k = 2 $ being the only non-trivial ones. 
 In $Y=[2,1,1]$, the additional labels are fixed to have 
 $\Lambda_1 = [2] , k=1$ and $\Lambda_2 = [1] , k=2 $ as the only non-trivial 
ones. The state labelled by $Y=[2,1,1]$ exists at $ N >3 $ and 
disappears at $N=2$. The state labelled by $Y=[2,2]$ exists 
at $N \ge 2$. Orthogonal BPS states with these properties in 
the sector $\Lambda = [2,2]$ were constructed in \cite{pgp} (see equation (C.18) in Appendix~C). Judging by exclusion properties, $\cO^{\rm BPS}_{(2)}$ corresponds to $Y=[2,1,1]$, as it disappears at $N=2$, and $\cO^{\rm BPS}_{(1)}$ corresponds to $Y=[2,2]$.
So in this case we have the explicit construction of operators 
with the group theoretic labels in (\ref{decompYNd}).

For studies of the perturbations away from half-BPS states, the $R$ label 
has been tremendously powerful \cite{bbfh} \cite{npint1106,rdgm}. 
A lot of detailed combinatorics has been found compatible with 
the Gauss Law expected from D-brane world-volume physics. 
The problem of exhibiting the implications of the Gauss Law 
for the perturbations of general eighth-BPS operators (which are not small 
deformations of half-BPS ones) is a fascinating one.

One is tempted to conjecture that the labels $Y$ will play an 
identical role to that played by $R$ for half-BPS.  
A word of caution is in order. For the half-BPS case, the label $Y$ 
indeed becomes $R$. It comes from a decomposition of the space of polynomials 
into those of fixed degree. In this case, we are dealing with polynomials in 
one variable and a polynomial of degree $d$ can always be factorized
into factors of degree $1$. Hence \emph{degree} $d$ is equivalent to \emph{multiplicity} 
$d$ of branes. A polynomial of the form $ ( z - a_1 ) ( z - a_2) \cdots $ 
leads to a brane world-volume consisting of multiple branes 
localized at multiple points $a_1, a_2 , \cdots $ (when $a_i \le 1 $).  For the multi-variable case, 
a polynomial of degree $d$ in $x, y, z $ is not always factorizable 
(i.e. reducible).  Our decomposition of the Hilbert space 
has been done according to degree 
not according to reducibility, and the Young diagram $Y$ naturally 
encodes finite $N$ constraints, in that $ \sum_{i=0} N_i (Y ) = N$, 
It is not  clear if such a simple $U(3)$ covariant 
decomposition of the Hilbert space into degrees of reducibility 
is even possible. 
Naively, it would appear that the multiplicity of branes ought to 
be related to factorizability rather than degree of the polynomial. 
If this correct, then $Y$ will not play a directly analogous role 
in terms of Gauss Law. Yet the simplicity of the decomposition 
in terms of $Y$ and the way it encodes the finite $N$ cutoffs, 
suggests that there will be very interesting 
physics to be found in describing the eighth-BPS operators according to 
this basis. If $Y$ is found to play the same role as $R$ in connection 
with Gauss Law, then it would mean that, at the level of quantum states, 
brane multiplicity is correctly captured by something closer to 
the ordinary degree of the polynomial in $x,y,z$ rather than 
by the more complicated notion of reducibility 
which may not have a simple expression in the Hilbert space.

From the results of \cite{ehs} we would expect that
 hidden Casimirs of a  $U(N)$ will also be relevant 
as an explanation of the $Y$ labels. 
 Another  approach to understanding $Y$ from the gauge theory side 
is to construct the 
 non-zero eigenstates of the group theoretic operator $\cP_{I} \cG \cP_{I} $ 
of \cite{pgp}. AdS/CFT predicts that we will find diagonal states labelled 
simultaneously by $ \Lambda $ of $U(3)$ and $ Y $ of $U(N)$ (along with the 
extra group-theoretic labels in (\ref{decompYNd})).

\section{World-volume excitations of maximal giants  }
\label{sec:states2geom}

\subsection{Maximal giant states}\label{maxwc}

In this section we will use the picture established in Section~\ref{sec:cp} to analyze the physics of some concrete brane geometries. 
We will be able to identify configurations of maximal giants
 with specific states in the Hilbert space, and to classify
 the spectrum of excitations of these configurations.

First, let us make some general remarks about the correspondence 
between quantum  states and points $p$  in the phase space $\cM_C$, 
where each point is associated to   supersymmetric brane geometry. 
A quantum state $|\Psi_p\rangle\in\cH_C$ which  ``corresponds'' to $p$ is  a state whose wavefunction is maximally localized around $p$. Such states are called \emph{coherent states} and can be built, for example, by translating a localized  ground state in the phase space \cite{woodhouse}.   In our case we have $SU(n_C)$ acting transitively on $\cM_C$, so we can write
\begin{equation}
\label{eq:coherent_state}
	|\Psi_p\rangle = e^{i \theta_{IJ} \hat{E}_{IJ}}|\Psi_0\rangle, \quad \text{if}\;
	p = e^{i \theta_{IJ} E_{IJ}} p_0 , \quad
	e^{i \theta_{IJ} E_{IJ}} \in SU(n_C)
\end{equation}
The states $|\Psi_p\rangle$ span the Hilbert space 
but are generically over-complete, since the Hilbert space 
 $\cH_C$ is finite dimensional. 

On the other hand, if we take the basis such as (\ref{eq:sunc_basis}), 
formed by eigenstates of the maximal set of commuting operators $E_{II}$ 
in $SU(n_C)$, it is complete and orthogonal.  Such states generically 
do not correspond to points on phase space. As we saw in Section~\ref{sec:cp}  
they are localized in the base simplex and spread out along the 
 the torus fibers of a toric fibration structure of 
the phase space. However, there are special corner points (vertices 
of the simplex) in the base where the torus fiber degenerates. The energy eigenstates localized near these corners \emph{are} in fact maximally localized and thus are also coherent states. Recall the wavefunction for such state is
\begin{equation}
\label{eq:Psi_max}
	\Psi^{\rm max}_{m_1,m_2,m_3} = (w_{m_1,m_2,m_3})^N
\end{equation}
and it is localized near the point $w_{n_1,n_2,n_3} = \delta_{m_1n_1}\delta_{m_2n_2}\delta_{m_3n_3}$ in $w$ coordinates. Let us denote these points as $p^{\rm max}_{m_1,m_2,m_3}$. From the point of view of $\CP^{n_C-1}$, all these states look the same as the vacuum state $(w_{0,0,0})^N$, but in different inhomogeneous coordinate patches.

It turns out the corner points enjoy special $U(3)$ transformation properties, that make them particularly accessible to analysis. 
If we take a generic point in the phase space, its time evolution is generated by $U(1) \subset U(1)^3 \subset U(3)$ which acts as $w_{n_1,n_2,n_3} \rightarrow e^{i (n_1+n_2+n_3) \alpha} w_{n_1,n_2,n_3}$. This forms an orbit in the toric fiber. There are special regions in the phase space where this $U(1)$ degenerates to a point. They correspond to static solutions which map to the homogeneous Mikhailov's polynomials. Among these, there are points invariant under the whole $U(1)^3 \subset U(3)$ action  $w_{n_1,n_2,n_3} \rightarrow e^{i n_1 \alpha_1} e^{i n_2 \alpha_2} e^{i n_3 \alpha_3} w_{n_1,n_2,n_3}$. If the points are labelled by homogeneous coordinates $\{ w_{n_1,n_2,n_3} \}$, each coordinate transforms under a different phase under $U(1)^3$. Then the only points that can be $U(1)^3$ invariant are those where only single a coordinate $w_{m_1,m_2,m_3}$ is non-zero, and the transformation acts with an overall phase. But these are precisely the discrete set of corner points $p^{\rm max}_{m_1,m_2,m_3}$. 
Thus we see that the corner points can be uniquely identified by their $U(3)$ transformation properties.

We now turn to the question of how to identify explicit brane geometries corresponding to the states $\Psi^{\rm max}_{m_1,m_2,m_3}$ at the corner points $p^{\rm max}_{m_1,m_2,m_3}$.
In order to map a generic state on the moduli space $\cM_C$ parametrized by $w_{n_1,n_2,n_3}$ to an explicit brane geometry defined by a polynomial $P(z)$ we need to know the map between coordinates $w_{n_1,n_2,n_3}$ and polynomial coefficients $c_{n_1,n_2,n_3}$. This is map is highly non-trivial and we do not know the precise mapping functions. However, the special $U(3)$ transformation properties of the corner points allows us to bypass this difficulty. 

We know $p^{\rm max}_{m_1,m_2,m_3}$ is a fixed point of the $U(1)^3$ action. Since $U(3)$ transforms $c_{n_1,n_2,n_3}$ and $w_{n_1,n_2,n_3}$ coordinates in the same way, we can analogously argue that the corresponding point must have a single non-zero $c_{n_1,n_2,n_3}$ coefficient, in order to be invariant under $c_{n_1,n_2,n_3} \rightarrow e^{i n_1 \alpha_1} e^{i n_2 \alpha_2} e^{i n_3 \alpha_3} c_{n_1,n_2,n_3}$. That is, the point in moduli space corresponds to a \emph{monomial}
\begin{equation}
\label{eq:Pz_maximal}
	P(z) = x^{m_1}y^{m_2}z^{m_3} = 0 \, .
\end{equation}
The condition of being a fixed point does not uniquely pick the degrees $(m_1,m_2,m_3)$ to match those in $p^{\rm max}_{m_1,m_2,m_3}$. We can complete this identification by checking the charges. The brane geometry specified by (\ref{eq:Pz_maximal}) consists of $m_1,m_2,m_3$ number of maximal giants wrapped on $x=0,y=0,z=0$ surfaces respectively. It will have charges
\begin{equation}
	L_i = N m_i, \quad E = N (m_1+m_2+m_3)
\end{equation}
which does exactly match the charges of quantum state (\ref{eq:Psi_max}). So we conclude the correspondence
\begin{equation}
\boxed{ 
	P(z)=x^{m_1}y^{m_2}z^{m_3}=0 
	\quad \leftrightarrow \quad
	\Psi^{\rm max}_{m_1,m_2,m_3} = (w_{m_1,m_2,m_3})^N
} 
\end{equation}
Note that this is a very plausible  correspondence: 
$w_{n_1,n_2,n_3}$ in some sense ``corresponds'' to $c_{n_1,n_2,n_3}$, 
and this is the state at the maximal value of the diagonal $E_{II}$ generator 
corresponding to $w_{m_1,m_2,m_3}$.

\subsection{Spectrum of excitations}

Having constructed the states corresponding to maximal giants, 
we now proceed to study physical properties of these 
configurations by looking at the spectrum of excitations.

On general grounds, we expect two kinds of low energy excitations in the 
background of D-branes: open and closed strings. The closed strings
 are  bulk gravitons while the open strings 
 correspond to world-volume excitations such as shape deformations
 of the branes. The eighth-BPS supersymmetric part of the excitation spectrum
  must already be included in our Hilbert space $\cH_C$, and so we want to
 identify these states. From the point of view of classical phase space,
 the neighborhood of the point $p^{\rm max}_{m_1,m_2,m_3}$ can be interpreted 
as small perturbations of the original configuration, as long as the
 difference in energy is $O(1)$. Therefore, the quantum states that 
live in this nearby region should be precisely the quantized supersymmetric
 excitations. We will return to  this picture
from the point of view of a local analysis of the symplectic 
form in Section~\ref{sec:struct_of_pert}.

It is easy  to identify  states in the full Hilbert space are ``near'' the configuration of maximal giants: they will be the ones that differ from the background  $(w_{m_1,m_2,m_3})^N$ in $O(1)$ number of quanta. Applying the same notion 
to the trivial background $w_{0,0,0}^N$ gives the usual Kaluza-Klein spectrum 
of small graviton multiplets. We expect that the location of 
states nearby in this sense, computed with expectation values of $\langle X_{II} \rangle $ as in Section~\ref{sec:cp} will be nearby in terms of the K\"ahler metric
on the moduli space. Expressing this more quantitatively would be interesting 
but is beyond the scope of this paper.  For example:
\begin{equation}
	w_{i_1,i_2,i_3} (w_{m_1,m_2,m_3})^{N-1}, \quad
	w_{i_1,i_2,i_3} w_{j_1,j_2,j_3} (w_{m_1,m_2,m_3})^{N-2}, \quad
	\ldots
\end{equation}
are nearby states in this sense. These nearby states  form a nice Fock space 
structure\footnote{ 
Strictly speaking, the commutators between the raising and lowering operators of different modes do not commute $[A^\dagger_{k_1,k_2,k_3},A_{-l_1,-l_2,-l_3}] \neq \delta_{k_1,l_1}\delta_{k_2,l_2}\delta_{k_3,l_3}$, so it is not literally a Fock space algebra of independent oscillators. Nevertheless, for our purposes this does not have any effect, and we will take the liberty of referring to $A^\dagger_{k_1,k_2,k_3}$ as Fock space generators.
} generated by operators:
\begin{equation}
\label{eq:Adagger_defn}
	A^\dagger_{k_1,k_2,k_3} \equiv w_{m_1+k_1,m_2+k_2,m_3+k_3} \partial_{m_1,m_2,m_3}, \quad
	k_i \ge -m_i, \; k_i \neq (0,0,0)
\end{equation}
Around a fixed background the commutators between different modes vanish: $[A^\dagger_{k_1,k_2,k_3}, A^\dagger_{l_1,l_2,l_3}] = 0$. Individual $A^\dagger_{k_1,k_2,k_3}$ should then correspond to the spectrum of allowed single-particle (string) excitations, as long as $k_i \sim O(1)$.
Note that the $U(1)^3$ charges of $A^\dagger_{k_1,k_2,k_3}$ are just $(k_1,k_2,k_3)$, and some have negative contribution to energy. This is natural since we are not in the vacuum and there are directions in phase space which decrease the energy.

We make the following proposal for interpretation of the excitations:
\begin{itemize}
	\item $A^\dagger_{k_1,k_2,k_3}$ with all $k_i\ge 0$ are the {\bf closed string} excitations in the bulk. This is supported by the fact that the spectrum is the same as the graviton spectrum in the vacuum, which is what we expect. That is, there is a single mode for each choice of non-negative $(k_1,k_2,k_3)$ except $(0,0,0)$, which reproduces the large $N$ partition function $\cZ(x_1,x_2,x_3)=\prod_{n_1,n_2,n_3} \frac{1}{1-x_1^{n_1}x_2^{n_2}x_3^{n_3}}$.
	
	\item $A^\dagger_{k_1,k_2,k_3}$ with at least one $k_i < 0$ are the {\bf open string} excitations on the world-volume. This is a novel spectrum, 
not visible in the perturbations around the vacuum,
but which can be extracted from full partition function of eighth-BPS sector
by looking at states near a giant graviton background. 
The spectrum is dependent on $m_i$, and carries information 
about the geometry of the branes.
\end{itemize}

We will dedicate most of the remaining sections to give further support for our proposed open string spectrum, and to work out some physical implications. In this section we go through a list of examples of backgrounds and take a closer look at the spectra. 

\paragraph{Single giant $z=0$} ~

As discussed in Section~\ref{sec:ex-sphere}, the equation $z=0$ 
describes a giant extended along the 
intersections of the $x,y$-planes with the $S^5$ in space-time.
By the arguments of Section~\ref{sec:cp} and \ref{maxwc}, the background is 
\begin{equation}
\label{eq:Psi_w001N}
\Psi^{\rm max}_{0,0,1} = (w_{0,0,1})^N\,.
\end{equation}
The bulk gravitons are generated by $A^\dagger_{k_1,k_2,k_3}=w_{k_1,k_2,k_3+1} \partial_{0,0,1}$ with non-negative $k_i$. 
The world-volume excitations are generated by 
\begin{equation}
\label{eq:Adagger_z}
  ~~~~~~~~~~~
	A^\dagger_{k_1,k_2,-1} = w_{k_1,k_2,0}\,\partial_{0,0,1}, 
	\quad k_1,k_2\ge 0
\end{equation}
which is only a \emph{two} parameter family. We interpret them as BPS wave modes on $S^3$ brane. Note that these excitations only carry momenta in $L_1,L_2$ but not in $L_3$ direction. In fact this makes sense. The world-volume is the intersection of $z=0$ with the sphere, so it stretches in the $x,y$ plane. Waves on the world-volume will have $L_1,L_2$ angular momenta but not $L_3$. 
We learn from the counting that there is in fact one BPS wave 
solution on the world-volume for each pair of charges $ (k_1, k_2 )$. 

There is a special excitation $A^\dagger_{0,0,-1}$ which decreases the energy and does not add momenta. It can be interpreted as the shrinking mode of the giant. Also $A^\dagger_{1,0,-1}$ and $A^\dagger_{0,1,-1}$ which do not change energy are just $U(3)$ generators for rotations in $(x,z)$ and $(y,z)$ directions.

\paragraph{Two giants $xy=0$} ~

This describes a composite of giants along $x=0$ and $y=0$. 
 As argued in Sections~\ref{sec:cp} and \ref{maxwc}
the corresponding background quantum state is 
\begin{equation}
\Psi^{\rm max}_{1,1,0} = (w_{1,1,0})^N
\end{equation}
This is a configuration of two intersecting branes, as the equation $xy=0$ has two branches 
\begin{equation}
\begin{split}
	x = 0  ~~ &: ~~  |y|^2 + |z|^2 =1 \\
	y = 0 ~~ &: ~~  |z|^2 + |x|^2 =1
\end{split}
\end{equation}
There is also an interesting region where the branes intersect along $S^1$ at $|z|^2=1$. We can distinguish the bulk modes
\begin{equation}
	A^\dagger_{k_1,k_2,k_3} = w_{k_1+1,k_2+1,k_3}\,\partial_{1,1,0}
\end{equation}
and three types of world-volume excitations:
\begin{equation}
\begin{split}
  A^\dagger_{-1,k_2,k_3} &= w_{0,k_2+1,k_3}\,\partial_{1,1,0} \\
  A^\dagger_{k_1,-1,k_3} &= w_{k_1+1,0,k_3}\,\partial_{1,1,0} \\
  A^\dagger_{-1,-1,k_3} &= w_{0,0,k_3}\,\partial_{1,1,0}
\end{split}
\end{equation}
with $k_i\ge 0$. The first two types have analogous spectrum as (\ref{eq:Adagger_z}) and can be interpreted as waves on $x=0$ and $y=0$ giants respectively. There is an additional one-parameter tower $A^\dagger_{-1,-1,k_3}$ which can be interpreted as modes living on the one-dimensional intersection of the $x=0$ and $y=0$ branes, or as strings stretching between the two branes. This intersection extends in the ${\rm arg}(z)$ direction and indeed this tower has $z$-charge. 

Note that $A^\dagger_{-1,-1,k_3}$ is related to the classical deformation $x y = \epsilon z^{k_3}$. The deformation is indeed most significant near the intersection $x=y=0$, where $x,y \sim \sqrt{\epsilon}$, consistent with the interpretation that that's where these open strings live. Since in that region $|z|\approx 1$, we can approximate $z=e^{i \psi}$ and so the intersection gets deformed as:
\begin{equation}
	x = \frac{\epsilon\, e^{i k_3 \psi}}{y}
\end{equation}
that is, with a twist around $S^1$.

Again we should point out that two modes $A^\dagger_{-1,-1,0}$ and $A^\dagger_{-1,-1,1}$ in the one-parameter family have negative energy. They are related to the deformations $xy=\epsilon$ or $xy=\epsilon z$.

\paragraph{Three giants $xyz=0$} ~

This is a composite involving three giants $x=0,y=0,z=0$. 
Following previous arguments, the background state 
\begin{equation}
\Psi^{\rm max}_{1,1,1} = (w_{1,1,1})^N
\end{equation}
The excitations can be summarized in the following table:
\begin{equation}
\label{eq:Atable_xyz}
	\begin{tabular}{c|l}
		Operator & Interpretation \\
		\hline				
		$A^\dagger_{k_1,k_2,k_3}$ & Bulk gravitons \\
		$A^\dagger_{-1,k_2,k_3}$ & Waves on $x=0$ giant \\
		$A^\dagger_{k_1,-1,k_3}$ & Waves on $y=0$ giant \\
		$A^\dagger_{k_1,k_2,-1}$ & Waves on $z=0$ giant \\
		$A^\dagger_{k_1,-1,-1}$ & Waves on $y=z=0$ intersection \\
		$A^\dagger_{-1,k_2,-1}$ & Waves on $x=z=0$ intersection \\
		$A^\dagger_{-1,-1,k_3}$ & Waves on $x=y=0$ intersection \\
		$A^\dagger_{-1,-1,-1}$ & Composite deformation 
	\end{tabular}
\end{equation}
Again in this classification $k_i\ge 0$. The first three families of world-volume excitations, each with two-parameter infinite sequences of modes, can be associated with each of the $S^3$ world-volume branches. The next three one-parameter families naturally correspond to each of the three $S^1$ intersections.

There is a single extra mode $A^\dagger_{-1,-1,-1}$ that does not fall into the categories discussed so far. It corresponds to the deformation $xyz = \epsilon$. Since $x=y=z=0$ is not part of the brane world-volume, the largest effect of the deformation is again near the brane intersections. If we look at each pairwise intersection:
\begin{equation}
\begin{split}
	\text{near } x=y=0, \; z = e^{i \psi_3}&: \qquad  x y = \epsilon e^{-i \psi_3}, \\
	\text{near } y=z=0, \; x = e^{i \psi_1}&: \qquad  y z = \epsilon e^{-i \psi_1}, \\
	\text{near } z=x=0, \; y = e^{i \psi_2}&: \qquad  z x = \epsilon e^{-i \psi_2}.
\end{split}	
\end{equation}
So all three intersections are simultaneously deformed with an ``anti-holomorphic'' twist, something that would locally come from $xy = \epsilon \bar{z}$, $yz = \epsilon \bar{x}$, $zx = \epsilon \bar{y}$. We are led to the conclusion that even though modes like $xy = \epsilon \bar{z}$ are individually not BPS, the particular \emph{composite} described by $xyz=\epsilon$ is BPS. In other words, we interpret $A^\dagger_{-1,-1,-1}$ as a BPS state involving  three open strings living on each $S^1$ intersection, which are individually not BPS.

\paragraph{Stack of branes $z^m=0$} ~

This can be interpreted as a limit of multiple branes located at
$ \prod_{i=1}^m (z-a_i) $ where the $a_i$ tend to zero. 
Background is 
\begin{equation}
	\Psi^{\rm max}_{0,0,m} = (w_{0,0,m})^N
\end{equation}
The bulk excitations are again $A^\dagger_{k_1,k_2,k_3}=w_{k_1,k_2,m+k_3}$ and we have world-volume excitations
\begin{equation}
\label{eq:Atable_zm}
	A^\dagger_{k_1,k_2,-p} = w_{k_1,k_2,m-p}\,\partial_{0,0,m}
\end{equation}
where $p = 1 \cdots m$. It contains excitation spectrum of a single brane $A^\dagger_{k_1,k_2,-1}$, but there are in total $m$ towers parametrized by $p$, with varying $L_3$ charge. This is related to the fact that there are $m$ branes which we can excite, but the identification is not as straightforward, because the branes are coincident ant indistinguishable. We will provide a more detailed interpretation in Section~\ref{sec:half-bps}. 

The fact that we do not have an infinite tower of states 
in the $z$-direction again makes sense, because we do not 
have world-volume extension in that direction. The fact that 
we do get an increasing number of excitations as $m$ increases 
suggests 
that the non-abelian nature of the branes effectively 
blows up an internal circle in the world-volume of the brane 
in the $z$-plane, a circle which in ordinary 
 geometry  looks to be of vanishing 
size.

The excitation spectrum above can be viewed as a prediction, based on 
the eighth-BPS spectrum known from the chiral ring of the $\cN =4 $ 
$U(N)$ SYM. It should match calculations starting from the point of view of 
the non-abelian $U(m)$ gauge theory on coincident branes. 
The arguments for the spectrum of giant graviton physics developed so far have been based largely on the symplectic form derived from the abelian DBI. 
Unraveling excitations of specific geometries allows the possibility 
of using gauge-string duality to predict non-abelian physics of coincident
 branes.  The use of dualities to predict non-abelian brane physics has been 
illuminating in the past  \cite{sendyons,wittenbsp}

\paragraph{Three stacks $(xyz)^m=0$} ~

Finally, let us take a look at the configuration with $m$ coincident branes wrapping each $S^3$. This will in fact display all features of a generic $x^{m_1}y^{m_2}z^{m_3}=0$. The background state is
\begin{equation}
	\Psi^{\rm max}_{m,m,m} = (w_{m,m,m})^N
\end{equation}
and we can classify excitations similarly like before:
\begin{equation}
\label{eq:Atable_xyzm}
	\begin{tabular}{c|l}
		Operator & Interpretation \\
		\hline		
		$A^\dagger_{k_1,k_2,k_3}$ & Bulk gravitons \\
		$A^\dagger_{-p,k_2,k_3}$ & Waves on $x=0$ stack \\
		$A^\dagger_{k_1,-p,k_3}$ & Waves on $y=0$ stack \\
		$A^\dagger_{k_1,k_2,-p}$ & Waves on $z=0$ stack \\
		$A^\dagger_{k_1,-p,-q}$ & Waves on $y=z=0$ intersection \\
		$A^\dagger_{-p,k_2,-q}$ & Waves on $x=z=0$ intersection \\
		$A^\dagger_{-p,-q,k_3}$ & Waves on $x=y=0$ intersection \\
		$A^\dagger_{-p,-q,-r}$ & Composite deformations
	\end{tabular}
\end{equation}
There are now extra parameters $0<p,q,r\le m$.

The structure is similar as for $xyz=0$. First we get modes $A^\dagger_{-p, k_2, k_3}$, etc., living on each stack of branes.

Next, $A^\dagger_{k_1,-p,-q}$ are states living on the $S^1$ intersection. Now they are labeled by extra parameters $(p,q)$ which can take $m^2$ values. This relates to the fact that between two stacks of $m$ branes we have $m^2$ intersections. Just like for modes $A^\dagger_{-p, k_2, k_3}$, here the interpretation is obscured by the fact that the $m^2$ intersections are in fact identical, and $(p,q)$ does not really label the intersection. But we will see in Section~\ref{sec:eighth-bps} when we analyze non-coincident branes, that this multiplicity is indeed related to the number of intersections.

Finally, we have $m^3$ modes $A^\dagger_{-p,-q,-r}$, which are extensions of the composite BPS mode $A^\dagger_{-1,-1,-1}$ to the case of multiple branes.


\section{Excitations from partition function}
\label{sec:single_partition}

In this section we will show how the results of Section~\ref{sec:states2geom}, 
which were interpreted in terms of the physics of branes, are reflected 
in the partition function. Since the partition 
function of BPS states is known from the dual $U(N)$ SYM  field theory side, 
we can view the calculations in this section as recovering,  
from the dual field theory, without a priori information from branes,
the  factorization into bulk and world-volume   states
which is expected from AdS/CFT duality. Specifically
we will  extract the spectrum of BPS excitations around a single 
half-BPS sphere giant. The factorization of the spectrum into bulk graviton states and world-volume excitations (\ref{eq:Adagger_z}) will be obtained here 
by considering a limit of the partition function which isolates the 
states discussed as being ``near'' the single giant $z=0$ in Section~\ref{sec:states2geom}. 

Recall the partition function of the bosonic eighth-BPS sector in $SU(N)$ $\cN=4$ SYM graded by $R$-charges $L_1, L_2, L_3$ is:
\begin{align}
\label{eq:ZN_defn}
  &\cZ_N(x_i) = \Tr_{\cH}\left( x_1^{L_1}x_2^{L_2}x_3^{L_3} \right)
  = \sum_{n_1,n_2,n_3} x_1^{n_1}x_2^{n_2}x_3^{n_3} \cZ_{N;n_1n_2n_3}
  \\
\label{eq:Znu_defn}  
	&\cZ(\nu;x_i) = \sum_{N=0}^\infty \nu^N \cZ_N(x_i)
	= \prod_{n_1,n_2,n_3=0}^{\infty} \frac{1}{1-\nu x_1^{n_1}x_2^{n_2}x_3^{n_3}}.
\end{align}
That is, $\cZ_N(x_i)$ is the partition function counting operators at a fixed $N$, while $\cZ(\nu;x_i)$ is the ``grand canonical'' partition function with chemical potential $\nu$ for $N$. Then $\cZ_N(x_i)$ can be calculated as the coefficient of $\nu^N$ in the RHS of (\ref{eq:Znu_defn}). This result can be derived by counting symmetric polynomials of the $3N$ values of the three diagonal complex scalar matrices $X_i$, after enforcing F-term constraints $[X_i,X_j]=0$. 

We already saw that $\cZ_N(x_i)$ is also reproduced by the Hilbert space (\ref{eq:HC_definition}), which is generated by oscillators $w_{n_1,n_2,n_3}$ with the restriction that the total number of excitations is $N$. This structure of the partition function is clearly seen from the RHS of (\ref{eq:Znu_defn}). Thus from the perspective of SYM, the Hilbert space (\ref{eq:HC_definition}) can be seen as just a formal construction of states accounting for the partition function. One common interpretation of this space is as the Hilbert space of $N$ bosons in a 3D harmonic oscillator. Then $w_{n_1,n_2,n_3}$ puts a single boson in the state $(n_1,n_2,n_3)$.

In order to use $\cZ_N(x_i)$ to extract the BPS spectrum around a half-BPS sphere giant\footnote{
Here we consider the giant to be of any size $E\le N$, not necessarily maximal}
we must first find a way to isolate the state corresponding to the giant itself. The charges are obviously not enough, because once we fix $L_1=L_2=0,\; L_3=E$ we get \emph{all} of the half-BPS states, and only one of them is a sphere giant (assuming $E\le N$). Recall the half-BPS states can be labelled by Young diagrams with $E$ boxes and height $\le N$. The giant states that we want to focus on are those labelled by the single-column Young diagrams.

In order to find the single-column state we introduce an extra quantum number \emph{size} $S$ by which we label the eighth-BPS states. We can do this by using the $N$-dependence of the Hilbert spaces $\cH_N$. It is natural to consider the sequence of subspaces
\begin{equation}
	\cH_1 \subset \cH_2 \subset \ldots \subset \cH_{N-1} \subset \cH_N
\end{equation}
For example, an operator like $\tr(X^2)$ is considered ``the same state'' for any $N$. Then if we pick an operator $\cO$ we can ask at what $N$ it gets \emph{excluded}. We label the operator to have size $S(\cO)$ if it gets excluded below $S$:
\begin{equation}
	\cO \in \cH_N, \quad \text{iff} ~~ N\ge S
\end{equation}
In the half-BPS sector the state $R$ in the Schur basis gets excluded when $N$ is below the height of the Young diagram $c_1(R)$, so $S=c_1(R)$. In the eighth-BPS sector if we represent states as (\ref{eq:HC_definition}), then $S$ is just the number of excitations different from $w_{0,0,0}$. Or, in terms of $N$ bosons, it is the number of bosons in excited states. It is reasonable that $S$ has a physical interpretation in both gauge and the gravity side. It measures how close the state is to the ``exclusion bound''. For example a sphere giant has $S=E$, and maximal giants are those with $S=N$. A dual giant, on the other hand, has $S=1$.

The partition function for number of states refined by $(S,L_1,L_2,L_3)$ is easy to get. If $Z_{S;n_1n_2n_3}$ is the number of such states then:
\begin{align}
\label{eq:ZS_from_ZN}
	Z_{S;n_1,n_2,n_3} &= \cZ_{S;n_1,n_2,n_3} - \cZ_{S-1;n_1,n_2,n_3}
\end{align}
and
\begin{equation}
\label{eq:Z_nu_xi}
\begin{split}
	Z(\nu;x_i) &\equiv \sum_{S;n_1,n_2,n_3} \nu^S x_1^{n_1} x_2^{n_2} x_3^{n_3} Z_{S;n_1,n_2,n_3} = (1-\nu)\cZ(\nu;x_i) \\
	&= \prod_{n_1+n_2+n_3>0} \frac{1}{1-\nu x_1^{n_1}x_2^{n_2}x_3^{n_3}}
\end{split}
\end{equation}
The only difference from (\ref{eq:Znu_defn}) is that we do not have a term $1/(1-\nu)$, so we only count bosons in excited states.

Now we can uniquely identify the single sphere giant with energy $E$ by specifying $(S,L_1,L_2,L_3) = (E,0,0,E)$. In terms of oscillators this is $(w_{0,0,1})^S$ as in (\ref{eq:Psi_w001N}), but not necessarily maximal. The excitations around this state should have charges differing by $O(1)$ from the background. Let us fix the size, and look at states with charges $(S,L_1,L_2,L_3)=(S,n_1,n_2,S+n_3)$ for small $n_i$. The number of such states is:
\begin{equation}
	\wt{Z}_{S;n_1,n_2,n_3} \equiv Z_{S;n_1,n_2,n_3+S}
\end{equation}
We can write the corresponding partition function
\begin{equation}
	\wt{Z}(\nu;x_i) = \sum_{S;n_1,n_2,n_3} \nu^S x_1^{n_1} x_2^{n_2} x_3^{n_3} Z_{S;n_1,n_2,n_3+S} = Z\left(\frac{\nu}{x_3};x_i\right)
\end{equation}
where the RHS is known explicitly (\ref{eq:Z_nu_xi}). Furthermore, we expect the counting $\wt{Z}_{S;n_1,n_2,n_3}$ to be independent of $S$ if $n_i \ll S$. That is, the spectrum of excitations should not depend on the size of the giant. We can confirm this by taking $S\rightarrow \infty$ limit, which in terms of the partition function reads as
\begin{equation}
	\wt{Z}(x_i) = \lim_{\nu\rightarrow 1} (1-\nu) \wt{Z}(\nu;x_i)
	= \prod_{\substack{n_1+n_2+n_3>0 \\ (n_1,n_2,n_3)\neq(0,0,1)}} \frac{1}{1- x_1^{n_1}x_2^{n_2}x_3^{n_3-1}}
\end{equation}
This produces finite counting for $O(1)$ charges. The partition function can be conveniently factored into pieces where $n_3=0$ and $n_3>0$:
\begin{equation}
\label{eq:tildeZ}
	\wt{Z}(x_i) = \left(
		\prod_{n_1+n_2>0} \frac{1}{1- x_1^{n_1}x_2^{n_2}x_3^{-1}}
	\right)
	\left(
		\prod_{n_1+n_2+n_3>0} \frac{1}{1- x_1^{n_1}x_2^{n_2}x_3^{n_3}}
	\right)
\end{equation}
where in the second term we renamed $n_3-1 \rightarrow n_3$.

The spectrum (\ref{eq:tildeZ}) that we found is almost exactly (\ref{eq:Adagger_defn}). The first factor is generated by $A^\dagger_{n_1,n_2,-1}$ and interpreted as world-volume excitations. The second factor corresponds to $A^\dagger_{n_1,n_2,n_3}$ with non-negative $n_i$ and generates the background graviton spectrum. We are missing here the negative energy mode $A^\dagger_{0,0,-1}$, but that's just because we fixed the size $S$ to be constant in the derivation, while $(0,0,-1)$ is precisely the mode that decreases size by 1.

We have now demonstrated how to take a limit of the partition 
function to achieve a factorization into closed and open strings. 
The same factorization was obtained  in Section~\ref{sec:states2geom}
by explicitly looking at the Fock space structure of the states. The 
quantum number $S$ related to the exclusion of states with varying $N$, was the additional data beside $R$-charges we needed to accomplish this. 
For more general brane configurations discussed in 
Section~\ref{sec:states2geom} we would need additional quantum numbers 
such as the higher conserved charges which determine a Young diagram 
in the half-BPS case \cite{cjr,ehs}. These higher charges which exist 
in the oscillator Hilbert space have not yet been exhibited from 
the gauge theory point of view at weak coupling. The story at zero 
coupling in the eighth-BPS sector is developed in \cite{ehs}.
The strategy of extracting the expected open-closed
factorization of states from the partition function should be specially instructive for unraveling 
the giant graviton physics in more general examples of AdS/CFT 
where the $S^5$ is replaced by a Sasaki-Einstein geometry. In these cases, 
the partition function is known from the dual quiver gauge theory 
but the matching of these states with giant gravitons extended in the 
Sasaki-Einstein space is a largely unexplored subject.


\section{Excitations from local quantization}
\label{sec:single_symplectic}

In this section we will see how to derive
 the spectrum (\ref{eq:Adagger_z}) of the world-volume excitations on a sphere giant directly from the brane action. This provides a non-trivial check of the analysis in Section~\ref{sec:states2geom} without relying on the fact that the phase space is isomorphic to $\CP^{n_C-1}$ with Fubini-Study symplectic form.

We will mostly focus on the case of maximal giant $P(z)=z=0$ as in (\ref{eq:Adagger_z}), but the analysis here works for non-maximal sphere giants $P(z)=z - c_0=0$ too, see Appendix~\ref{app:w_derivation}. This in fact provides evidence that generic analysis in Section~\ref{sec:states2geom} should also work for non-maximal giants.


\subsection{Structure of perturbations}
\label{sec:struct_of_pert}

We start in this section by revisiting the space of perturbations of a spherical giant.

The polynomial defining the unperturbed maximal giant is
\begin{equation}
	P_0(z) = z  = 0 .
\end{equation}
This is a point in the space of polynomials $\cP$ and also in the phase space $\cM$.
In order to study perturbations, we need to identify the neighborhood of $P_0$ in $\cM$. Naively, one might guess that it is the image of the neighborhood in $\cP$, so a nearby point in $\cM$ corresponds to
\begin{equation}
\label{eq:P_deltac}
	P_{\delta c}(z)= z + \sum_{n_1,n_2,n_3=0}^\infty \delta c_{n_1,n_2,n_3} x^{n_1} y^{n_2} z^{n_3} \, .
\end{equation}
First, we note that any perturbation involving a non-zero power of $z$ in fact does not deform the giant at all. This can be seen from the following factorization:
\begin{equation}
\begin{split}
	P_{\delta c}(z) = 
	\left(z + \sum_{n_1,n_2} \delta c_{n_1,n_2,0}\, x^{n_1} y^{n_2} \right) 
	\left(1 + \sum_{n_3>0,n_1,n_2} \delta c_{n_1,n_2,n_3} \, x^{n_1} y^{n_2} z^{n_3-1} \right) 
\end{split}
\end{equation}
where we drop $O(\delta c^2)$ terms.	The second factor does not intersect $S^5$, so under the map $\cP\rightarrow \cM$, $P_{\delta c}$ has to be identified with just the first factor:
\begin{equation}
\label{eq:Pdeltab}
	P_{\delta b}(z) = z + \sum_{n_1,n_2} \delta b_{n_1,n_2}\, x^{n_1} y^{n_2} 
\end{equation}
This is the subspace of $P_0$ neighborhood in $\cP$ that corresponds to neighborhood in the actual phase space $\cM$ .

There is another problem with the guess (\ref{eq:P_deltac}) in that it does not, in fact, explore the whole neighborhood of $P_0$ in $\cM$. Intuitively the reason is that we should be able to add an infinitesimally small disconnected surface by e.g. $P(z)=z(c\,z - 1)$ with $|c|^2=1+\epsilon$, but this polynomial is not a small deformation of $P_0$ in $\cP$. We can handle this case by recalling that there are many polynomials identified with the same point $P_0$ in $\cM$, namely, any
\begin{equation}
	\tilde P_0(z) = z\, Q(z)
\end{equation}
where $Q(z)=0$ does not intersect $S^5$. Any polynomials which are near $\tilde P_0(z)$ then also correspond to points in $\cM$ near $P_0$. In particular, if we consider $Q_0(z)$ which just touches $S^5$ then a deformation
\begin{equation}
	\tilde P(z) = z\, \left( Q_0(z) + \delta Q(z) \right)
\end{equation}
corresponds to a new point in $\cM$ near $P_0$, not included in (\ref{eq:P_deltac}). The physical interpretation of this class of deformations is clear from the factorized form of $\tilde P(z)$: with $\delta Q(z)$ we are adding infinitesimally small disconnected surfaces rather than deforming the shape of the original sphere giant.

The final conclusion of this section is then that the most general perturbation of a sphere giant $z=0$ is given by
\begin{equation}
\label{eq:P_deltabQ}
	P(z) = \left( 
		z + \sum_{n_1,n_2=0}^\infty \delta b_{n_1,n_2}\,x^{n_1} y^{n_2}
	\right) \,
	\left( Q_0(z) + \delta Q(z) \right)
\end{equation}
such that $Q_0(z)=0$ touches $S^5$ and $Q_0(z)+\delta Q(z)$ intersects it. The first factor involving $\delta b_{n_1,n_2}$ deforms the surface $z=0$, while the second factor adds infinitesimally small disconnected surfaces. The action and the symplectic form for the two pieces is independent, because it involves an integral over each surfaces separately. That means we have a product structure to the phase space in the neighborhood of $P_0$
\begin{equation}
	\cM_{P_0} = \cM_{P_0}^{\rm wv} \times \cM_{P_0}^{\rm bulk}
\end{equation}
which has a natural interpretation as world-volume and bulk excitations.

If we perform the quantization locally, we will get a product of Hilbert spaces $\cH_{P_0} = \cH_{P_0}^{\rm wv} \times \cH_{P_0}^{\rm bulk}$, as long as the excitation number is small so we stay in the neighborhood. Furthermore, note that $\cM_{P_0}^{\rm bulk}$ is exactly the same as the full phase space $\cM$ around the vacuum point. That is, we might as well be considering quantization of $Q(z)=0$ which barely intersects $S^5$, the existence of $z=0$ brane does not have an effect. That means, we know what $\cH_{P_0}^{\rm bulk}$ is -- it matches the low-energy spectrum of the full $\cH$ and describes bulk gravitons, generated by Fock space of $w_{n_1,n_2,n_3}$. We identify $A^\dagger_{k_1,k_2,k_3}$ in (\ref{eq:Adagger_defn}) with non-negative $k_i$ as the operators generating this ``closed string'' Fock space around a giant.

The remaining problem is then to get the world-volume
spectrum $\cH_{P_0}^{\rm wv}$ arising from perturbations (\ref{eq:Pdeltab}).


\subsection{Quantization of world-volume excitations}
\label{sec:general_pert}

We now turn to the analysis of the world-volume deformations of the maximal giant
\begin{equation}
\label{eq:P_deltab}
	P(z) = z + \sum_{m,n\ge0} \delta b_{m,n}\,x^m y^n	
\end{equation}
We want to explicitly calculate the symplectic form on this slice of phase space (assuming $|\delta b|^2 \ll 1$) and subsequently quantize it. This process of quantizing the phase space ``locally'' around a solution is analogous to canonical quantization of first-order perturbations using quadratic effective action \cite{Rychkov:2005nk}.

We are deforming an $S^3$ at $z=0$:
\begin{equation}
\begin{split}
	|x|^2 + |y|^2 &= 1 
\end{split}
\end{equation}
Let us introduce some world-volume coordinates $(\sigma^1, \sigma^2, \sigma^3)$ on $S^3$, then $x(\sigma^i)$, $y(\sigma^i)$ are embedding functions. Small time-dependent perturbations around the spherical shape can be parametrized by the function $z(\sigma^i, t)$. Effectively these are the 2 real transverse coordinates to $S^3$ in $S^5$, which is a single complex scalar field on the world-volume. In principle for non-zero $z(\sigma^i, t)$ we need to modify $x(\sigma^i,t)$, $y(\sigma^i,t)$ such that $|x|^2+|y|^2+|z|^2=1$ still holds, however, for $|z|\ll 1$ this effect is second order in perturbation, and we can ignore it. In that case the full symplectic form (\ref{eq:omega_defined}) simplifies to (see Appendix~\ref{app:w_derivation}):
\begin{equation}
\label{eq:w_zz}
	\omega = \frac{2N}{\pi^2} \int_{S^3} \d^3 \sigma \,
	\left(
		\frac{\delta \bar z \wedge \delta z}{2i}
		- \frac{\delta \dot{\bar z} \wedge \delta z}{8}
		+ \frac{\delta \bar z \wedge \delta \dot z}{8}
	\right)
\end{equation}
where the integral $\d^3 \sigma$ is over unit $S^3$ with its standard volume form.

If we put the time-dependence back in (\ref{eq:P_deltab}) according to (\ref{eq:Pxyz_timedep}) we get
\begin{equation}
\label{eq:z_bmn}
\begin{split}
	z &= \delta z = - \sum_{m,n\ge0} \delta b_{m,n} e^{(m+n-1)it}  x^m y^n \, .
\end{split}
\end{equation}
Plugging this in (\ref{eq:w_zz}) we find
\begin{equation}
	\omega = \frac{2N}{2 \pi^2} \int_{S^3} \d^3 \sigma \,
	\sum_{m,n\ge0}
	(m+n+1)
	|x|^{2m} |y|^{2n}
	\frac{\delta \bar b_{m,n} \wedge \delta b_{m,n}}{2i}	
\end{equation}
The integral is easy to do:
\begin{equation}
	\int_{S^3} \d^3 \sigma \, |x|^{2m} |y|^{2n}
	= 2\pi^2 \frac{m! \, n!}{(m+n+1)!}
\end{equation}
Note that we never needed the explicit choice of the coordinate $\sigma^i$ on the sphere. The final symplectic form evaluated at $P(z)=z$ is thus
\begin{equation}
\label{eq:w_bmn}
	\omega = 2N \sum_{m,n\ge0} 
		\frac{m! \, n!}{(m+n)!}		
		\frac{\delta \bar b_{m,n} \wedge \delta b_{m,n}}{2i}	
\end{equation}

Symplectic form (\ref{eq:w_bmn}) is just that of a flat $\mC^{n_C-1}$, and has a simple structure of decoupled harmonic oscillators $\delta b_{m,n}$. Quantization of these perturbations has a straightforward Fock space structure
\begin{equation}
\label{eq:psi_bmn}
	\Psi = \prod_{m,n} (b_{m,n})^{k_{m,n}}
\end{equation}
The $U(1)^3$ charges of the oscillators can be inferred from the transformation of $\delta b_{m,n}$ in (\ref{eq:P_deltab}) under $z^i \rightarrow e^{i\alpha_i} z^i$:
\begin{equation}
	P(z)=z + \sum_{m,n\ge0} \delta b_{m,n} x^m y^n
	\;\rightarrow\;
	e^{i\alpha_3} \left( z + \sum_{m,n\ge0} \delta b_{m,n}\,e^{i m \alpha_1 + i n\alpha_2 - i \alpha_3} x^m y^n \right)
\end{equation}
We have factored out an overall irrelevant phase, to keep $z$ term unchanged. This means $b_{m,n}$ have charges $(L_1,L_2,L_3) = (m, n, -1)$.
This does precisely match the spectrum of world-volume excitations $A^\dagger_{k_1,k_2,-1}$ proposed in (\ref{eq:Adagger_z}).

One way to see this result, is as the derivation of the relationship between $c_{n_1,n_2,n_3}$ coordinates on $\cP$ and $w_{n_1,n_2,n_3}$ on $\cM$ in this particular region. If we expand the Fubini-Study form (\ref{eq:omega_fs}) in the inhomogeneous coordinate patch $w_{0,0,1}=1$, then we know the symplectic form around $P(z)=z=0$ must be
\begin{equation}
\label{eq:omega_w_approx}
	\omega = 2 N \sum_{(n_1,n_2,n_3)\neq(0,0,1)} \frac{\d \bar w_{n_1,n_2,n_3} \wedge \d w_{n_1,n_2,n_3}}{2i}
\end{equation}
for $|w|^2 \ll 1$. The $U(1)^3$ charges of $w_{n_1,n_2,n_3}$ in this patch are $(n_1,n_2,n_3-1)$. Comparing with (\ref{eq:w_bmn}) we can thus identify the coordinates\footnote{
Perhaps it is clearer in terms of homogeneous coordinates: $\frac{w_{m,n,0}}{w_{0,0,1}} \approx \sqrt{\frac{m! \, n!}{(m+n)!}} \frac{c_{m,n,0}}{c_{0,0,1}}$, where $w_{0,0,1},c_{0,0,1}\rightarrow\infty$
}:
\begin{equation}
	w_{m,n,0} = \sqrt{\frac{m! \, n!}{(m+n)!}} \delta b_{m,n}
\end{equation}
up to corrections of order $O(|\delta b|^2)$.
The remaining coordinates $w_{n_1,n_2,n_3}$ with $n_3\ge 1$ must be associated with the directions in the phase space which add disconnected surfaces. Note, however, following the discussion in the previous section, we can not say that $w_{n_1,n_2,n_3}$ is proportional to $\delta c_{n_1,n_2,n_3}$ in (\ref{eq:P_deltac}), although it does have the same charges.

Finally, let us say a word about the limits of approximation in this section. Given the symplectic form (\ref{eq:w_bmn}) in $\delta b_{m,n}$ coordinates, a single quantum state occupies an area in phase space
\begin{equation}
	|\Delta b_{m,n}|^2 \sim \frac{1}{2N}\frac{(m+n)!}{m!\,n!}
\end{equation}
If we require to stay in the region $\delta b_{m,n} \ll 1$, there is only a finite number of states available to fill, and so the number of excitations in state $\ref{eq:psi_bmn}$ should obey
\begin{equation}
	k_{m,n} \ll 2N \frac{m!\,n!}{(m+n)!}
\end{equation}
Note if both $m,n$ are non-zero, the right-hand side could be much less than $N$. This limit is misleading, however. More precisely, the requirement for approximation (\ref{eq:w_zz}) to be valid is that $\delta z, \delta \dot{z} \ll 1$ in (\ref{eq:z_bmn}). We can just as well require the whole integral over $S^3$ to be small, which, looking at (\ref{eq:w_bmn}) boils down to
\begin{equation}
	\frac{m! \, n!}{(m+n)!} \, |\delta b_{m,n}|^2 \ll 1
\end{equation}
So the approximation can be valid even if $\delta b_{m,n} \gg 1$, given $m,n$ are large. In fact, it is valid precisely where $w_{m,n,0} \ll 1$. This is just what we expect from the global picture, because at $w_{m,n,0} \sim O(1)$ the phase space starts looking like $\CP^{n_C-1}$ rather than just local $\mbC^{n_C-1}$. In $w_{m,n,0}$ coordinates (\ref{eq:omega_w_approx}) a single quantum state occupies area $|w_{m,n,0}|^2 \sim \frac{1}{N}$ so the true limit is
\begin{equation}
	k_{m,n} \ll N
\end{equation}
independent of the mode. This is consistent with the requirement $\sum k_{m,n} \le N$, which we know from the global quantization.

The limit on the mode numbers $m,n$ would be set not by the approximations in our derivation, but rather by the validity of DBI action itself. Since we are in the BPS sector, the string length does not play a role, but we can certainly worry if the waves on the brane have wavelengths of less than Planck length. Recall the Planck length is $N^{-1/4}$ in units of $AdS$ radius, while the wavelengths for mode $\delta b_{m,n}$ are $m^{-1}$ and $n^{-1}$. Requiring them to be longer than Planck length sets a limit
\begin{equation}
	m,n \ll N^{1/4}
\end{equation}
For states with higher quantum numbers the interpretation as waves on the brane will not hold.


\section{World-volume excitations beyond maximal giants}
\label{sec:non-maximal}

In this section we further generalize the spectrum of excitations around giants found in (\ref{eq:Adagger_defn}). In particular, we wish to study more general backgrounds, including non-maximal giants. The discussion here will be more qualitative, nevertheless it will allow us to make connections to previous work in the literature, and also suggest possible directions for future work.

\subsection{Half-BPS backgrounds}
\label{sec:half-bps}

\begin{figure}[h]
\begin{center}
\includegraphics{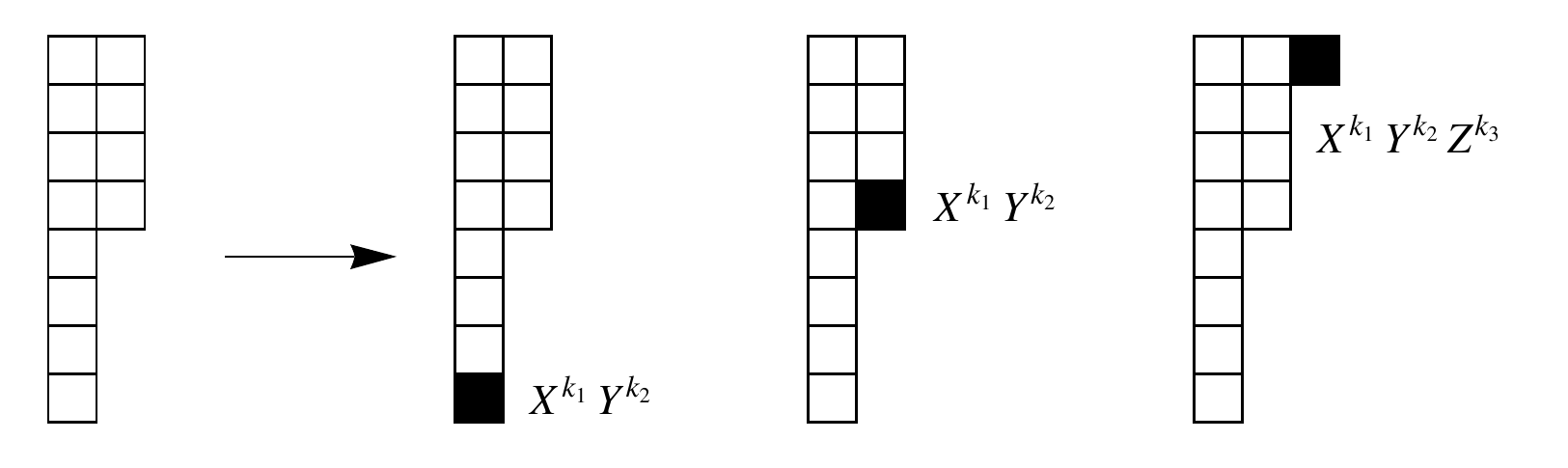}
\caption{Young diagram representation of a two giant state $(w_{0,0,2})^{r_2}(w_{0,0,1})^{r_1} (w_{0,0,0})^{N-r_1-r_2}$ and its BPS excitations. Each white row of length $k_3$ corresponds to a factor $w_{0,0,k_3}$ in the wavefunction and a row with impurity of charge $(k_1,k_2,0)$ and $k_3$ white boxes corresponds to factor $w_{k_1,k_2,k_3}$. }
\label{fig:impurities-zz-abel}
\end{center}
\end{figure}

Consider a general half-BPS state
\begin{equation}
\label{eq:Psi_halfbps}
	\Psi = (w_{0,0,p})^{r_p} (w_{0,0,p-1})^{r_{p-1}} \ldots (w_{0,0,2})^{r_2} (w_{0,0,1})^{r_1} (w_{0,0,0})^{N-\sum r_i}
\end{equation}
In the Schur polynomial basis it corresponds to a Young diagram with $r_i$ rows of length $i$. Recall the individual sphere giants in the state can be associated with \emph{columns} of the diagram. For starters we want to pick a background which has a few well-separated giants, for example 
\begin{equation}
\label{eq:Psi_z1z2_giants}
	\Psi = (w_{0,0,2})^{r_2}(w_{0,0,1})^{r_1} (w_{0,0,0})^{N-r_1-r_2}
\end{equation}
with $r_1,r_2\sim O(N)$. This state has a bigger giant of size $r_1+r_2$ and a smaller one of size $r_2$, and is dual to operator $\chi_{[2^{r_2}1^{r_1}]}(Z)$. See Figure~\ref{fig:impurities-zz-abel}. This quantum state corresponds to the classical configuration\footnote{
Remember in this case, unlike Section~\ref{sec:states2geom}, the quantum state $\Psi$ is not actually localized at a particular $c_1,c_2$ but rather the wavefunction is spread out along the torus $(c_1,c_2)\rightarrow(e^{i\theta_1}c_1,e^{i\theta_2}c_2)$. See the discussion in Section~\ref{sec:cp}.
}:
\begin{equation}
	P(z) = (z - c_1) (z - c_2) .
\end{equation}

We want to find the spectrum of excitations around this configuration. From previous sections we know the open string spectrum on a single brane is $A^\dagger_{k_1,k_2,-1}$, and this applies to non-maximal branes too\footnote{
One difference for non-maximal giants is that besides the shrinking mode $A^\dagger_{0,0,-1}=w_{0,0,0}\,\partial_{0,0,1}$ we can also act on the background with its conjugate $A_{0,0,1}= w_{0,0,1}\, \partial_{0,0,0}$, which will grow the brane as long as it is not maximal}. 
Since the two giants are separated, it is reasonable to expect that each one should carry excitations of a single giant. In fact that is what the symplectic form tells us -- integrate over each surface separately -- and we can consider deforming each giant as in (\ref{eq:P_deltabQ}) and quantizing those deformations. So we expect Fock space generators with charges:
\begin{equation}
	A^{(1)\,\dagger}_{k_1,k_2,-1} \;,\quad
	A^{(2)\,\dagger}_{k_1,k_2,-1} \;,\quad
	A^{(c)\,\dagger}_{k_1,k_2,k_3} 
\end{equation}
generating world-volume excitations on each brane and the bulk closed string states (hence superscript $c$) respectively. The representations of these operators when acting on the state (\ref{eq:Psi_z1z2_giants}) can be constructed as:
\begin{equation}
\begin{split}
	A^{(1)\,\dagger}_{k_1,k_2,-1} &= w_{k_1,k_2,0}\,\partial_{0,0,1} \\
	A^{(2)\,\dagger}_{k_1,k_2,-1} &= w_{k_1,k_2,1}\,\partial_{0,0,2} \\
	A^{(c)\,\dagger}_{k_1,k_2,k_3} &= w_{k_1,k_2,2+k_3}\,\partial_{0,0,2}
\end{split}
\end{equation}
Note that any state differing in $O(1)$ oscillators can be build from these, and other generators we might consider, such as $w_{k_1,k_2,0}\,\partial_{0,0,2}$ or $w_{k_1,k_2,2}\,\partial_{0,0,1}$, are not independent.

A suggestive way to visualize this spectrum of excitations is shown in Figure~\ref{fig:impurities-zz-abel}.
The rule adopted here to go from diagrams to states is the following: each $w_{0,0,k_3}$ in the state is represented by $k_3$ length row of white boxes. If, however, we have a generic $w_{k_1,k_2,k_3}$, it is represented by a row with $k_3$ white boxes and one ``impurity'' carrying charge $(k_1,k_2,0)$. The diagrams for excited states, in fact, look very much like the ``restricted Schur'' operators \cite{bbfh,dMSSI}, which are dual to giants with open string excitations. 

The correspondence with the restricted Schur operators can indeed be made quite precise. First, consider states with a single attached string of $Y$-charge $k_2$. There are two such states and the dual SYM operators are constructed by adding a $Y^{k_2}$ impurity to either corner of the Young diagram. They can be expressed as
\begin{equation}
	\cO_{R, R_1}(Y^{k_2}) = 
	\sum_{\sigma \in S_n} 
	\Tr_{R_1}(\Gamma_R(\sigma))
	\Tr\left(\sigma Z^{\otimes n-1} \otimes (Y^{k_2}) \right)	
\end{equation}
with $R$ the background Young diagram and $R_1$ the diagram after removing the impurity box. In group theoretic terms, $R$ labels a representation of $S_n$, and $R_1$ labels an irreducible component of $R$ under reduction to subgroup $S_n\rightarrow S_{n-1}$. See \cite{dMSSI} for details.
The states with $Y^{k_2}$ impurity are known to be ``near-BPS'', and can be corrected by $O(1/N)$ terms to be BPS, without affecting the counting \cite{Berenstein:2003ah}. In our Hilbert space we also have two single-particle world-volume excitations
 $A^{(i)^\dagger}_{0,k_2,-1}$ with $i\in\{1,2\}$, and so we can identify them with the dual operators\footnote{Here we are making comparison at the level of counting, so the dual operators might actually be some linear combinations of $A^\dagger$ operators with the same charges. Also, by the dual operator we mean the exact BPS state, for which the restricted Schur operator (being near-BPS) is the leading term.}. 
If we consider a more general single string excitation with $(X,Y)$ charges $(k_1,k_2)$, we can construct many distinct impurities, corresponding to different orderings of $X,Y$ in $X^{k_1}Y^{k_2}$, and attach them to either of the two corners. Most of these are dual to excited states of the attached string spin chain\cite{bcv}, which are not BPS. Among the states, there are precisely two (near-)BPS combinations, those where $X,Y$ are symmetrized. That is because the symmetrized combination is related by a $U(2)$ transformation to the $Y^{k_1+k_2}$ impurity. The two BPS states can thus be identified with $A^{(i)\dagger}_{k_1,k_2,-1}$ (Figure~\ref{fig:impurities-zz-abel}).

Now let us compare states with multiple attached strings. For simplicity take the attached strings to be distinct. We fix the $i$'th attached string to have charges $(k_i,l_i)$, and also to be in the ground state. That means the dual state will have $n$ impurities $(X^{k_1}Y^{l_1}), (X^{k_2}Y^{l_2}), \ldots (X^{k_n}Y^{l_n})$, with $X,Y$ again symmetrized. According to \cite{bbfh,dMSSI}, the state is then labelled by pairs of Young diagrams with $n$ labels, where the labeling indicates the order in which $n$ boxes are removed (same boxes in each pair). For example, with two distinct strings on two giants we have 6 states:
\begin{equation}
\label{eq:rest_schur_ops}
	({\tiny \young(\,\,,\,\,,\,\,,\,\,,\,,2,1)},
	 {\tiny \young(\,\,,\,\,,\,\,,\,\,,\,,2,1)}),\quad
	({\tiny \young(\,\,,\,\,,\,2,\,1,\,,\,,\,)},
	 {\tiny \young(\,\,,\,\,,\,2,\,1,\,,\,,\,)}),\quad
	({\tiny \young(\,\,,\,\,,\,\,,\,2,\,,\,,1)},
	 {\tiny \young(\,\,,\,\,,\,\,,\,2,\,,\,,1)}),\quad
	({\tiny \young(\,\,,\,\,,\,\,,\,1,\,,\,,2)},
	 {\tiny \young(\,\,,\,\,,\,\,,\,1,\,,\,,2)}),\quad	
	({\tiny \young(\,\,,\,\,,\,\,,\,2,\,,\,,1)},
	 {\tiny \young(\,\,,\,\,,\,\,,\,1,\,,\,,2)}),\quad	
	({\tiny \young(\,\,,\,\,,\,\,,\,1,\,,\,,2)},
	 {\tiny \young(\,\,,\,\,,\,\,,\,2,\,,\,,1)}) 	 	 	 
\end{equation}
In group theoretic terms these states correspond to
\begin{equation}
	|R \rightarrow R_1,i\rangle \langle R\rightarrow R_1,j|
\end{equation}
where $|R \rightarrow R_1,i\rangle$ labels the irreducible component of $R$ under $S_n \rightarrow S_{n-2}$. Extra label $i$ is necessary because $R_1$ can occur multiple times.
The question is, which of the operators are BPS, and thus can be matched to the BPS excitations we consider. The precise answer in general is not known, since the one-loop dilatation action is complicated, and it is hard to find the BPS states annihilated by it. Nevertheless, the states where the labels of two diagrams are different (also called ``off-diagonal'' restricted Schurs) have a natural interpretation as containing strings stretched between branes, whereas if the two diagrams are the same (``on-diagonal''), all strings are attached individually to one of the branes. Stretched strings can not be BPS, since the tension energy would be much higher than the charge, so only the on-diagonal restricted Schur operators are candidates for BPS states. In the example (\ref{eq:rest_schur_ops}) these would be the first 4 operators. It is easy to see that with $n$ impurities on two giants there are $2^n$ such states (given $n<r_1,r_2$, which we assume to be true). We simply pick from which column to remove each of the $n$ boxes, and get a labelling. Incidentally, there are also $2^n$ BPS excitations in our Hilbert space, if we specify charges of individual open strings to be $(k_i,l_i)$. These excitations are constructed as:
\begin{equation}
	A^{(i_1)^\dagger}_{k_1,l_1,-1} A^{(i_2)^\dagger}_{k_2,l_2,-1} \ldots A^{(i_n)^\dagger}_{k_n,l_n,-1}
\end{equation}
with each $i_a\in\{1,2\}$. This suggests a correspondence between arbitrary BPS world-volume excitations and on-diagonal restricted Schur operators, at least at the level of counting. Still, the precise BPS operators annihilated by one-loop dilatation operator can include corrections from arbitrary other operators with the same charges, and constructing them explicitly is an important unsolved problem

We have not carried out a detailed comparison of counting in the case where not all impurities are distinct, but the correspondence with restricted Schurs should still hold. In fact, in one special case, where all impurities are single-letter $X$ or $Y$, we can find an exact agreement with the results of \cite{npint1106}. The authors of the paper found the number of BPS states with $k_1$ $X$ impurities and $k_2$ $Y$ impurities to be
\begin{equation}
d_{00} = (k_1 + 1)(k_2 + 1).
\end{equation}
We expect the $X$ and $Y$ impurities to be generated by
\begin{equation}
A^{(i)^\dagger}_{1,0,-1}, \quad A^{(i)\dagger}_{0,1,-1}, \quad i \in \{1,2\}
\end{equation} 
The excitations with $(k_1,k_2)$ charges are then
\begin{equation}
(A^{(1)^\dagger}_{1,0,-1})^{i_1} (A^{(2)^\dagger}_{1,0,-1})^{k_1-i_1}
(A^{(1)^\dagger}_{0,1,-1})^{i_2} (A^{(2)^\dagger}_{0,1,-1})^{k_2-i_2}
\end{equation}
with $0\le i_1 \le k_1$ and $0 \le i_2 \le k_2$, giving rise to precisely $(k_1+1)(k_2+1)$ excitations. The comparison for this case is even more robust, since the BPS states in \cite{npint1106} were found explicitly, by looking for eigenstates of one-loop dilatation operator, so we know precisely the dual BPS operators, that we are matching the excitations with.

The construction described here can be extended to the background (\ref{eq:Psi_halfbps}) with any number $p$ of separated branes. There will be a BPS open string spectrum generated by
\begin{equation}
\label{eq:Adagger_halfbps}
	A^{(i)\,\dagger}_{k_1,k_2,-1} = w_{k_1,k_2,i-1}\,\partial_{0,0,i},
\end{equation}
with $i$ running over the number of branes $1 \le i \le p$.
The correspondence with the restricted Schur operators can be established in exactly the same way, where now the impurities are attached to any of the $p$ corners in the Young diagram, and we get $p^n$ states for $n$ distinct attached strings.

\begin{figure}[h]
\begin{center}
\includegraphics{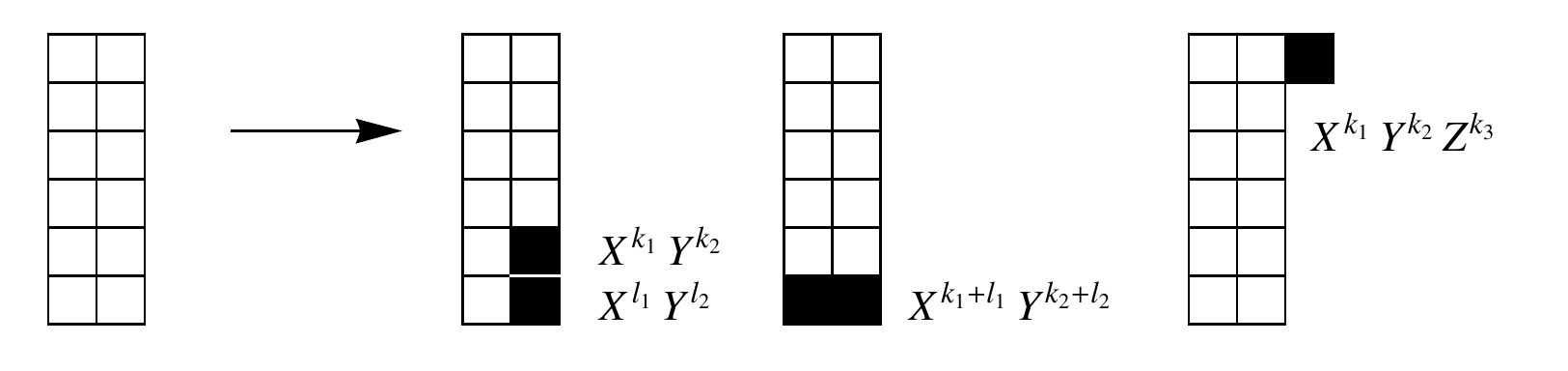}
\caption{Young diagram representation of a two coincident giant state $(w_{0,0,2})^{r_2}$ and its BPS excitations.}
\label{fig:impurities-zz-nonabel}
\end{center}
\end{figure}

Finally, let us consider the case where the branes are coincident (Figure~\ref{fig:impurities-zz-nonabel}).
The state is
\begin{equation}
	\Psi = (w_{0,0,2})^{r_2} (w_{0,0,0})^{N-r_2}
\end{equation}
corresponding to a Young diagram with two columns of the same height.
As already discussed in Section~\ref{sec:states2geom} for the case of maximal branes, the open string excitations are:
\begin{equation}
\begin{split}
	A^\dagger_{k_1,k_2,-1} &= w_{k_1,k_2,1}\, \partial_{0,0,2} \\
	A^\dagger_{k_1,k_2,-2} &= w_{k_1,k_2,0}\, \partial_{0,0,2}
\end{split}
\end{equation}
Because of identical branes it is harder to find an intuitive interpretation for these states: there is no clear distinction between strings stretching between branes, or attached on one or the other. Also there is no easy way to see which of the restricted Schur operators should not be BPS (such as stretched strings).
Nevertheless, we can still attempt to match some of the states. A single excitation $A^\dagger_{k_1,k_2,-1}$ clearly still corresponds to a single attached string, dual to a restricted Schur operator with a single impurity on the corner. Next consider attaching two distinct strings $X^{k_1}Y^{k_2}$ and $X^{l_1}Y^{l_2}$. There are two such restricted Schur operators that can be built, corresponding to two ways of taking out two boxes. According to Figure~\ref{fig:impurities-zz-nonabel} we could roughly match the two options to the excitations:
\begin{equation}
\begin{split}
	A^\dagger_{k_1,k_2,-1}\, A^\dagger_{l_1,l_2,-1}, \quad
	A^\dagger_{k_1+l_1,k_2+l_2,-2}
\end{split}
\end{equation}
But note that the operator $A^\dagger_{k_1+l_1,k_2+l_2,-2}$ would be the same for different choices of two impurities, as long as they have the same total charge. 
Thus there is no obvious one-to-one mapping between a subclass of restricted Schur operators and the BPS spectrum of excitations. One possible scenario is that the states with impurities put in the bottom two boxes are generically not BPS. However, there is one combination of two impurities with charges $(k_1+l_1,k_2+l_2)$ that does produce a BPS state in the bottom two boxes, and that corresponds to $A^\dagger_{k_1+l_1,k_2+l_2,-2}$.

It would be very interesting to confirm the results in this section by explicitly constructing the BPS combinations of the restricted Schur operators. 


\subsection{Eighth-BPS backgrounds} 
\label{sec:eighth-bps}

In Section~\ref{sec:states2geom} we derived the spectrum of excitations around maximal giants, and in the previous section we discussed the case of general (non-maximal) half-BPS giants. Guided by these examples we now consider a general eighth-BPS configuration of intersecting sphere giants.

The background we wish to analyze is the following classical configuration
\begin{equation}
\label{eq:Pz_xyz_multi}
	P(z) = \prod_{i=1}^{p} (x - c^{(1)}_i) \prod_{i=1}^q (y - c^{(2)}_i) \prod_{i=1}^r (z - c^{(3)}_i)
\end{equation}
with $p,q,r\sim O(1)$. We assume the giants are separated enough to each carry individual world-volume excitations. Additionally, we take the regime where the giants are not far from maximal, that is $c_i^{(j)} \ll 1$. That means that all of them will still have pairwise intersections along $S^1$'s. We expect the spectrum of excitations to consist of
\begin{enumerate}
	\item Bulk gravitons
	\item $(p+q+r)$ two-parameter towers of open string modes on each brane
	\item $(pq + qr +rp)$ one-parameter towers of open string modes on $S^1$ intersections
	\item $(pqr)$ extra individual modes of the type $A^\dagger_{-1,-1,-1}$ in (\ref{eq:Atable_xyz}) for each choice of three branes.
\end{enumerate}

The first question is what is the quantum state corresponding to this background. More precisely, we look for an energy eigenstate which is spread out on a torus of complex phases of $c_i^{(a)}$, as discussed in Section~\ref{sec:cp}. Let us construct it starting from the maximal giant configuration $(w_{p,q,r})^N$ containing $p,q,r$ of the $x=0,y=0,z=0$ maximal giants respectively. First note, that if we start from a maximal half-BPS state with $r$ giants $(w_{0,0,r})^N$ then the general half-BPS state (\ref{eq:Psi_halfbps}) can be achieved by shrinking the branes
\begin{equation}
\label{eq:psi_ri}
	\Psi = \left( \prod_{i=1}^r (A^\dagger_{0,0,-i})^{m_i} \right) (w_{0,0,r})^N
\end{equation}
with $A^\dagger_{0,0,-i} = w_{0,0,r-i}\,\partial_{0,0,r}$ as in (\ref{eq:Atable_zm}). In the background $(w_{p,q,r})^N$ we expect operators
\begin{equation}
\label{eq:Adagger_i00}
\begin{split}
	A^\dagger_{-i,0,0} &= w_{p-i,q,r}\,\partial_{p,q,r} \\
	A^\dagger_{0,-i,0} &= w_{p,q-i,r}\,\partial_{p,q,r} \\
	A^\dagger_{0,0,-i} &= w_{p,q,r-i}\,\partial_{p,q,r}
\end{split}	
\end{equation}
to have the same physical interpretation on each stack of branes, i.e. correspond to the shrinking modes. So we propose that the state corresponding to (\ref{eq:Pz_xyz_multi}) in the near-maximal regime $c_i \ll 1$ is:
\begin{equation}
\label{eq:Psi_eighth_bps}
\begin{split}
	\Psi &= 
		\left( \prod_{i=1}^p (A^\dagger_{-i,0,0})^{k_i} \right) 
		\left( \prod_{i=1}^q (A^\dagger_{0,-i,0})^{l_i} \right) 
		\left( \prod_{i=1}^r (A^\dagger_{0,0,-i})^{m_i} \right) 
		(w_{p,q,r})^N
\\		
	&= 
		\left( \prod_{i=1}^p (w_{p-i,q,r})^{k_i} \right) 
		\left( \prod_{i=1}^q (w_{p,q-i,r})^{l_i} \right) 
		\left( \prod_{i=1}^r (w_{p,q,r-i})^{m_i} \right) 
		(w_{p,q,r})^{N-\sum k_i - \sum l_i - \sum m_i}
\end{split}
\end{equation}
Note the condition that we have no coincident giants remaining translates into the requirement that $k_i,l_i,m_i$ are all non-zero. In fact, for them to be well-separated but still near-maximal we pick the parameters in the regime
\begin{equation}
	1 \ll k_i,l_i,m_i \ll N
\end{equation}
Note that we could establish the relationship between $k_i,l_i,m_i$ and $c_i^{(a)}$ by recalling that a single giant $P(z)=(z-c)$ has energy $E=N(1-|c|^2)$ and the corresponding state $(w_{0,0,0})^{m_1}(w_{0,0,1})^{N-m_1}$ has $E=N-m_1$, so $m_1 = N |c|^2$. For multiple giants we consider the Young diagram after taking away $m_i$ rows of length $i$, by comparison with (\ref{eq:psi_ri}), and interpret each column as a separate giant. We do this for each direction $k_i,l_i,m_i$ individually. 

The excitations around (\ref{eq:Psi_eighth_bps}) can still be generated similarly like for maximal giants (\ref{eq:Adagger_defn}):
\begin{equation}
  ~~~~~~~~~~~~~~~~~~~~~~~~~~~~~~~~~
	A^\dagger_{i,j,k} \equiv w_{p+i,\,q+j,\,r+k}\,\partial_{p,q,r}
	\qquad (i\ge -p,\; j\ge -q,\; k\ge -r)
\end{equation}
except now we can also act with conjugates of (\ref{eq:Adagger_i00})
\begin{equation}
	A_{i,0,0}=w_{p,q,r}\,\partial_{p-i,q,r}, \quad 
	A_{0,i,0}=w_{p,q,r}\,\partial_{p,q-i,r}, \quad 
	A_{0,0,i}=w_{p,q,r}\,\partial_{p,q,r-i}
\end{equation}
The $A^\dagger_{i,j,k}$ and their conjugates still form a complete basis for generators of excitations, that is, any ``nearby state'' can be constructed by their action.
The fact that we are free to act with conjugates, however, has an important consequence. We can define a new basis for the generators of excitations
\begin{equation}
\begin{tabular}{ll}
  $B^\dagger_{i,j,k} = A^\dagger_{i,j,k}$ &
  	$i,j,k \ge 0, \;\; i+j+k>0$ \\
  $B^{(i)^\dagger}_{-1,j,k} = A^\dagger_{-i,j,k}\,A_{i-1,0,0}$ &
  	$j,k \ge 0, \;\; i \in \{1\ldots p\}$ \\
  $B^{(i,j)^\dagger}_{-1,-1,k} = A^\dagger_{-i,-j,k}\,A_{i-1,0,0} \, A_{0,j-1,0}$ &
    $k \ge 0, \;\; i \in \{1\ldots p\}, \; j \in \{1\ldots q\} $ \\
  $B^{(i,j,k)^\dagger}_{-1,-1,-1} = A^\dagger_{-i,-j,-k}\,A_{i-1,0,0}\, A_{0,j-1,0}\, A_{0,0,k-1}$ &
    $i \in \{1\ldots p\}, \; j \in \{1\ldots q\}, \; k \in \{1\ldots r\}$
\end{tabular}
\end{equation}
plus analogous excitations in other directions:
\begin{equation}
	B^{(j)^\dagger}_{i,-1,k}, \quad B^{(k)^\dagger}_{i,j,-1}, \quad
	B^{(i,k)^\dagger}_{-1,j,-1}, \quad B^{(j,k)^\dagger}_{i,-1,-1}
\end{equation}
Note that $B^{(i)^\dagger}_{-1,0,0}, B^{(i)^\dagger}_{0,-1,0}, B^{(i)^\dagger}_{0,0,-1}$ are the only modes that are in non-ground state, that is, we can also act with their conjugates
\begin{equation}
	B^{(i)}_{1,0,0} = A^\dagger_{-i+1,0,0} A_{i,0,0}, \quad 
	B^{(i)}_{0,1,0} = A^\dagger_{0,-i+1,0} A_{0,i,0},\quad 
	B^{(i)}_{0,0,1} = A^\dagger_{0,0,-i+1} A_{0,0,i}
\end{equation}
on the background. 

The point of this construction is that we can consistently ``fix'' each excitation $A^\dagger_{i,j,k}$ by some extra conjugate generators $A_{i',j',k'}$ to get the desired charges. Most importantly, one can convince oneself that every excited state can be uniquely generated by some combination of $B^\dagger$ and their conjugates $B$, so this basis is as good as $A^\dagger$ and $A$. In other words, we can invert the relationship to express each $A^\dagger$ and $A$ in terms of $B^\dagger$ and $B$. One may worry that by doing this transformation we ruin the commutation relationships. However, for current purposes we are not using the full algebra, instead we just use the operators to freely generate the spectrum. Thus the requirement for the operators is only that by acting with arbitrary combinations of raising operators (or lowering, if the mode is not in the ground state), ordered in some canonical way, we generate each state exactly once.

The goal of all this is, of course, that we can now easily interpret these excitations as expected in the beginning of the section:
\begin{equation}
	\begin{tabular}{l|l}
		Operator & Interpretation \\
		\hline			
		$B^\dagger_{i,j,k}$ & Bulk closed strings \\	
		$B^{(i)\dagger}_{-1,j,k}$ & Modes on $x$ giants ($p$ of them) \\
	  $B^{(j)\dagger}_{i,-1,k}$ & Modes on $y$ giants ($q$ of them) \\
		$B^{(k)\dagger}_{i,j,-1}$ & Modes on $z$ giants ($r$ of them)\\
		$B^{(i,j)^\dagger}_{-1,-1,k}$ & Modes on $(x,y)$ giant  intersections ($pq$ of them) \\		
		$B^{(i,k)^\dagger}_{-1,j,-1}$ & Modes on $(x,z)$ giant  intersections ($pr$ of them) \\	
		$B^{(j,k)^\dagger}_{i,-1,-1}$ & Modes on $(y,z)$ giant intersections ($qr$ of them) \\		
		$B^{(i,j,k)^\dagger}_{-1,-1,-1}$ & Composite mode on each triplet of giants ($pqr$ of them)
	\end{tabular}
\end{equation}
Note the superscript $(i)$ in e.g. $B^{(i)\dagger}_{-1,j,k}$ does not \emph{exactly} label the giant which we excite -- our main point is that there is the right multiplicity of modes. The actual excitations of each giant could be some linear combinations of those.

As a consistency check note that in the half-BPS case with $r$ non-coincident branes (\ref{eq:Psi_halfbps}) we get:
\begin{equation}
\begin{split}
	B^{(k)\dagger}_{i,j,-1} &= A^\dagger_{i,j,-k}\,A_{0,0,k-1} \\
	&= \left( w_{i,j,r-k}\,\partial_{0,0,r} \right) \left( w_{0,0,r}\,\partial_{0,0,r-k+1} \right)
	\\
	&= w_{i,j,r-k}\,\partial_{0,0,r-k+1}
\end{split}
\end{equation}
which matches (\ref{eq:Adagger_halfbps}).

Note the construction in this section clarifies the question raised in Section~\ref{sec:states2geom}, how exactly the label $k$ in $A^\dagger_{i,j,-k}$ corresponds to the number of branes. If the branes are separate, we can combine this operator with $A_{0,0,k-1}$ to build the excitation $B^{(k)\dagger}_{i,j,-1}$ with the right charge, in which case $k$ corresponds directly to the multiplicity of branes that we can excite. When the branes coincide, however, we lose such individual states since we can not distinguish the branes. Then we can only act with $A^\dagger_{i,j,-k}$, since $A_{0,0,k-1}$ annihilates the state, and the label $k$ is better interpreted as \emph{how many} out of the $r$ branes we excite.


\section{Discussion} 
\label{sec:discussion}

\subsection{From oscillators to gauge invariant BPS operators  } 

The analysis we have given in this paper of the 
connection between specific states in the Hilbert space
of eighth-BPS states at strong coupling ($\lambda = g_{YM}^2 N $ 
large)  and the moduli space of brane geometries is a step in 
addressing the broader problem  of associating explicit gauge invariant 
eighth-BPS  operators to brane geometries. From the gauge theory, 
the weak coupling limit is the natural starting point, but encounters 
an important subtlety that the BPS spectrum jumps from zero to weak coupling. 
It is believed not to jump further as the coupling is tuned to become strong. 
For the Hilbert space of BPS operators at zero coupling 
a nice group theoretic orthogonal basis was found in \cite{bhrI,bhrII}. 
The labels of this basis are $ ( \Lambda , M_{ \Lambda  } , R , \tau  ) $.
Here  $ \Lambda $ is a representation of $U(3)$, $R$ is a  
Young diagram with a finite number of rows. $R$ and $ \Lambda $ have the 
same number of boxes, say this is $n$, then $\tau $ is a Clebsch-Gordan 
multiplicity for $S_n$ which couples $ R \otimes R \rightarrow \Lambda $.
The problem of a $U(3)$ covariant construction of BPS operators 
at weak coupling was considered in \cite{pgp}. 
We defined  the Hilbert space $ \cH^{\infty}$, 
spanned by $ |  R , \Lambda, \tau \rangle $ for all $n$, which 
can be viewed as an inductive  ($N \rightarrow \infty $) limit 
of the free Hilbert space associated with CFT operators (equally states 
by the operator-states correspondence) constructed from holomorphic
 combinations of three complex matrices $X, Y ,  Z $ of size $N$. 
The construction of BPS operators at finite $N$ is achieved 
by considering $ \cP_{I}^{ (N)} $ which is a projector for the intersection 
space of $ \Im ( \cP )  \cap \Im ( \cI_{N} ) $. 
The projector $\cP$ acts on the free Hilbert space of operators 
by symmetrizing traces 
and the projector $ \cI_{N}$ implements the finite $N$ constraint
(see \cite{pgp} for more details). 

The derivation from the weakly coupled gauge theory 
of the Hilbert space in terms of oscillators 
such as arises from the quantization of giant graviton moduli spaces is a 
highly non-trivial problem. At the level of counting states, there is already 
a clear argument for the matching of states. However, the matching at
the level of a map between oscillator states and 
explicit BPS gauge invariant operators is a non-trivial problem. 
Solving it would allow the study of fluctuations around the eighth 
BPS configurations from the gauge theory point of view, extending 
the detailed understanding that is available for fluctuations 
around the half-BPS configurations.

 One aspect of the oscillator Hilbert space that has played a role 
 in this paper (Section~\ref{sec:single_partition}) is the way the number of states changes 
 as a function of $N$. At some fixed $N$, we can grade the Hilbert space 
 according to the states that exist at $N$ but not at $N-1$, at $N-1$ 
 but not at $N-2$, etc. A similar construction is possible on the gauge theory side. There is a chain of subspaces 
\bea 
 \Im ( \cP_I^{ ( N=2 )}  ) \subset \Im ( \cP_I^{ ( N=3  )} )  \subset \cdots 
\subset  \Im ( \cP_I^{ ( N-1   )}  )   \subset  \Im ( \cP_I^{ ( N   )} ) 
 \eea
By applying $\cG^{(N)} $ to these subspaces, we can construct 
BPS operators for the Hilbert space at rank $N$. We can start from 
$\cG^{(N)}\cP_I^{ ( N=2 )} $ and then construct $ \cG^{(N)}\cP_I^{ ( N=3 )} $ 
ensuring that the components parallel to 
$\cG^{(N)}\cP_I^{ ( N=2 )} $,  in the finite $N$ free field 
inner product,  are projected out. Then inductively by a 
Gram-Schmidt procedure we can obtain a decomposition 
\bea 
\cH_{BPS}^{(N)} = \cH_{BPS}^{ (N ; 2)} \oplus  \cH_{BPS}^{ (N ; 3 )} 
\cdots \oplus 
 \cH_{BPS}^{ (N ; N  )}
\eea 
The space  $\cH_{BPS}^{ (N ; i )}$ is the subspace of BPS states 
at rank $N$ which exists at  rank $i$ but not above. These states can be 
matched with the subspace of the finite $N$  oscillators which have precisely 
$i $ excited oscillators and no more.

The above construction  provides an algorithm for identifying the decomposition 
of the Hilbert space corresponding to number of oscillators. From
the oscillator Hilbert space, we have a more complete group theoretic 
basis including labels $ \Lambda , Y $. The label $\Lambda $ is well 
understood using the $U(3)$ covariant description of the free Hilbert space 
and of the action of the one-loop dilatation operator. An algorithm 
 for constructing the label $Y$ is an interesting problem for the 
near future. More ambitiously, we would like to find the gauge theory construction for all the labels appearing in (\ref{decompYNd}).

It is  instructive to express the problem at hand in more geometric 
language. The Fock space states are in one to one correspondence with 
functions on the symmetric product $ S^N ( \mC^3 )$. This is in fact 
 the description that arises naturally from the quantization of AdS 
giants. The correspondence is 
\bea 
\prod_{p=1}^L  w_{i_p j_p k_p } 
\leftrightarrow \sum_{ \sigma \in S_{N}  }
 \prod_{p=1}^L   x_{\sigma (p) }^{i_p}  y_{ \sigma (p)}^{j_p} 
z_{ \sigma(p)}^{k_p}  
\eea
The gauge theory involves the zero-eigenstates of the one-loop 
dilatation operator 
\bea 
\cH_2 = \sum_{ i \ne j =1 }^3 
 tr [ X_{i } , X_j ]  [ \check X_{i} , \check X_j ] 
\eea
acting on the space of gauge-invariant 
holomorphic functions of the matrices $X , Y , Z $. 
The gauge symmetry $U(N)$ acts as 
\bea 
X_i \rightarrow U X_i U^{\dagger} 
\eea
We may say that the precise understanding of the gauge theory construction 
of the operators is a question of understanding the mapping 
\bea 
\Fun( S^N ( \mC^3 ) )  \leftrightarrow 
	\Ker( \cH_2 ) \hbox{ on } 
 \Fun\left( gl(N; \mC ) ^{ \times 3 } // U(N) \right)
\eea

\subsection{ Localization in moduli space versus space-time }

One of the most fascinating  aspects of a single half-BPS giant graviton 
is that its growth in size as a function of angular momentum 
has  been interpreted as a consequence of space-time uncertainty \cite{mst}. 
This may be related to non-commutative deformations of space-time 
\cite{jevram}.  
Our investigations highlight an  interesting variation on this discussion. 
When we view giant graviton states from the perspective of 
quantization of their moduli space, there is naturally some 
delocalization on this moduli space due to conventional quantum 
mechanics. Our analysis showed how oscillator 
eigenstates are localized on the base simplex of a toric fibration of the 
moduli space, while being spread out on the torus fibers. 
Special states were identified as being localized at the corners 
of the simplex, where the torus fibers degenerate and the uncertainty 
on the moduli space is minimized. We then found that these extremal 
states correspond to composites of maximal giants $x=0, y=0, z=0$
described by monomials $x^{m_1} y^{m_2} z^{m_3} = 0$. In the simple 
case of single giant in the half-BPS sector the maximal state corresponds to 
the end of an interval. Here the $S^1$ fiber of the toric description of $S^2$ 
degenerates. So delocalization (maximally large brane) is accompanied 
by maximal localization in the moduli space. At the opposite end 
of the interval, there is also localization in moduli space, but 
here the small giant is better described as a Kaluza-Klein wave of low 
momentum, so again spread out in space, albeit not in $S^3 \in S^5$ 
but along $S^1$ in $S^5$. It is natural to wonder how general 
is this complementarity between delocalization in space and localization 
in moduli space.

\subsection{From Brane geometries to space-time geometries } 

It is known that operators with dimensions of order $N$
correspond to giant graviton branes. Stacks with large numbers of 
branes back-react and produce LLM-type geometries when the dimensions
are of order $N^2$.   

While it is true that one can construct coherent states 
localized near any point of the moduli space, this does not 
give any detailed information about how a complete set of states 
is localized on the moduli space. In the example of eighth-BPS 
giant gravitons described by Mikhailov polynomials,
it has been argued that after closing appropriate 
holes etc, the physical moduli space is made of projective spaces. 
The map between the coefficients of the polynomials 
and projective spaces is non-trivial. The holomorphic wavefunctions 
in this case have a localization, and lack thereof, which can be understood
using fuzzy geometry. They are localized in the base of 
the toric fibration description of the projective spaces 
 and they are constant along the torus fibers. 

The oscillator states (equivalently holomorphic wavefunctions 
on the projective space) are eigenstates of various Casimir-like operators 
e.g 
\bea 
Q_{n_1 , n_2 , n_3 } = \sum_{ i , j , k } i^{n_1} j^{n_2} k^{n_3} w_{i,j,k} \partial_{ i , j ,  k } 
\eea
which commute with the Hamiltonian. In any given physical situation, 
certain observables will be measurable, and they will determine 
the right basis. It is then of interest to find the degree of localization 
of the states in this basis. 

In  Mathur's fuzzball program for understanding black holes 
\cite{mathur05review}, one obtains the states of black holes by 
quantizing an appropriate moduli space of solutions. The validity of 
semiclassical quantization around a fixed background, as an account 
of Hawking radiation or of infall, is subject to the question of 
the extent to which the appropriate physics can be solved 
in a local analysis in the moduli space. The precise set-up and 
the questions asked will determine the observables. These 
 will be analogous to the Casmirs mentioned above, 
and this in turn will constrain the appropriate  bases of states. 
The degree of localization of 
these states will then affect the validity of semiclassical quantization 
around a fixed background. While eighth-BPS backgrounds do not admit finite horizon areas, they have a substantial layer of complexity beyond the half-BPS 
case, and can provide a valuable laboratory for quantitative 
studies of states, localization and dynamics. Some of the lessons 
and techniques can be expected to carry over to the sixteenth BPS 
case where finite horizon area black holes are possible and efforts 
at state counting have been initiated \cite{index}. Turning 
on perturbations to go away from extremality is another way to obtain 
finite horizon area \cite{BBJS}. The way fuzzy geometry techniques 
provide the tools to connect states to points or regions of the moduli space 
is one lesson we expect will have applicability beyond the eighth-BPS
 set-up here.

\subsection{Fuzzy spaces and correlators}
We have focused attention on the quantum state space here, and used 
fuzzy geometry to connect to points and regions of moduli space. 
The fuzzy projective spaces will also provide a way to approach
the gauge theory correlators. Consider for example 
the half-BPS operators $\cO_R$, which have 2 and 3-point 
functions \cite{cjr}
\begin{equation}\label{cjrres}
\begin{split}
 \langle \cO_R \cO_S \rangle &= \delta_{RS} f_R \\
 \langle \cO_R \cO_S \cO_T^{\dagger}  \rangle &= g(R,S,T) f_T 
\end{split}
\end{equation}
where $f_R, g(R,S,T)$ are appropriate group-theoretic quantities. 
If we restrict to Young diagrams with no more than $d$ columns, 
we would be looking at the quantization of polynomials 
in one variable of degree up to $d$. The Hilbert space consists of
holomorphic wavefunctions on $\mC \mP^d $ and is isomorphic to 
$\Sym^N ( V_{d+1}  )$. The holomorphic coordinates are $W_I = w_{0,0,I}$ 
for $I = 0 \cdots d $.  The states $ W_I^{n_I}  | 0 \rangle  $ 
can be mapped to Young diagrams with $n_I $ rows of length $I$. 
If the inner product of the holomorphic wave functions is 
$\tilde f_R $, then 
\bea\label{statemap}  
 \prod_I  W_I^{n_I(R)}   | 0 \rangle \rightarrow 
{ \tilde f_R \over  f_R } | R \rangle  
\eea
If we associate an operator $\cO_R$ to each state $|R \rangle $
according to the formula 
\bea\label{stateop}  
\cO_R | S \rangle = \sum_T f_T g ( R , S , T ) | T \rangle 
\eea
then $ \langle T | \cO_R | R \rangle $ will reproduce 
the three-point function in (\ref{cjrres}).
Since $| S \rangle , | R \rangle $ have been expressed 
in terms of the Hilbert space of the fuzzy $\mC \mP$ using 
(\ref{statemap}), the  operators is (\ref{stateop}) 
are elements in the fuzzy $\mC \mP$ algebra, which reproduce
the correct 3-point function by construction. The challenge 
is to understand how to derive (\ref{stateop}) from the dynamics
of the moduli space of gravitons. Such a derivation
may require an effective action taking into account 
the effects of integrating out the non-BPS excitations. 
Certainly for the non-extremal correlators, an appropriate
account of non-BPS states in intermediate channels  would  
be necessary. An interesting discussion of the space-time computation of the half-BPS 
correlators, from another point of view, can be found in \cite{Bissi:2011dc}.


\section{Summary and outlook}

We have developed techniques to map BPS states to 
BPS brane configurations. Fuzzy geometry  and toric fibrations
(Section~\ref{sec:cp}), factorization properties of  
partition functions(Section~\ref{sec:single_partition}), 
and local analysis of the symplectic form  on the moduli space
(Section~\ref{sec:single_symplectic}) have all been useful in shedding 
light on this problem.  

Our first main result is an  identification 
of the spectrum of BPS world-volume excitations for 
specific brane geometries. The second main result is 
a group theoretic labelling of the states in the 
eighth-BPS sector at finite $N$, which comes from the structure of
giant graviton moduli spaces.

We expect the results of this paper to give useful 
information towards the construction of gauge theory 
operators for eighth-BPS states corresponding to 
specific giant graviton geometries, which in turn 
should lead to operators for both BPS and non-BPS 
excitations of these states. The restricted Schur technology 
gives a way to modify half-BPS operators, guided by 
group theoretic labels (Young diagrams) characterizing the 
background operator. The group theoretic labels we developed in Section 
\ref{complete-group-labels} for the eighth-BPS sector, 
involving $U(3)$ and $U(N)$ Young diagrams 
along with other group-theoretic multiplicities, would be expected 
to play a similar role.  The specific BPS excitation spectrum around 
various geometries of branes constitute predictions for a variety 
of brane systems, including non-abelian systems of coincident parallel branes
as well as branes which are composites intersecting along a circle. 
Recovering these predictions from non-abelian DBI actions or 
world-sheet string methods is a fascinating direction for the future. 

We also expect that the techniques for mapping quantum states 
to brane geometries can be applicable in the context of bulk geometries 
related to black holes. 
These can involve generalizing the considerations 
of this paper to quarter and eighth-BPS space-time geometries  and their 
non-extremal finite horizon deformations, which have been studied in \cite{BBJS,deBoer:2011zt}. 
It would be extremely interesting to develop a general map between quarter- or eighth-BPS bulk geometries and operators, analogous to LLM \cite{LLM} in the half-BPS case. Progress in characterization of such quarter-BPS geometries has been made in \cite{Lunin:2008tf}, and a study of corresponding operators in Brauer basis was initiated in \cite{Kimura:2011df}. 
Another direction is to consider sixteenth BPS states. 

The techniques for mapping quantum states to geometries 
should admit application to more general cases of AdS/CFT
where  the $S^5$ is replaced by a more general Sasaki-Einstein 
manifold. The connection between dual giant gravitons 
(large in the AdS) and the description of 
quantum states available from the gauge theory has been developed
\cite{martsparks}. But the analogous connection with 
giants which are large in the Sasaki-Einstein space is 
a very interesting area for future research.


\section*{Acknowledgements}
We thank Robert de Mello Koch, Joan Simon, Vishnu Jejjala
 for very useful discussions and email communications. 
S.R is supported by  STFC Standard Grant ST/J000469/1 "String Theory, Gauge
Theory and Duality". JP is supported by a Queen Mary, University of London studentship.


\appendix




\section{Symplectic form for perturbations of sphere giant}
\label{app:w_derivation}

In this appendix we derive the symplectic form for arbitrary perturbations of a non-maximal half-BPS giant. Our derivation is along the lines of that found in Appendix~F in \cite{minwalla}, and we use some results from there.
The unperturbed solution is defined by the polynomial:
\begin{equation}
	P(z) = z - c_{0}
\end{equation}
The surface in $S^5$ is:
\begin{equation}
\begin{split}
	|x|^2 + |y|^2 &= 1 - |c_0|^2 \\
	z &= e^{it} c_0
\end{split}
\end{equation}
where we have also put the time-dependence back in. We pick world-volume coordinates $(\sigma^1, \sigma^2, \sigma^3)$ to be some coordinates on a unit $S^3$ embedded in $\mbC^2$, so that we have functions $x_0(\sigma^i)$ and $y_0(\sigma^i)$ satisfying
\begin{equation}
	|x_0(\sigma^i)|^2 + |y_0(\sigma^i)|^2 = 1.
\end{equation}
The unperturbed surface in terms of the world-volume coordinates is
\begin{equation}
\label{eq:xyz_undeformed}
\begin{split}
	x(\sigma^i,t) &= \sqrt{1-|c_0|^2}\, x_0(\sigma^i) \\
	y(\sigma^i,t) &= \sqrt{1-|c_0|^2}\, y_0(\sigma^i) \\
	z(\sigma^i,t) &= e^{it} c_0
\end{split}
\end{equation}
Small perturbations around the spherical shape are parametrized by a complex function 
\begin{equation}
	\delta z(\sigma^i, t) = z(\sigma^i,t) - e^{it} c_0	
\end{equation}
Effectively these are the 2 real transverse coordinates to $S^3$ in $S^5$.

The general expression for symplectic form is (\ref{eq:omega_defined}):
\begin{equation}
\begin{split}
	\omega &= \omega_{\rm BI} + \omega_{\rm WZ} \\
	\omega_{\rm BI}	&= \frac{N}{2\pi^2} \int_\Sigma \d^3 \sigma \,
		\delta\left(
			\sqrt{-g}g^{0\alpha} \frac{\pd x^\nu}{\pd \sigma^\alpha}
			G_{\mu\nu}
		\right) \wedge \delta x^\mu
	\\
	\omega_{\rm WZ} &= 
		\frac{2N}{\pi^2} \int_\Sigma \d^3 \sigma
		\frac{\delta x^\lambda \wedge \delta x^\mu}{2}
		\left( 
			\frac{\pd x^\nu}{\pd \sigma^1}
			\frac{\pd x^\rho}{\pd \sigma^2}
			\frac{\pd x^\sigma}{\pd \sigma^3}
		\right)
		\epsilon_{\lambda \mu \nu \rho \sigma}
\end{split}		
\end{equation}
$G_{\mu \nu}$ is the metric on unit $S^5 \times \mbR$ and $g_{\alpha \beta}$ is the induced metric on the world-volume $\Sigma\times \mbR$. Note
\begin{equation}
	\omega_{\rm BI}	= \frac{N}{2\pi^2} \int_\Sigma \d^3 \sigma \,
		\delta p_\mu \wedge \delta x^\mu
\end{equation}
with the definition of conjugate momentum
\begin{equation}
	p_\mu = \sqrt{-g}g^{0\alpha} \frac{\pd x^\nu}{\pd \sigma^\alpha} G_{\mu\nu} \,.
\end{equation}

We will see now that these expressions can be simplified significantly for the case at hand. First, the only perturbation of the surface $\delta x^\mu$ can be taken to be $\delta z$. In principle the surface has to be deformed in $\delta x$ and $\delta y$ away from (\ref{eq:xyz_undeformed}), but those are higher order in $\delta z$ and can be dropped. That results in:
\begin{equation}
\label{eq:wbi_pzz}
\begin{split}
	\omega_{\rm BI}	&= \frac{N}{2\pi^2} \int_\Sigma \d^3 \sigma \,
		\left(
			\delta p_z \wedge \delta z +
			\delta \bar{p}_z \wedge \delta \bar{z}			
		\right) 
	\\
	p_z &= \sqrt{-g}\, g^{00} \left( 
		G_{zz} \dot{z} +
		G_{z\bar{z}} \dot{\bar{z}}
	\right)
\end{split}		
\end{equation}
and the Wess-Zumino piece:
\begin{equation}
	\omega_{\rm WZ} = 
		\frac{2N}{\pi^2} \int_\Sigma \d^3 \sigma
		\sqrt{g^{S^3}}\,
		(1 - z \bz)
		\frac{\delta \bar{z} \wedge \delta z}{2i}		
\end{equation}
The $S^5$ is now represented as a $S^3$ fibered over a unit disk in $z$, so
\begin{equation}
	(\d s^2)_G = -\d t^2 
	+ \frac{\bz^2 \d z^2 + 2(2-z \bz)\d z\d \bz + z^2 \d \bz^2}{4(1-z\bz)}
	+ (1-z \bz) (\d s^2)_{S^3}
\end{equation}
and the relevant components:
\begin{equation}
\label{eq:Gzz}
	G_{zz} = \frac{\bz^2}{4(1-z\bz)}, \quad G_{z\bz} = \frac{2-z\bz}{4(1-z\bz)}
\end{equation}
The induced metric on the unperturbed solution is
\begin{equation}
	(\d s^2)_g = - (1-|c_0|^2) \d t^2 + (1-|c_0|^2) (\d s^2)_{S^3}
\end{equation}
and so
\begin{equation}
	\sqrt{-g} = (1-|c_0|^2)^2 \, \sqrt{g^{S^3}}
\end{equation}

The bit that requires some work is the evaluation of $\delta p_z$ in (\ref{eq:wbi_pzz}) under the deformation $\delta z$. We need to vary all components:
\begin{equation}
\begin{split}
	\delta p_z &= 
	  \delta(\sqrt{-g})\, g^{00} \left( G_{zz} \dot{z} + G_{z\bz} \dot{\bz} \right) \\
	&+ \sqrt{-g}\, \delta g^{00} \left( G_{zz} \dot{z} + G_{z\bz} \dot{\bz} \right) \\
	&+ \sqrt{-g}\, g^{00} \left( \delta G_{zz} \dot{z} + \delta G_{z\bz} \dot{\bz} \right) \\
	&+ \sqrt{-g}\, g^{00} \left( G_{zz} \delta \dot{z} + G_{z\bz} \delta \dot{\bz} \right)
\end{split}
\end{equation}
First we reexpress $\delta \sqrt{-g} = \frac{1}{2}\sqrt{-g}\,g^{\mu\nu}\delta g_{\mu\nu}$ and $\delta g^{00} = - (g^{00})^2 \delta g_{00}$. Then we need variations of the induced metric:
\begin{equation}
\begin{split}
	\delta g_{00} &= \delta \left( \dot z \dot z G_{zz} + 2 \dot z \dot \bz G_{z\bz} + \dot \bz \dot \bz G_{\bz\bz} \right) \\
	\delta g_{ij} &= - (g^{S^3})_{ij} \delta (z \bz)
\end{split}
\end{equation}
And the $\delta G_{zz}$, $\delta G_{z\bz}$ are calculated by varying (\ref{eq:Gzz}). Putting everything together we find a simple result:
\begin{equation}
	\delta p_z = - \sqrt{g^{S^3}} \left(
		\frac{1}{2} \delta \dot \bz + i |c_0|^2 \delta \bz + \frac{i}{2} \bar{c}_0 \, \delta z
	\right)
\end{equation}
Now we can evaluate (\ref{eq:wbi_pzz}):
\begin{equation}
	\omega_{\rm BI} = \frac{2N}{\pi^2} \int_{S^3} \d^3 \sigma \,
	\left(
	|c_0|^2 \, \frac{\delta \bar z \wedge \delta z}{2i}
		- \frac{\delta \dot{\bar z} \wedge \delta z}{8}
		+ \frac{\delta \bar z \wedge \delta \dot z}{8}
	\right)
\end{equation}
We have dropped the explicit measure on the unit sphere $\sqrt{g^{S^3}}$ and consider it part of the definition of $\int_{S^3} \d^3 \sigma$. Finally, combining this with $\omega_{\rm WZ}$ we arrive at the final result
\begin{equation}
\label{eq:w_zz_app}
\boxed{
	\omega = \frac{2N}{\pi^2} \int_{S^3} \d^3 \sigma \,
	\left(
		\frac{\delta \bar z \wedge \delta z}{2i}
		- \frac{\delta \dot{\bar z} \wedge \delta z}{8}
		+ \frac{\delta \bar z \wedge \delta \dot z}{8}
	\right)
}
\end{equation}
where the integral $\d^3 \sigma$ is now over unit $S^3$ with its standard volume form.

Now let us use (\ref{eq:w_zz_app}) to evaluate the symplectic form in a particular basis of world-volume perturbations:
\begin{equation}
\label{eq:P_deltab_app}
	P(z) = z - c_0 + \sum_{m,n\ge0} \delta b_{m,n}\,x^m y^n	
\end{equation}
With time dependence as $P(e^{-it}x, e^{-it}y, e^{-it}z)=0$ it is
\begin{equation}
\begin{split}
	z &= e^{it} c_0 - \sum_{m,n\ge0} \delta b_{m,n} e^{(1-m-n)it}  x^m y^n \, .
\end{split}
\end{equation}
We have $x=\sqrt{1-|c_0|^2}\,x_0$ and $y=\sqrt{1-|c_0|^2}\,y_0$ as in (\ref{eq:xyz_undeformed}). To first order in $\delta b_{m,n}$ it remains unchanged and so:
\begin{equation}
	z = e^{it} c_0 - 
		\sum_{m,n\ge0} \delta b_{m,n} e^{(1-m-n)it} 
		(1 - |c_0|)^{(m+n)/2}  x_0^m y_0^n
\end{equation}
That gives us the variation in $z$ and $\dot z$:
\begin{equation}
\begin{split}
	\delta z &=  -\sum_{m,n\ge0} \delta b_{m,n} e^{(1-m-n)it} (1 - |c_0|)^{(m+n)/2}  x_0^m y_0^n	\\
	\delta \dot z &= i \sum_{m,n\ge0} \delta b_{m,n} (m+n-1) e^{(1-m-n)it} (1 - |c_0|)^{(m+n)/2}  x_0^m y_0^n
\end{split}
\end{equation}
Plugging this in (\ref{eq:w_zz_app}) we find
\begin{equation}
	\omega = \frac{2N}{2 \pi^2} \int_{S^3} \d^3 \sigma \,
	\sum_{m,n\ge0}
	(m+n+1)
	|x_0|^{2m} |y_0|^{2n}
	(1- |c_0|^2)^{m+n}
	\frac{\delta \bar b_{m,n} \wedge \delta b_{m,n}}{2i}	
\end{equation}
We have already dropped the cross-terms which depend on $x_0,y_0$ and not only on $|x_0|,|y_0|$, since the integral of such terms on $S^3$ is 0. The remaining terms are time-independent.
The integral is easy to do:
\begin{equation}
	\int_{S^3} \d^3 \sigma \, |x_0|^{2m} |y_0|^{2n}
	= 2\pi^2 \frac{m! \, n!}{(m+n+1)!}
\end{equation}
Note that we never needed the explicit choice of the coordinate $\sigma^i$ on the sphere. The final symplectic form evaluated at $P(z)=z-c_0$ is thus
\begin{equation}
\label{eq:w_bmn_app}
\boxed{
	\omega = 2N \sum_{m,n\ge0} 
		\frac{m! \, n!}{(m+n)!}
		(1- |c_0|^2)^{m+n}
		\frac{\delta \bar b_{m,n} \wedge \delta b_{m,n}}{2i}	
}
\end{equation}

Note from (\ref{eq:P_deltab_app}) that $\delta b_{0,0}$ is in fact the variation of $c_0$, that is, $\d c_0 = - \delta b_{0,0}$. Thus we can use the requirement that the symplectic form be closed 
\begin{equation}
	\d \omega = 0	
\end{equation}
to complete $\omega$ to an exact expression in $c_0$ and up to quadratic order in other $b_{m,n}$. The result is:
\begin{equation}
\label{eq:w_sbmn}
\begin{split}
	\omega = 2N &\left[
		\left(
			1
			- \sum_{m+n>0} 
				\frac{m! \, n!}{(m+n-1)!}
				(1- |c_0|^2)^{m+n-1}
				|b_{m,n}|^2
		\right.
		\right.
\\			
	&	\quad\quad\quad
	\left.
			+ \sum_{m+n>0} 
				\frac{m! \, n!}{(m+n-2)!}
				|c_0|^2 (1- |c_0|^2)^{m+n-2}
				|b_{m,n}|^2
		\right) 
		\frac{\d\bar c_0 \wedge \d c_0}{2i}		
\\ 
	&+ \sum_{m+n>0} 
			\frac{m! \, n!}{(m+n)!}
			(1- |c_0|^2)^{m+n}
			\frac{\d \bar b_{m,n} \wedge \d b_{m,n}}{2i}		
\\	
	&
	\left. -
		\sum_{m+n>0}
			\frac{m! \, n!}{(m+n-1)!}
			(1- |c_0|^2)^{m+n-1}
			\frac{c_0 \bar b_{m,n} \, \d \bar c_0 \wedge \d b_{m,n} + b_{m,n} \bar c_0 \, \d \bar b_{m,n} \wedge \d c_0}{2i}
	\right]
	+ O(|b|^4)
\end{split}
\end{equation}
This is the full symplectic form at any point $c_0$ expanded for small $b_{m,n}$.


\bibliography{eighthpaper}        

\providecommand{\href}[2]{#2}\begingroup\raggedright\begin{thebibliography}{10}

\bibitem{malda}
J.~M. Maldacena, ``{The Large N limit of superconformal field theories and
  supergravity},'' \href{http://dx.doi.org/10.1023/A:1026654312961}{{\em
  Adv.Theor.Math.Phys.} {\bfseries 2} (1998) 231--252},
\href{http://arxiv.org/abs/hep-th/9711200}{{\ttfamily arXiv:hep-th/9711200
  [hep-th]}}.

\bibitem{gkp}
S.~Gubser, I.~R. Klebanov, and A.~M. Polyakov, ``{Gauge theory correlators from
  noncritical string theory},''
  \href{http://dx.doi.org/10.1016/S0370-2693(98)00377-3}{{\em Phys.Lett.}
  {\bfseries B428} (1998) 105--114},
\href{http://arxiv.org/abs/hep-th/9802109}{{\ttfamily arXiv:hep-th/9802109
  [hep-th]}}.

\bibitem{witten}
E.~Witten, ``{Anti-de Sitter space and holography},'' {\em
  Adv.Theor.Math.Phys.} {\bfseries 2} (1998) 253--291,
\href{http://arxiv.org/abs/hep-th/9802150}{{\ttfamily arXiv:hep-th/9802150
  [hep-th]}}.

\bibitem{mst}
J.~McGreevy, L.~Susskind, and N.~Toumbas, ``{Invasion of the giant gravitons
  from Anti-de Sitter space},'' {\em JHEP} {\bfseries 0006} (2000) 008,
\href{http://arxiv.org/abs/hep-th/0003075}{{\ttfamily arXiv:hep-th/0003075
  [hep-th]}}.

\bibitem{bbns}
V.~Balasubramanian, M.~Berkooz, A.~Naqvi, and M.~J. Strassler, ``{Giant
  gravitons in conformal field theory},'' {\em JHEP} {\bfseries 0204} (2002)
  034,
\href{http://arxiv.org/abs/hep-th/0107119}{{\ttfamily arXiv:hep-th/0107119
  [hep-th]}}.

\bibitem{cjr}
S.~Corley, A.~Jevicki, and S.~Ramgoolam, ``{Exact correlators of giant
  gravitons from dual N=4 SYM theory},'' {\em Adv.Theor.Math.Phys.} {\bfseries
  5} (2002) 809--839,
\href{http://arxiv.org/abs/hep-th/0111222}{{\ttfamily arXiv:hep-th/0111222
  [hep-th]}}.

\bibitem{mikhailov}
A.~Mikhailov, ``{Giant gravitons from holomorphic surfaces},'' {\em JHEP}
  {\bfseries 0011} (2000) 027,
\href{http://arxiv.org/abs/hep-th/0010206}{{\ttfamily arXiv:hep-th/0010206
  [hep-th]}}.

\bibitem{hasitzh}
A.~Hashimoto, S.~Hirano, and N.~Itzhaki, ``{Large branes in AdS and their field
  theory dual},'' {\em JHEP} {\bfseries 0008} (2000) 051,
\href{http://arxiv.org/abs/hep-th/0008016}{{\ttfamily arXiv:hep-th/0008016
  [hep-th]}}.

\bibitem{myers}
M.~T. Grisaru, R.~C. Myers, and O.~Tafjord, ``{SUSY and goliath},'' {\em JHEP}
  {\bfseries 0008} (2000) 040,
\href{http://arxiv.org/abs/hep-th/0008015}{{\ttfamily arXiv:hep-th/0008015
  [hep-th]}}.

\bibitem{Balasubramanian:2002sa}
V.~Balasubramanian, M.-x. Huang, T.~S. Levi, and A.~Naqvi, ``{Open strings from
  N=4 superYang-Mills},'' {\em JHEP} {\bfseries 0208} (2002) 037,
\href{http://arxiv.org/abs/hep-th/0204196}{{\ttfamily arXiv:hep-th/0204196
  [hep-th]}}.

\bibitem{Aharony:2002nd}
O.~Aharony, Y.~E. Antebi, M.~Berkooz, and R.~Fishman, ``{'Holey sheets':
  Pfaffians and subdeterminants as D-brane operators in large N gauge
  theories},'' {\em JHEP} {\bfseries 0212} (2002) 069,
\href{http://arxiv.org/abs/hep-th/0211152}{{\ttfamily arXiv:hep-th/0211152
  [hep-th]}}.

\bibitem{Berenstein:2003ah}
D.~Berenstein, ``{Shape and holography: Studies of dual operators to giant
  gravitons},'' \href{http://dx.doi.org/10.1016/j.nuclphysb.2003.10.004}{{\em
  Nucl.Phys.} {\bfseries B675} (2003) 179--204},
\href{http://arxiv.org/abs/hep-th/0306090}{{\ttfamily arXiv:hep-th/0306090
  [hep-th]}}.

\bibitem{bbfh}
V.~Balasubramanian, D.~Berenstein, B.~Feng, and M.-x. Huang, ``{D-branes in
  Yang-Mills theory and emergent gauge symmetry},''
  \href{http://dx.doi.org/10.1088/1126-6708/2005/03/006}{{\em JHEP} {\bfseries
  0503} (2005) 006},
\href{http://arxiv.org/abs/hep-th/0411205}{{\ttfamily arXiv:hep-th/0411205
  [hep-th]}}.

\bibitem{dMSSI}
R.~de~Mello~Koch, J.~Smolic, and M.~Smolic, ``{Giant Gravitons - with Strings
  Attached (I)},'' \href{http://dx.doi.org/10.1088/1126-6708/2007/06/074}{{\em
  JHEP} {\bfseries 0706} (2007) 074},
\href{http://arxiv.org/abs/hep-th/0701066}{{\ttfamily arXiv:hep-th/0701066
  [hep-th]}}.

\bibitem{dMSSII}
R.~de~Mello~Koch, J.~Smolic, and M.~Smolic, ``{Giant Gravitons - with Strings
  Attached (II)},'' \href{http://dx.doi.org/10.1088/1126-6708/2007/09/049}{{\em
  JHEP} {\bfseries 0709} (2007) 049},
\href{http://arxiv.org/abs/hep-th/0701067}{{\ttfamily arXiv:hep-th/0701067
  [hep-th]}}.

\bibitem{dMSSIII}
D.~Bekker, R.~de~Mello~Koch, and M.~Stephanou, ``{Giant Gravitons - with
  Strings Attached. III.},''
  \href{http://dx.doi.org/10.1088/1126-6708/2008/02/029}{{\em JHEP} {\bfseries
  0802} (2008) 029},
\href{http://arxiv.org/abs/0710.5372}{{\ttfamily arXiv:0710.5372 [hep-th]}}.

\bibitem{npint1103}
W.~Carlson, R.~d.~M. Koch, and H.~Lin, ``{Nonplanar Integrability},''
  \href{http://dx.doi.org/10.1007/JHEP03(2011)105}{{\em JHEP} {\bfseries 1103}
  (2011) 105},
\href{http://arxiv.org/abs/1101.5404}{{\ttfamily arXiv:1101.5404 [hep-th]}}.

\bibitem{npint1106}
R.~d.~M. Koch, B.~A.~E. Mohammed, and S.~Smith, ``{Nonplanar Integrability:
  Beyond the SU(2) Sector},''
\href{http://arxiv.org/abs/1106.2483}{{\ttfamily arXiv:1106.2483 [hep-th]}}.

\bibitem{rdgm}
R.~d.~M. Koch, M.~Dessein, D.~Giataganas, and C.~Mathwin, ``{Giant Graviton
  Oscillators},'' \href{http://dx.doi.org/10.1007/JHEP10(2011)009}{{\em JHEP}
  {\bfseries 1110} (2011) 009},
\href{http://arxiv.org/abs/1108.2761}{{\ttfamily arXiv:1108.2761 [hep-th]}}.

\bibitem{Das:2000st}
S.~R. Das, A.~Jevicki, and S.~D. Mathur, ``{Vibration modes of giant
  gravitons},'' \href{http://dx.doi.org/10.1103/PhysRevD.63.024013}{{\em
  Phys.Rev.} {\bfseries D63} (2001) 024013},
\href{http://arxiv.org/abs/hep-th/0009019}{{\ttfamily arXiv:hep-th/0009019
  [hep-th]}}.

\bibitem{bhrI}
T.~W. Brown, P.~Heslop, and S.~Ramgoolam, ``{Diagonal multi-matrix correlators
  and BPS operators in N=4 SYM},''
  \href{http://dx.doi.org/10.1088/1126-6708/2008/02/030}{{\em JHEP} {\bfseries
  0802} (2008) 030},
\href{http://arxiv.org/abs/0711.0176}{{\ttfamily arXiv:0711.0176 [hep-th]}}.

\bibitem{tomquarter}
T.~Brown, ``{Cut-and-join operators and N=4 super Yang-Mills},''
  \href{http://dx.doi.org/10.1007/JHEP05(2010)058}{{\em JHEP} {\bfseries 1005}
  (2010) 058},
\href{http://arxiv.org/abs/1002.2099}{{\ttfamily arXiv:1002.2099 [hep-th]}}.

\bibitem{yusuke}
Y.~Kimura, ``{Quarter BPS classified by Brauer algebra},''
  \href{http://dx.doi.org/10.1007/JHEP05(2010)103}{{\em JHEP} {\bfseries 1005}
  (2010) 103},
\href{http://arxiv.org/abs/1002.2424}{{\ttfamily arXiv:1002.2424 [hep-th]}}.

\bibitem{pgp}
J.~Pasukonis and S.~Ramgoolam, ``{From counting to construction of BPS states
  in N=4 SYM},'' \href{http://dx.doi.org/10.1007/JHEP02(2011)078}{{\em JHEP}
  {\bfseries 1102} (2011) 078},
\href{http://arxiv.org/abs/1010.1683}{{\ttfamily arXiv:1010.1683 [hep-th]}}.

\bibitem{Kimura:2011df}
Y.~Kimura and H.~Lin, ``{Young diagrams, Brauer algebras, and bubbling
  geometries},'' \href{http://dx.doi.org/10.1007/JHEP01(2012)121}{{\em JHEP}
  {\bfseries 1201} (2012) 121},
  \href{http://arxiv.org/abs/1109.2585}{{\ttfamily arXiv:1109.2585 [hep-th]}}.
40 pages, 2 figures/ journal version.

\bibitem{index}
J.~Kinney, J.~M. Maldacena, S.~Minwalla, and S.~Raju, ``{An Index for 4
  dimensional super conformal theories},''
  \href{http://dx.doi.org/10.1007/s00220-007-0258-7}{{\em Commun.Math.Phys.}
  {\bfseries 275} (2007) 209--254},
\href{http://arxiv.org/abs/hep-th/0510251}{{\ttfamily arXiv:hep-th/0510251
  [hep-th]}}.

\bibitem{Mandal:2006tk}
G.~Mandal and N.~V. Suryanarayana, ``{Counting 1/8-BPS dual-giants},''
  \href{http://dx.doi.org/10.1088/1126-6708/2007/03/031}{{\em JHEP} {\bfseries
  0703} (2007) 031},
\href{http://arxiv.org/abs/hep-th/0606088}{{\ttfamily arXiv:hep-th/0606088
  [hep-th]}}.

\bibitem{beasley}
C.~E. Beasley, ``{BPS branes from baryons},'' {\em JHEP} {\bfseries 0211}
  (2002) 015,
\href{http://arxiv.org/abs/hep-th/0207125}{{\ttfamily arXiv:hep-th/0207125
  [hep-th]}}.

\bibitem{minwalla}
I.~Biswas, D.~Gaiotto, S.~Lahiri, and S.~Minwalla, ``{Supersymmetric states of
  N=4 Yang-Mills from giant gravitons},''
  \href{http://dx.doi.org/10.1088/1126-6708/2007/12/006}{{\em JHEP} {\bfseries
  0712} (2007) 006},
\href{http://arxiv.org/abs/hep-th/0606087}{{\ttfamily arXiv:hep-th/0606087
  [hep-th]}}.

\bibitem{bdlmo}
A.~Balachandran, B.~P. Dolan, J.-H. Lee, X.~Martin, and D.~O'Connor, ``{Fuzzy
  complex projective spaces and their star products},''
  \href{http://dx.doi.org/10.1016/S0393-0440(02)00020-7}{{\em J.Geom.Phys.}
  {\bfseries 43} (2002) 184--204},
\href{http://arxiv.org/abs/hep-th/0107099}{{\ttfamily arXiv:hep-th/0107099
  [hep-th]}}.

\bibitem{abiy}
G.~Alexanian, A.~Balachandran, G.~Immirzi, and B.~Ydri, ``{Fuzzy CP**2},''
  \href{http://dx.doi.org/10.1016/S0393-0440(01)00070-5}{{\em J.Geom.Phys.}
  {\bfseries 42} (2002) 28--53},
\href{http://arxiv.org/abs/hep-th/0103023}{{\ttfamily arXiv:hep-th/0103023
  [hep-th]}}.

\bibitem{gs99}
H.~Grosse and A.~Strohmaier, ``{Towards a nonperturbative covariant
  regularization in 4-D quantum field theory},''
  \href{http://dx.doi.org/10.1023/A:1007518622795}{{\em Lett.Math.Phys.}
  {\bfseries 48} (1999) 163--179},
\href{http://arxiv.org/abs/hep-th/9902138}{{\ttfamily arXiv:hep-th/9902138
  [hep-th]}}.

\bibitem{saefuztor}
C.~Saemann, ``{Fuzzy toric geometries},''
  \href{http://dx.doi.org/10.1088/1126-6708/2008/02/111}{{\em JHEP} {\bfseries
  0802} (2008) 111},
\href{http://arxiv.org/abs/hep-th/0612173}{{\ttfamily arXiv:hep-th/0612173
  [hep-th]}}.

\bibitem{vaidtriv}
S.~P. Trivedi and S.~Vaidya, ``{Fuzzy cosets and their gravity duals},'' {\em
  JHEP} {\bfseries 0009} (2000) 041,
\href{http://arxiv.org/abs/hep-th/0007011}{{\ttfamily arXiv:hep-th/0007011
  [hep-th]}}.

\bibitem{BBJS}
V.~Balasubramanian, J.~de~Boer, V.~Jejjala, and J.~Simon, ``{Entropy of
  near-extremal black holes in AdS(5)},''
  \href{http://dx.doi.org/10.1088/1126-6708/2008/05/067}{{\em JHEP} {\bfseries
  0805} (2008) 067},
\href{http://arxiv.org/abs/0707.3601}{{\ttfamily arXiv:0707.3601 [hep-th]}}.

\bibitem{malstrom}
J.~M. Maldacena and A.~Strominger, ``{AdS(3) black holes and a stringy
  exclusion principle},'' {\em JHEP} {\bfseries 9812} (1998) 005,
\href{http://arxiv.org/abs/hep-th/9804085}{{\ttfamily arXiv:hep-th/9804085
  [hep-th]}}.

\bibitem{woodhouse}
N.~Woodhouse, {\em {Geometric Quantization}}.
\newblock Clarendon Press (Oxford and New York),
1980.
\newblock

\bibitem{heckver}
J.~J. Heckman and H.~Verlinde, ``{Evidence for F(uzz) Theory},''
  \href{http://dx.doi.org/10.1007/JHEP01(2011)044}{{\em JHEP} {\bfseries 1101}
  (2011) 044},
\href{http://arxiv.org/abs/1005.3033}{{\ttfamily arXiv:1005.3033 [hep-th]}}.

\bibitem{furok}
K.~Furuuchi and K.~Okuyama, ``{D-branes Wrapped on Fuzzy del Pezzo Surfaces},''
  \href{http://dx.doi.org/10.1007/JHEP01(2011)043}{{\em JHEP} {\bfseries 1101}
  (2011) 043},
\href{http://arxiv.org/abs/1008.5012}{{\ttfamily arXiv:1008.5012 [hep-th]}}.

\bibitem{vafaiqmyst}
A.~Iqbal, A.~Neitzke, and C.~Vafa, ``{A Mysterious duality},'' {\em
  Adv.Theor.Math.Phys.} {\bfseries 5} (2002) 769--808,
\href{http://arxiv.org/abs/hep-th/0111068}{{\ttfamily arXiv:hep-th/0111068
  [hep-th]}}.

\bibitem{wiki-simplex}
``Simplex Wikipedia article.''
\newblock \url{http://en.wikipedia.org/wiki/Simplex}.

\bibitem{madore}
J.~Madore, ``{The Fuzzy sphere},''
\href{http://dx.doi.org/10.1088/0264-9381/9/1/008}{{\em Class.Quant.Grav.}
  {\bfseries 9} (1992) 69--88}.

\bibitem{martsparks}
D.~Martelli and J.~Sparks, ``{Dual Giant Gravitons in Sasaki-Einstein
  Backgrounds},'' \href{http://dx.doi.org/10.1016/j.nuclphysb.2006.10.008}{{\em
  Nucl.Phys.} {\bfseries B759} (2006) 292--319},
\href{http://arxiv.org/abs/hep-th/0608060}{{\ttfamily arXiv:hep-th/0608060
  [hep-th]}}.

\bibitem{hmp}
A.~Hamilton, J.~Murugan, and A.~Prinsloo, ``{Lessons from giant gravitons on
  $AdS_{5}\times T^{1,1}$},''
  \href{http://dx.doi.org/10.1007/JHEP06(2010)017}{{\em JHEP} {\bfseries 1006}
  (2010) 017},
\href{http://arxiv.org/abs/1001.2306}{{\ttfamily arXiv:1001.2306 [hep-th]}}.

\bibitem{fulhar}
W.~Fulton and J.~Harris, {\em {Representation theory: a first course}}.
\newblock Springer, 1991.

\bibitem{bhrII}
T.~W. Brown, P.~Heslop, and S.~Ramgoolam, ``{Diagonal free field matrix
  correlators, global symmetries and giant gravitons},''
  \href{http://dx.doi.org/10.1088/1126-6708/2009/04/089}{{\em JHEP} {\bfseries
  0904} (2009) 089},
\href{http://arxiv.org/abs/0806.1911}{{\ttfamily arXiv:0806.1911 [hep-th]}}.

\bibitem{ehs}
Y.~Kimura and S.~Ramgoolam, ``{Enhanced symmetries of gauge theory and
  resolving the spectrum of local operators},''
  \href{http://dx.doi.org/10.1103/PhysRevD.78.126003}{{\em Phys.Rev.}
  {\bfseries D78} (2008) 126003},
\href{http://arxiv.org/abs/0807.3696}{{\ttfamily arXiv:0807.3696 [hep-th]}}.

\bibitem{tom-oneloop}
T.~W. Brown, ``{Permutations and the Loop},''
  \href{http://dx.doi.org/10.1088/1126-6708/2008/06/008}{{\em JHEP} {\bfseries
  0806} (2008) 008},
\href{http://arxiv.org/abs/0801.2094}{{\ttfamily arXiv:0801.2094 [hep-th]}}.

\bibitem{bcv}
D.~Berenstein, D.~H. Correa, and S.~E. Vazquez, ``{A Study of open strings
  ending on giant gravitons, spin chains and integrability},''
  \href{http://dx.doi.org/10.1088/1126-6708/2006/09/065}{{\em JHEP} {\bfseries
  0609} (2006) 065},
\href{http://arxiv.org/abs/hep-th/0604123}{{\ttfamily arXiv:hep-th/0604123
  [hep-th]}}.

\bibitem{sendyons}
A.~Sen, ``{Dyon - monopole bound states, selfdual harmonic forms on the multi -
  monopole moduli space, and SL(2,Z) invariance in string theory},''
  \href{http://dx.doi.org/10.1016/0370-2693(94)90763-3}{{\em Phys.Lett.}
  {\bfseries B329} (1994) 217--221},
\href{http://arxiv.org/abs/hep-th/9402032}{{\ttfamily arXiv:hep-th/9402032
  [hep-th]}}.

\bibitem{wittenbsp}
E.~Witten, ``{Bound states of strings and p-branes},''
  \href{http://dx.doi.org/10.1016/0550-3213(95)00610-9}{{\em Nucl.Phys.}
  {\bfseries B460} (1996) 335--350},
\href{http://arxiv.org/abs/hep-th/9510135}{{\ttfamily arXiv:hep-th/9510135
  [hep-th]}}.

\bibitem{Rychkov:2005nk}
L.~Maoz and V.~S. Rychkov, ``{Geometry quantization from supergravity: The Case
  of 'Bubbling AdS'},''
  \href{http://dx.doi.org/10.1088/1126-6708/2005/08/096}{{\em JHEP} {\bfseries
  0508} (2005) 096},
\href{http://arxiv.org/abs/hep-th/0508059}{{\ttfamily arXiv:hep-th/0508059
  [hep-th]}}.

\bibitem{jevram}
A.~Jevicki and S.~Ramgoolam, ``{Noncommutative gravity from the AdS / CFT
  correspondence},'' {\em JHEP} {\bfseries 9904} (1999) 032,
\href{http://arxiv.org/abs/hep-th/9902059}{{\ttfamily arXiv:hep-th/9902059
  [hep-th]}}.

\bibitem{mathur05review}
S.~D. Mathur, ``{The Fuzzball proposal for black holes: An Elementary
  review},'' \href{http://dx.doi.org/10.1002/prop.200410203}{{\em
  Fortsch.Phys.} {\bfseries 53} (2005) 793--827},
\href{http://arxiv.org/abs/hep-th/0502050}{{\ttfamily arXiv:hep-th/0502050
  [hep-th]}}.

\bibitem{Bissi:2011dc}
A.~Bissi, C.~Kristjansen, D.~Young, and K.~Zoubos, ``{Holographic three-point
  functions of giant gravitons},''
  \href{http://dx.doi.org/10.1007/JHEP06(2011)085}{{\em JHEP} {\bfseries 1106}
  (2011) 085},
\href{http://arxiv.org/abs/1103.4079}{{\ttfamily arXiv:1103.4079 [hep-th]}}.

\bibitem{deBoer:2011zt}
J.~de~Boer, M.~Johnstone, M.~Sheikh-Jabbari, and J.~Simon, ``{Emergent IR Dual
  2d CFTs in Charged AdS5 Black Holes},''
  \href{http://arxiv.org/abs/1112.4664}{{\ttfamily arXiv:1112.4664 [hep-th]}}.
37 page, 3 .eps figures.

\bibitem{LLM}
H.~Lin, O.~Lunin, and J.~M. Maldacena, ``{Bubbling AdS space and 1/2 BPS
  geometries},'' \href{http://dx.doi.org/10.1088/1126-6708/2004/10/025}{{\em
  JHEP} {\bfseries 0410} (2004) 025},
\href{http://arxiv.org/abs/hep-th/0409174}{{\ttfamily arXiv:hep-th/0409174
  [hep-th]}}.

\bibitem{Lunin:2008tf}
O.~Lunin, ``{Brane webs and 1/4-BPS geometries},''
  \href{http://dx.doi.org/10.1088/1126-6708/2008/09/028}{{\em JHEP} {\bfseries
  0809} (2008) 028},
\href{http://arxiv.org/abs/0802.0735}{{\ttfamily arXiv:0802.0735 [hep-th]}}.

\end{thebibliography}\endgroup
\bibliographystyle{utphys}   

\end{document}